%% file: ms.tex
\documentclass{jfm}

\usepackage{graphicx}
\usepackage{xcolor}
\usepackage{natbib}
\usepackage{amsmath}
\usepackage{amssymb}
\usepackage{bm}

\usepackage{floatrow}
\usepackage{subfig}
\newcommand{\ie}{i.e.,\ }
\newcommand{\eg}{e.g.,\ }

\newcommand{\secref}[1]{\mbox{section \ref{#1}}}

\newcommand{\equrefC}[1]{\mbox{Equation (\ref{#1})}}
\newcommand{\equref}[1]{\mbox{equation (\ref{#1})}}

\newcommand{\figrefC}[2][]{\mbox{Figure \ref{#2}(#1)}}
\newcommand{\figref}[2][]{\mbox{figure \ref{#2}(#1)}}
\newcommand{\figrefSC}[1]{\mbox{Figure \ref{#1}}}
\newcommand{\figrefS}[1]{\mbox{figure \ref{#1}}}

\title[Turbulent channel flow of an elastoviscoplastic fluid]
{Turbulent channel flow of an elastoviscoplastic  fluid}

\author[M. E. Rosti, D. Izbassarov, O. Tammisola, S. Hormozi and L. Brandt]{Marco E. Rosti$^1$\thanks{Email address for correspondence: merosti@mech.kth.se}, Daulet Izbassarov$^1$, Outi Tammisola$^1$, \\Sarah Hormozi$^2$ and Luca Brandt$^1$}

\affiliation{$^1$ Linn\'{e} Flow Centre and SeRC, KTH Mechanics, Stockholm, Sweden \\$^2$ Department of Mechanical Engineering, Ohio University, Athens, OH, 45701-2979, USA}

\begin{document}

\maketitle

\begin{abstract}
We present numerical simulations of laminar and  turbulent channel flow of an elastoviscoplastic fluid. The non-Newtonian flow is simulated by solving the full incompressible Navier-Stokes equations coupled with the evolution equation for the elastoviscoplastic stress tensor. The laminar simulations are carried out for a wide range of Reynolds numbers, Bingham numbers and ratios of the fluid and total viscosity, while the turbulent flow simulations are performed at a fixed bulk Reynolds number equal to $2800$ and weak elasticity. We show that in the laminar flow regime the friction factor increases monotonically with the Bingham number (yield stress) and decreases with the viscosity ratio, while in the turbulent regime the the friction factor is almost independent of the viscosity ratio and decreases with the Bingham number, until the flow eventually returns to a fully laminar condition for large enough yield stresses. Three main regimes are found in the turbulent case, depending on the Bingham number: for low values, the friction Reynolds number and the turbulent flow statistics only slightly differ from those of a Newtonian fluid; for intermediate values of the Bingham number, the fluctuations increase and the inertial equilibrium range is lost. Finally, for higher values the flow completely laminarises. These different behaviors are associated with a progressive increases of the volume where the fluid is not yielded, growing from the centerline towards the walls as the Bingham number increases. The unyielded region interacts with the near-wall structures, forming preferentially above the high speed streaks. In particular, the near-wall streaks and the associated quasi-streamwise vortices are strongly enhanced in an highly elastoviscoplastic fluid and the flow becomes more correlated in the streamwise direction.
\end{abstract}

\section{Introduction} \label{sec:introduction}
Many fluids in nature and industrial applications exhibit a non-Newtonian behavior, \ie a non-linear relation between the shear stress and the shear rate, such as shear thinning, shear thickening, yield stress, thixotropic, shear banding and viscoelastic behaviors. Moreover, several non-Newtonian features are often present simultaneously. Here, we focus on elastoviscoplastic fluids, \ie complex non-Newtonian fluids that can exhibit simultaneously elastic, viscous and plastic properties. In particular, they behave as solids when the applied stress is below a certain threshold $\tau_0$, \ie the yield stress, while for stresses above it, they start to flow as liquids. In this context, the aim of this work is to explore and better understand the laminar and turbulent flow of an elastoviscoplastic fluid by means of numerical simulations. Indeed, turbulent flows of elastoviscoplastic fluids occur in many industrial settings, such as petroleum, paper, mining and sewage treatment \citep{hanks_1963a, hanks_1967a, maleki_hormozi_2017a}.

\subsection{Stability of yield stress fluids}
Several studies have been devoted to the stability of yield stress fluids \citep{nouar_frigaard_2001a, metivier_nouar_brancher_2005a, nouar_kabouya_dusek_mamou_2007a, nouar_bottaro_2010a, bentrad_esmael_nouar_lefevre_ait-messaoudene_2017a}. The first study on the stability of viscoplastic fluid flows was reported by \citet{frigaard_howison_sobey_1994a}, who studied the linear stability of a Bingham fluid in a plane channel flow. More recently, \citet{nouar_kabouya_dusek_mamou_2007a} performed a modal and non-modal linear stability analysis of the flow of a Bingham fluid also in a plane channel; these authors showed that the flow is always linearly stable and that the optimal disturbance for moderate/high Bingham number is oblique, \ie not aligned with the Cartesian coordinate axes as in Newtonian fluids. A key results arising from the linear stability analysis is that the regions where the stress is below the yield stress value remain unyielded for linear perturbations, a fact that can lead to interesting mathematical anomalies. For example, \citet{metivier_nouar_brancher_2005a} showed that the critical Reynolds number for linear stability is different when the Bingham number tends to zero, compared to a Newtonian fluid with a null value of Bingham number. Thus, the authors suggest that the passage to the Newtonian limit of a yield stress fluid is ill-defined in terms of stability. Besides linear analysis, fully nonlinear (energy) stability results were derived in \citet{nouar_frigaard_2001a}. These authors showed that the critical Reynolds number for transition increases with the Bingham number; however they also observed that the energy stability results are very conservative. Moreover, since for yield stress fluids the nonlinearity of the problem is not simply in the inertial terms, but also in the shear stress and in the existence of unyielded plug regions, the gap between linear and nonlinear theories is much wider than with Newtonian fluids. While in Newtonian fluids weakly nonlinear theories provide useful insights, in the case of viscoplastic fluids, these methods are algebraically more complicated and only \citet{metivier_nouar_brancher_2010a} has performed this type of analysis for a Rayleigh-Benard-Poiseuille flow finding that the range of validity of an amplitude equation is fairly limited. Only a small number of studies on the stability of more complicated geometry exist: recently, nonlinear (energy) stability analysis has been extended to multi-layer flows of yield stress and viscoelastic fluids by \citet{moyers-gonzalez_frigaard_nouar_2004a, hormozi_frigaard_2012a}. Recently, in order to identify possible paths to transition \citet{nouar_bottaro_2010a} perturbed the base flow slightly, and found that very weak defects are indeed capable to excite exponentially amplified streamwise traveling waves. Finally, \citet{kanaris_kassinos_alexandrou_2015a} performed numerical simulations of a Bingham fluid flowing past a confined circular cylinder to study the viscoplastic effects in the wake-transition regime.

\subsection{Friction losses and drag reduction}
Turbulent flows of generalized Newtonian fluids occur in many industrial process. Despite the numerous applications, it has not been possible to estimate the force needed to drive a complex fluid yet, while in a Newtonian flow the pressure drop can be accurately predicted as a function of the Reynolds number, both in laminar and turbulent flows \citep{pope_2001a}, and for different properties of the wall surface, \eg roughness \citep{orlandi_leonardi_2008a}, porosity \citep{breugem_boersma_uittenbogaard_2006a, rosti_cortelezzi_quadrio_2015a, rosti_brandt_pinelli_2018a} and elasticity \citep{rosti_brandt_2017a}. This is due to the complexity of such flows where additional parameters become relevant, such as the yield stress value (above which the material flows), the relaxation time, the ratio of the solvent to the total viscosity, \ldots; each of these parameters may affect the overall flow dynamics  in different and sometimes surprising ways. Some work has been done on measuring and trying to estimate the hydraulic pressure losses in practical applications \citep{hanks_1963a, hanks_1967a, hanks_dadia_1971a, ryan_johnson_1959a}, with the most popular phenomenological approach suggested by \citet{metzner_reed_1955a}. These authors provide a closure for the pressure drop as a function of a generalized Reynolds number defined using  the local power-law parameters, subsequently extended to yield stress fluids by \citet{pinho_whitelaw_1990a, founargiotakis_kelessidis_maglione_2008a}. \citet{rudman_blackburn_graham_pullum_2004a} performed numerical simulations of a turbulent pipe flow of shear-thinning fluids and compared their results with the pressure drop closure discussed above, finding a decent agreement although with some differences.

There exists a large literature on the turbulent flow with polymer additives, with the main focus being the drag reduction \citep{logan_1972a, pinho_whitelaw_1990a, escudier_presti_1996a, den-toonder_hulsen_kuiken_nieuwstadt_1997a, beris_dimitropoulos_1999a, warholic_massah_hanratty_1999a, escudier_presti_smith_1999a, escudier_smith_2001a, dubief_white_terrapon_shaqfeh_moin_lele_2004a, dubief_terrapon_white_shaqfeh_moin_lele_2005a, escudier_poole_presti_dales_nouar_desaubry_graham_pullum_2005a, escudier_nickson_poole_2009a,  xi_graham_2010a, owolabi_dennis_poole_2017a, shahmardi_zade_ardekani_poole_lundell_rosti_brandt_2018a}. The interested reader is refereed to the work by \citet{berman_1978a} and \citet{white_mungal_2008a} for a through review on the subject.

\subsection{Elastoviscoplastic fluid}
Despite the numerous studies performed to analyze viscoelastic turbulent flows, much less attention has been given to viscoplastic and elastoviscoplastic fluids. Indeed, very few numerical works exist on fully turbulent flows of an elastoviscoplastic fluid, and to the best of our knowledge the only direct numerical simulations of the effect of a yield stress on a turbulent non-Newtonian flow were performed by \citet{rudman_blackburn_2006a} and \citet{guang_rudman_chryss_slatter_bhattacharya_2011a}. These authors simulated a yield-pseudoplastic fluid using the Herschel-Bulkley model and compared the results with experimental measurements. Although qualitative agreement was found, the simulation results strongly over-predict the flow velocity, and the authors were not able to find the source of the discrepancy. Their numerical results suggest that as the yield stress increases, the mean velocity profile deviates more and more from the Newtonian one, and that the turbulent flow will be fully developed only for low values of the yield stress.

Many materials used in experiments, such as Carbopol solutions (\ie a conventional yield stress test fluid) and liquid foams, exhibit simultaneously elastic, viscous and yield stress behavior. Thus, in order to properly predict the behavior of such materials, it is essential to model them as a fully elastoviscoplastic fluid, rather than an ideal yield stress fluid (\eg using the Bingham or Herschel-Bulkley model). Recently, \citet{saramito_2007a} proposed a new constitutive equation for elastoviscoplastic fluid flows, which reproduces a viscoelastic solid for stresses lower then the yield stress, and a viscoelastic Oldroyd-B fluid for stresses higher then the yield stress. Furthermore, in order to describe the yielding process it uses the von Mises yielding criterion, which has been also experimentally confirmed \citep{shaukat_kaushal_sharma_joshi_2012a, martinie_buggisch_willenbacher_2013a}. \citet{cheddadi_saramito_dollet_raufaste_graner_2011a} simulated the inertialess flow of an elastoviscoplastic fluid around a circular object using the model proposed by \citet{saramito_2007a}; these authors were able to capture the fore-aft asymmetry and also the overshoot of the velocity (negative wake) after the circular hindrance, which was previously observed experimentally by \citet{dollet_graner_2007a} for the flow of a liquid foam and by \citet{putz_burghelea_frigaard_martinez_2008a} who related this behaviour to the rheological properties of the fluid. Note that the Bingham model always predicts fore-aft symmetry and the lack of a negative wake, which is in contradiction with the aforementioned experimental observations. Recently, the loss of the fore-aft symmetry and the formation of the negative wake around a single particle sedimenting in a Carbopol solution was captured by the numerical calculations in \citet{fraggedakis_dimakopoulos_tsamopoulos_2016a} using the constitutive law by \citet{saramito_2007a}; their results are in a quantitative agreement with experimental observations obtained with a Carbopol gel.

The model proposed by \citet{saramito_2007a} was extended by the same author to account for shear-thinning effects \citep{saramito_2009a}. The new model combines the Oldroyd viscoelastic model with the Herschel-Bulkley viscoplastic model, with a power law index that allows a shear-thinning behavior in the yielded state. When the index is equal to unity, the model reduces to the one proposed in his previous work, \ie \citet{saramito_2007a}. Apart from the models proposed by Saramito, many others exist in the literature. The interested reader is referred to \citet{crochet_walters_1983a, balmforth_frigaard_ovarlez_2014a, saramito_wachs_2016a, saramito_2016a} for a through review of models and numerical methods.

\subsection{Outline}
In this work, we present the first direct numerical simulations of both laminar and turbulent channel flows of an incompressible elastoviscoplastic fluid. In the laminar regime, a wide range of Reynolds numbers is investigated, while in the turbulent regime, we consider the bulk Reynolds number $Re = 2800$. The non-Newtonian flow is simulated by solving the full unsteady incompressible Navier-Stokes equations coupled with the model proposed by \citet{saramito_2007a} for the evolution of the additional elastoviscoplastic stress tensor. In \secref{sec:formulation}, we first discuss the flow configuration and the governing equations, and then present the numerical methodology used. A validation of the numerical implementation is reported in \secref{sec:validation}, while the results on the laminar and on the fully developed turbulent channel flows are presented in \secref{sec:result}. In particular, we discuss the role of some of the parameters defining the elastoviscoplastic fluid, \ie the Bingham number $Bi$ and the viscosity ratio $\beta$. Finally, a summary of the main findings and conclusions are presented in \secref{sec:conclusion}.

\section{Formulation} \label{sec:formulation}
\begin{figure}
  \centering
  \sidesubfloat[]{\includegraphics[width=0.4\textwidth]{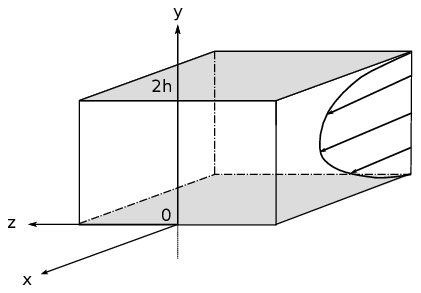} \hspace{0.5cm}}
  \sidesubfloat[]{\includegraphics[width=0.4\textwidth]{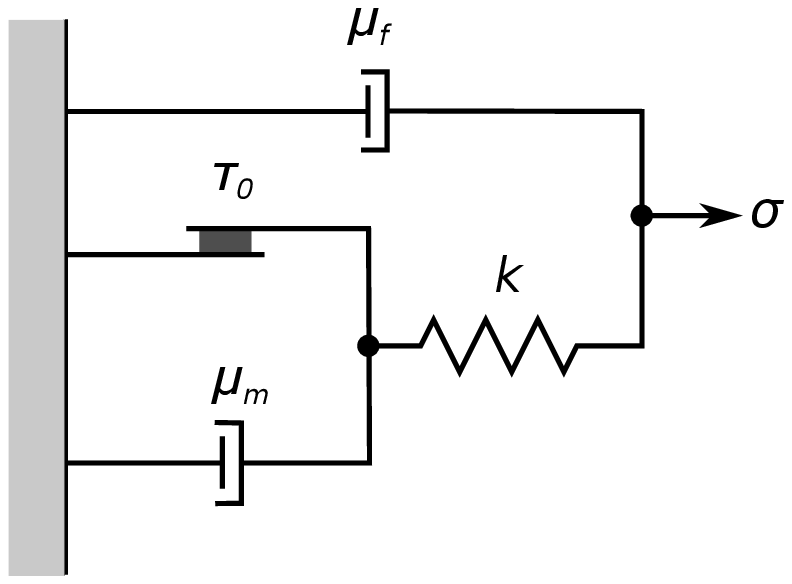} \vspace{0.4cm}} \\
  \caption{(a) Sketch of the computational domain. (b) Sketch of the mechanical model of the elastoviscoplastic fluid proposed by \citet{saramito_2007a} used in the present work.}
  \label{fig:sketch}
\end{figure}
We consider the laminar and turbulent flows of an incompressible elastoviscoplastic fluid through a plane channel with two impermeable rigid walls. \figrefC[a]{fig:sketch} shows a sketch of the geometry and the Cartesian coordinate system, where $x$, $y$ and $z$ ($x_1$, $x_2$, and $x_3$) denote the streamwise, wall-normal and spanwise coordinates, while $u$, $v$ and $w$ ($u_1$, $u_2$, and $u_3$) denote the respective components of the velocity field. The lower and upper stationary impermeable walls are located at $y=0$ and $2h$, respectively, where $h$ represents the channel half height.

The fluid motion is governed by the conservation of momentum and the incompressibility constraint:
\begin{subequations}
\label{eq:NS}
\begin{align}
\frac{\partial u_i}{\partial t} + \frac{\partial u_i u_j}{\partial x_j} &= \frac{1}{\rho} \frac{\partial \sigma_{ij}}{\partial x_j}, \\
\frac{\partial u_i}{\partial x_i} &= 0,
\end{align}
\end{subequations}
where $\rho$ is the fluid density and $\sigma_{ij}$ the total Cauchy stress tensor, which is written as
\begin{equation}
\label{eq:stress}
\sigma_{ij} = -p \delta_{ij} + 2 \mu_f D_{ij} + \tau_{ij},
\end{equation}
where $p$ is the pressure, $\mu_f$ the fluid molecular dynamic viscosity of the fluid (also called solvent viscosity), $\delta$ the Kronecker delta and $D_{ij}$ the strain rate tensor defined as
\begin{equation}
\label{eq:strainrate}
D_{ij}=\frac{1}{2} \left( \frac{\partial u_i}{\partial x_j} + \frac{\partial u_j}{\partial x_i} \right).
\end{equation}
In \equref{eq:stress}, $\tau_{ij}$ is the additional elastoviscoplastic stress tensor which accounts for the non-Newtonian behavior of the fluid, here described by the model proposed by \citet{saramito_2007a}. A $1D$ schematic of the mechanical behavior of the model is shown in \figref[b]{fig:sketch}: when the stress $\sigma$ is below the yield stress $\tau_0$, the friction element is rigid and the system predicts only recoverable Kelvin-Voigt viscoelastic deformation due to the spring $\kappa$ and the viscous element $\mu_f$. When the stress exceeds the yield value $\tau_0$, the friction element breaks and an additional viscous element $\mu_m$ activates; the fluid then behaves as an Oldroyd-B viscoelastic fluid. Thus, the total strain rate $\dot{\varepsilon}$ is shared between an elastic contribution $\dot{\varepsilon}_e$ and a plastic one $\dot{\varepsilon}_p$ \citep{cheddadi_saramito_dollet_raufaste_graner_2011a}. The following limits can be identified: the model reduces to the Oldroyd-B model for $\tau_0 = 0$, the Bingham model is recovered for $\lambda = 0$, and the fluid is Newtonian with a total viscosity $\mu$ equal to $\mu_f + \mu_m$ for $\tau_0 = 0$ and $\lambda = 0$. The instantaneous values of all the components of the stress $\tau_{ij}$ are found by solving the following objective and frame-independent transport equation
\begin{equation}
  \label{eq:EVP}
  \lambda \left( \frac{\partial \tau_{ij}}{\partial t} + \frac{\partial u_k \tau_{ij}}{\partial x_k} - \tau_{kj}\frac{\partial u_i}{\partial x_k} - \tau_{ik}\frac{\partial u_j}{\partial x_k} \right) + \max \left(0, \frac{\vert \tau_d \vert - \tau_0}{\vert \tau_d \vert} \right) \tau_{ij} = 2 \mu_m D_{ij}.
\end{equation}
Here, $\lambda$ is the relaxation time, $\mu_m$ is an additional viscosity, $\tau_0$ the yield stress and $\vert \tau_d \vert$ represents the second invariant of the deviatoric part of the added stress tensor, \ie $\vert \tau_d \vert = \sqrt{1/2 \tau^d_{ij} \tau^d_{ij}}$. Note that, the first four terms in the left hand side of the previous equation are the upper convected derivative of the elastoviscoplastic stress tensor \citep{gordon_schowalter_1972a}. The elastoviscoplastic parameters $\mu_f$, $\mu_m$, $\lambda$ and $\tau_0$ can be obtained by experimental data following the procedure detailed by \citet{fraggedakis_dimakopoulos_tsamopoulos_2016a}, based on the determination of the linear material functions, \ie the storage modulus $G'$ and the loss modulus $G''$.

The previous set of equations can be rewritten in a non-dimensional form as
\begin{subequations}
\label{eq:NSnondim}
\begin{align}
Re \left( \frac{\partial u_i}{\partial t} + \frac{\partial u_i u_j}{\partial x_j} \right) &= \frac{\partial}{\partial x_j} \left( -p \delta_{ij} + 2 \beta D_{ij} + \tau_{ij} \right), \\
\frac{\partial u_i}{\partial x_i} &= 0, \\
Wi \left( \frac{\partial \tau_{ij}}{\partial t} + \frac{\partial u_k \tau_{ij}}{\partial x_k} - \tau_{kj}\frac{\partial u_i}{\partial x_k} - \tau_{ik}\frac{\partial u_j}{\partial x_k} \right) &+ \max \left(0, \frac{\vert \tau_d \vert - Bi}{\vert \tau_d \vert} \right) \tau_{ij} = 2 \left( 1 - \beta \right) D_{ij},
\end{align}
\end{subequations}
where we have used the same symbols to define the non-dimensional variables for simplicity. Four non-dimensional numbers appear in the previous set of equations: the Reynolds number $Re$, the Weissenberg number $Wi$, the Bingham number $Bi$ and the viscosity ratio $\beta$. The Reynolds number is the ratio of inertia and viscous forces $Re = \rho U L / \mu_0$, the Bingham number the ratio of the yield and viscous stresses $Bi = \tau_0 L / \mu_0 U$, the Weissenberg number the ratio of the elastic and viscous forces $Wi = \lambda U / L$ \citep{poole_2012a}, and the viscosity ratio $\beta=\mu_f/\mu_0$ the ratio between the fluid viscosity $\mu_f$ and the reference one $\mu_0$. In the previous definitions, $U$ and $L$ are a characteristic velocity and length scales of the flow, $\rho$ the fluid density and $\mu_0$ a characteristic viscosity, set equal to the total viscosity, \ie $\mu_0=\mu_f+\mu_m$. Note that, the choice of the characteristic viscosity is an open topic of discussion in the community, with the most common choice being the total viscosity $\mu_0$.

\subsection{Numerical discretisation}
The equations of motion are solved with an extensively validated in-house code \citep{picano_breugem_brandt_2015a, rosti_brandt_2017a, rosti_brandt_mitra_2018a, rosti_brandt_2018a}. \equrefC{eq:NS} and \equref{eq:EVP} are solved on a staggered uniform grid with velocities located on the cell faces and all the other variables (pressure, stress and material component properties) at the cell centers.  All the spatial derivatives are approximated with second-order centered finite differences except for the advection term in \equref{eq:EVP} where the fifth-order WENO (weighted essentially non-oscillatory) scheme is adopted \citep{shu_2009a, sugiyama_ii_takeuchi_takagi_matsumoto_2011a}. The time integration is performed with a fractional-step method \citep{kim_moin_1985a}, where all the terms in the evolution equations are advanced in time with a third-order explicit Runge-Kutta scheme except for the elastoviscoplastic stress terms which are advanced with the Crank–Nicolson method; moreover, a Fast Poisson Solver is used to enforce the condition of zero divergence for the velocity field. Note that, the choice of an explicit time integration is typically preferred  for high Reynolds number turbulent flows. In particular, to solve the system of governing equations, we perform the following steps \citep[see also][]{dubief_terrapon_white_shaqfeh_moin_lele_2005a, min_yoo_choi_2001a}: i) the elastoviscoplastic stress tensor $\tau_{ij}$ is updated by solving \equref{eq:EVP}; ii) the NS equations (\equref{eq:NS}) are advanced in time by first solving the momentum equation (prediction step), then by solving a Poisson equation for the projection variable, and finally by correcting the velocity and pressure to make the velocity field divergence free (correction step).

\subsection{Code validation} \label{sec:validation}
\begin{figure}
  \centering
  \input{saramitoS}
  \input{saramitoO} \vspace{0.4cm} \\
  \caption{(a) Time evolution of $\tau_{11}-\tau_{22}$ (red) and $\tau_{12}$ (blue) in a stationary shear flow. The stress components are normalised with $\mu_0 \dot{\gamma}$. (b) Time evolution of the shear stress $\tau_{12}$ in an oscillating shear flow with $Bi=0$ (red) and $Bi=300$ (blue). The stress components are normalised with $\mu_0 \gamma_0 \omega_0$. In both panels, solid lines are used for our numerical results and symbols for the analytical solution reported by \citet{saramito_2007a}.}
  \label{fig:val}
\end{figure}
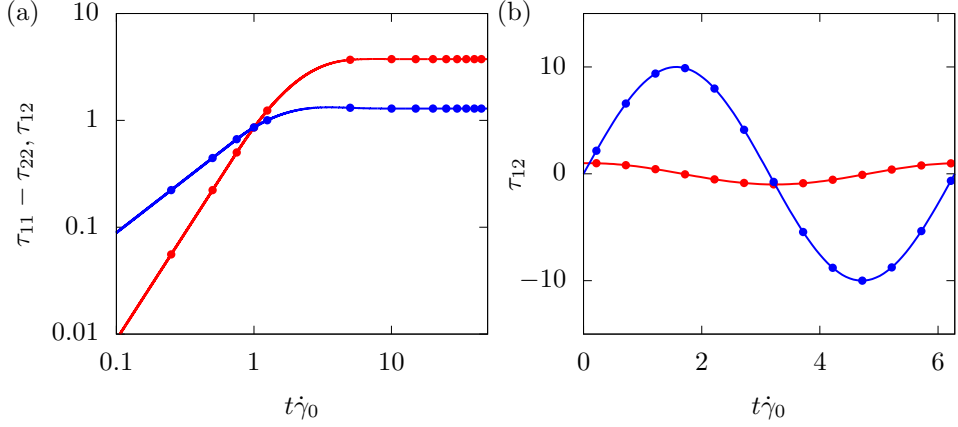

The present implementation for single and multiphase flows of an elastoviscoplastic fluid has been extensively validated in \citet{izbassarov_rosti_niazi-ardekani_sarabian_hormozi_brandt_tammisola_2018a}, where the details of the algorithm are discussed in further detail. Nonetheless, we report here two validation cases  for the sake of completeness.

First, we consider a simple constant shear flow, with the shear rate $\dot{\gamma}_0$: the Weissenberg number is fixed to $Wi=\lambda \dot{\gamma}_0=1$, the Bingham number $Bi=\tau_0/(\mu_0 \dot{\gamma})=1$ and the viscosity ratio $\beta=1/9$. The time evolution of $\tau_{11} - \tau_{22}$ (the normal stress difference) and $\tau_{12}$ (the wall-normal shear stress) are reported in \figref[a]{fig:val} with red and blue lines, respectively. We observe that, initially both the stress components grow linearly, but when the stress level is above a threshold, \ie the yield stress, the growth stops and they reach a plateau, as expected in the yielded state. As shown in the figure, we find a very good agreement with the analytical results by \citet{saramito_2007a} depicted with symbols of the same colors.

Next, we consider a time-periodic uniform shear flow, \ie $\gamma_0 sin(\omega_0 t)$,  where $\gamma_0$ is the strain amplitude and $\omega_0$ the angular frequency of the oscillations. The Weissenberg number is $Wi=\lambda \omega_0=0.1$ and two Bingham numbers $Bi= \tau_y/(\mu_0 \gamma_0 \omega_0)$ are considered: $Bi=0$ and $300$. Note that, when $Bi=0$, the material behaves like a viscoelastic fluid, and when $Bi=300$ as an elastic solid. The viscosity ratio $\beta$ is null in both cases, \ie $\mu_f=0$. The evolution of $\tau_{12}$ is plotted in \figref[b]{fig:val} for the two cases (red $Bi=0$ and blue $Bi=300$) and compared with the analytical solution provided by \citet{saramito_2007a}, shown with symbols in the figure. Again, an excellent agreement is found.

\subsection{Numerical set-up}
For all the cases considered hereafter, the equations of motion are discretised by using $1728 \times 576 \times 864$ grid points on a computational domain of size $6h \times 2h \times 3h$ in the streamwise, wall-normal and spanwise directions. The spatial resolution has been chosen in order to properly resolve the turbulent scales, as well as the unyielded plug regions which form intermittently in the domain. In the high Reynolds number simulations at $Re_b=2800$, the resolution satisfies the constraint $\Delta x^{+} = \Delta y^{+} = \Delta z^{+} < 0.6$, where the superscript $^+$ indicates the wall units defined in the next section. In one of the simulations at $Re_b=2800$, a grid refinement study was performed using $2160 \times 720 \times 1080$ grid points in the streamwise, wall-normal and spanwise directions (around $25\%$ more in each direction); the difference in the resulting friction coefficient $C_f$ was less than $2\%$. Note that, in the low Reynolds fully laminar cases, the spatial resolution was relaxed and the domain size in the homogeneous directions reduced.

In all the simulations, periodic boundary conditions are used in the streamwise and spanwise directions, while the no-slip and no-penetration boundary conditions are enforced on the solid walls. All the turbulent flows are initialized with a fully developed channel flow with zero elastoviscoplastic added stress ($\tau_{ij}=0$). After the flow has reached statistically steady state, the calculations are continued for an interval of $500 h/U_b$ time units, during which around $100$ full flow fields are stored for further statistical analysis. To verify the convergence of the statistics, we have computed them using a different number of samples and verified that the differences are negligible.

\section{Results} \label{sec:result}
We study both laminar and turbulent channel flows of an elastoviscoplastic fluid, together with the baseline Newtonian cases. All the simulations are performed at a constant flow rate, so that the flow Reynolds number based on the bulk velocity is fixed, \ie $Re=\rho U_b h/\mu_0$, where the bulk velocity $U_b$ is the average value of the mean velocity computed across the whole domain and $\mu_0$ is the total viscosity, \ie $\mu_0=\mu_f+\mu_m$. In the present work, consistently with choosing $U_b$ as the characteristic velocity, we opt for enforcing the constant flow rate condition; hence, the necessary value of the instantaneous streamwise pressure gradient is determined at every time step. This choice facilitates the comparison between the non-Newtonian and Newtonian flows. In the laminar regime, the bulk Reynolds number is varied between $0.1$ and $2800$, where the corresponding baseline Newtonian solutions are known analytically; in the turbulent regime, the bulk Reynolds number is fixed to $2800$, corresponding to a nominal friction Reynolds number $Re_\tau=\rho u_\tau h/ \mu_0=180$ for a Newtonian fluid, being $u_\tau$ the friction velocity defined later on. In the turbulent case, we compare our Newtonian solution with the seminal work of \citet{kim_moin_moser_1987a}.

The properties of the elastoviscoplastic  fluids are chosen as follows: the Weissenberg number $Wi=\lambda U_b/h$ is fixed in all the simulations to $0.01$ in order to limit the role of fluid elasticity in this first study of elastoviscoplastic flows; the Bingham number $Bi=\tau_0 h/ \mu_0 U_b$ is varied in the range between $0$ and $1000$ in the laminar cases ($0$, $0.1$, $1$, $10$, $100$ and $1000$), and between $0$ and $2.8$ ($0$, $0.28$, $0.7$, $1.4$ and $2.8$) in the turbulent cases, which are computationally significantly more expensive than Newtonian turbulence. Finally, all the cases have been studied for three different viscosity ratios: $\beta=\mu_f / \mu_0 =0.25$, $0.5$ and $0.95$. Overall, we have performed $108$ laminar and $15$ turbulent simulations.

Viscous units, used above to express the spatial resolution, will be often employed in the following; they are indicated by the superscript $^+$, and are built using the friction velocity $u_\tau$ as the velocity scale and the viscous length $\delta_\nu = \nu / u_\tau$ as the length scale. For a Newtonian turbulent channel flow, the dimensionless friction velocity is defined as
\begin{equation} 
\label{eq:friction_velocity}
u_\tau = \sqrt{\frac{1}{Re_b} \left. \frac{d \overline{u}}{dy} \right\vert_{y=0}},
\end{equation}
where $\overline{u}$ is the mean velocity, and the derivative is taken at $y=0$, the location of the wall. When the fluid is non-Newtonian, \equref{eq:friction_velocity} must be modified to account for the elastoviscoplastic shear stress that is in general non-zero at the wall. Similarly to previous works with polymers \citep{shahmardi_zade_ardekani_poole_lundell_rosti_brandt_2018a}, we define
\begin{equation} \label{eq:friction_velocity_total}
u_\tau = \sqrt{ \left. \left( \frac{1}{Re_b} \frac{d \overline{u}}{d y} + \overline{\tau}_{12} \right) \right\vert_{y=0}}.
\end{equation}
Note that, the actual value of the friction velocity in our simulations is computed from the friction coefficient, found by the driving streamwise pressure gradient, rather than from its definition, \ie $u_\tau = \sqrt{- \delta/\rho~d\overline{p}/dx}$.

\subsection{Laminar flow}
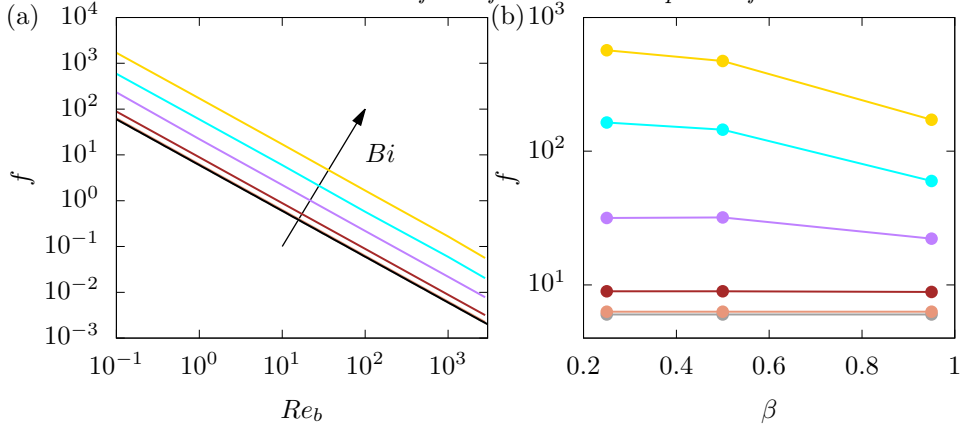
\begin{figure}
  \centering
  \input{fan}
  \input{fanB} \vspace{0.4cm} \\
  \caption{Fanning friction factor $f$ as a function of (a) the bulk Reynolds number $Re_b$ for $\beta=0.95$ and (b) the viscosity ratio $\beta$ for $Re=1$, for different Bingham numbers $Bi$. In all the elastoviscoplastic cases, the Weissenberg number is fixed to $Wi=0.01$. Grey, orange, brown, purple, cyan and gold colors are used for $Bi=0$, $0.1$, $1$, $10$, $100$ and $1000$, respectively, and the black line is the Newtonian analytical solution. Note that, the lines in the graphs are simple connections between the available data points.}
  \label{fig:laminar1}
\end{figure}

We start our analysis by considering the laminar flow of an elastoviscoplastic fluid. First, we consider the effect of the Bingham number on the frictional resistance of the flow quantified by the Fanning friction factor $f$, defined as $2 \tau_w/\rho U_b^2$ being $\tau_w$ the total wall shear stress including both the viscous and elastoviscoplastic contributions. \figrefC[a]{fig:laminar1} shows the Fanning friction factor $f$ as a function of the Reynolds number in the case with $\beta=0.95$, $Wi=0.01$, and for various Bingham numbers. In particular, the grey, orange, brown, purple, cyan and gold lines are used for $Bi=0$, $0.1$, $1$, $10$, $100$ and $1000$, respectively. In the figure we also show the Newtonian analytical solution $f=6/Re_b$ with a black line. The results clearly show that all the non-Newtonian fluids have the same slopes as the reference Newtonian case, but with increasing $f$ as the Bingham number $Bi$ increases. As expected, the case with $Bi=0$ (grey line) is almost indistinguishable from the Newtonian flow, since the elastic effects are small for the low value of the Weissenberg number $Wi$ chosen. The results shown here are consistent with the experimental measurements reported by \citet{guzel_frigaard_martinez_2009a}, who also found a linear relation between the Reynolds number and the friction factor in a laminar pipe flow.

\figrefC[b]{fig:laminar1} shows the effect of $\beta$ on the Fanning friction factor $f$ at $Re_b=1$. We find that $f$ decreases non-linearly with the viscosity ratio $\beta$, and that the dependency on $\beta$ increases with the Bingham number $Bi$. The increase in the friction factor, due to the increase of wall shear stress, comes from the change of the laminar streamwise velocity profile $u$ shown in \figref[a]{fig:laminar2} as a function of the wall-normal distance $y$, with $u$ and $y$ being normalized with the bulk velocity $U_b$ and $h$, respectively. 
Again, we observe that the viscoelastic flow ($Bi=0$) almost perfectly overlaps with the Newtonian solution due to the very low Weissenberg number considered in this study. As expected, as the Bingham number $Bi$ increases, we note the appearance of a region in the middle of the channel with a uniform velocity, \ie a plug is formed away from the walls flowing with uniform velocity; this corresponds to the region where the fluid is not yielded and behaves as an elastic solid. 
Consequently, the centerline velocity $U_c=u(y=h)$ reduces and the wall shear increases for mass conservation. The volume of the unyielded fluid, denoted $Vol_s$, grows with the Bingham number $Bi$ from $0\%$ for $Bi=0$ up to $87\%$ for $Bi=1000$, as shown in \figref[b]{fig:laminar2}.

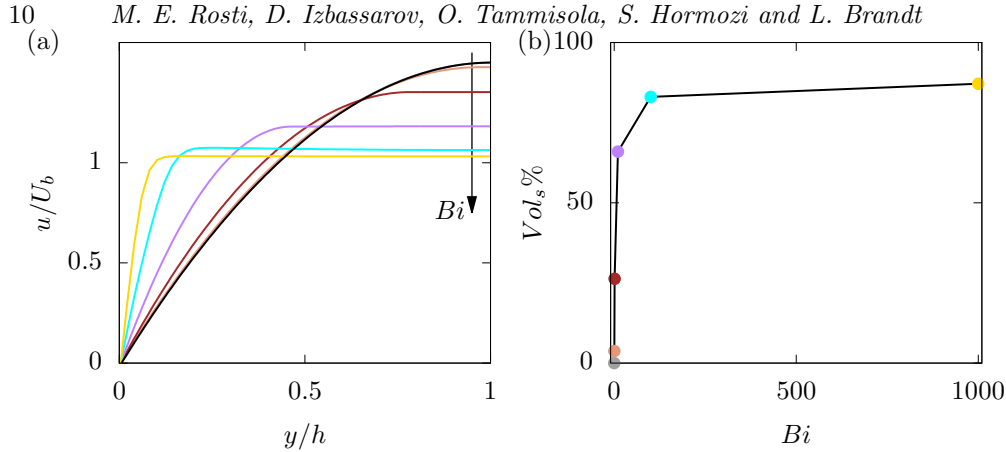
\begin{figure}
  \centering
  \input{meanL} \hspace{0.2cm}
  \input{volL} \vspace{0.4cm} \\
  \caption{(a) Mean streamwise velocity profile $\overline{u}$ as a function of the wall-normal distance $y$. (b) Percentage of the unyielded volume $Vol_s$ as a function of the Bingham number $Bi$. The Reynolds number is equal to $1$, and the color scheme is the same as in \figrefS{fig:laminar1}. Note that, the lines in the graphs are simple connections between the available data points.}
  \label{fig:laminar2}
\end{figure}

Finally, we provide a fit to our numerical data for the Fanning friction factor $f$ (\figrefS{fig:laminar1}). In general, the Fanning friction factor $f$ of an elastoviscoplastic fluid in a channel flow is a function of inertia ($Re_b$), elasticity ($Wi$), plasticity ($Bi$) and viscosity ratio $\beta$, \ie $f = \mathcal{F}(Re, Wi, Bi, \beta)$. In our study the Weissenberg number is fixed to a very low value ($Wi=0.01$), thus we drop its dependency. A very good agreement with our data is found when using the following expression
\begin{equation}
  \label{eq:fit}
  f = \frac{6 + \mathcal{C} \sqrt{Bi}}{Re_b},
\end{equation}
where $\mathcal{C}$ is a fit parameter which depends on $\beta$: $\mathcal{C}=2.47$ for $\beta=0.25$, $7.55$ for $\beta=0.5$ and $8.86$ for $\beta=0.95$. \equrefC{eq:fit} clearly recovers the Newtonian analytical solution for $Bi=0$, and provides an error below $2\%$ for all the elastoviscoplastic results. It is worth noticing, that an analogous expression was found by \citet{de-vita_rosti_izbassarov_duffo_tammisola_hormozi_brandt_2018a} for the flow of an elastoviscoplastic fluid through a porous media, with the same dependency on $Re_b$ and $Bi$.

\subsection{Turbulent flow}
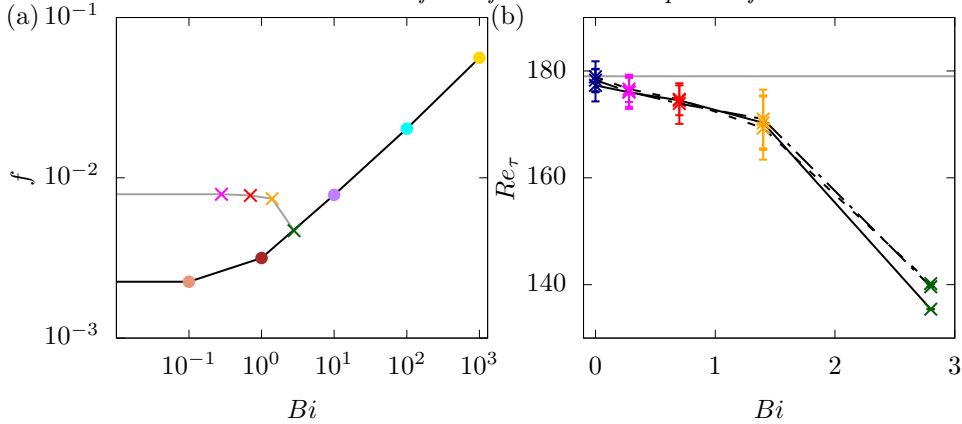
\begin{figure}
  \centering
  \input{fan2800}
  \input{reynolds} \vspace{0.4cm} \\
  \caption{(a) Fanning friction factor $f$ as a function of the Bingham number $Bi$  for $Re_b=2800$ and $\beta=0.95$. The points on the black line correspond to laminar flows, while those on the grey line to turbulent flows. (b) The friction Reynolds number $Re_\tau$ as a function of the Bingham number $Bi$, with the vertical errorbars measuring the variance of $Re_\tau$. The dashed, dash-dotted and solid lines are used for $\beta=0.25$, $0.5$ and $0.95$, respectively. The Reynolds number is equal to $Re_b=2800$ for all cases. The blue, magenta, red, orange and green colors are used for the turbulent cases with $Bi=0$, $0.28$, $0.7$, $1.4$ and $2.8$, respectively, while the color scheme for the laminar results is the same as in \figrefS{fig:laminar1}.}
  \label{fig:turbF}
\end{figure}

Next, we examine the turbulent flow cases, all at a fixed bulk Reynolds number $Re_b=2800$ and Weissenberg number $Wi=0.01$. The turbulent purely viscoelastic flow with $Bi=0$ has a Fanning friction factor $f$ higher than its laminar counterpart, raising by $300\%$ from $0.002$ in the laminar case to $0.008$ in the turbulent one, as shown in \figref[a]{fig:turbF} for the case with $\beta=0.95$. As the Bingham number increases, the Fanning friction factor progressively decreases, with an opposite trend than in the laminar elastoviscoplastic cases. The decrease of $f$ is small for the two lowest $Bi$ ($-1.4\%$ for $Bi=0.28$ and $-3.1\%$ for $Bi=0.7$), moderate for the intermediate $Bi$ ($-7.7\%$ for $Bi=1.4$), and large for the highest $Bi$ ($-41\%$ for $Bi=2.8$). For the highest $Bi$ considered here, the turbulence cannot be sustained and hence $f$ reaches its laminar value.

The friction Reynolds number $Re_\tau$ is depicted in \figref[b]{fig:turbF} as a function of the Bingham number. Again, we observe a progressive decrease of $Re_\tau$ with $Bi$, corresponding to a net drag reduction when compared to a Newtonian turbulent channel flow at the same flow rate (horizontal grey line). The error bar in the figure represents the variance of the value, which is initially small, then grows suddenly for $Bi=1.4$ (as discussed later), and finally becomes null for $Bi=2.8$. The different line styles used in \figref[b]{fig:turbF} correspond to different values of the viscosity ratio $\beta$. We observe that, in the turbulent regime the results are almost independent of the viscosity ratio, both in terms of mean and r.m.s.~values. To summarize, we identify three different regimes: \textit{i)} for low Bingham numbers ($\lesssim 1$) the friction Reynolds number decreases slowly, approximately linearly, with approximately constant r.m.s. values; \textit{ii)} for intermediate values, $Re_\tau$ decreases more than linearly and its r.m.s. increases; \textit{iii)} for high Bingham numbers ($\gtrsim 2$) the flow becomes stationary and fully laminar. Interestingly,  these three separate regimes are found to be independent of the value of the viscosity ratio $\beta$.

We can define a Bingham number in wall units, \ie $Bi^+$, as the ratio between the yield stress $\tau_0$ and the wall shear stress $\tau_w$ as $Bi^+ = \tau_0/\tau_w$: from the results of our simulations we found that $Bi^+=0$, $0.025$, $0.064$, $0.135$ and $0.425$ for the turbulent simulations at $Re=2800$ with $Bi=0$, $0.28$, $0.7$, $1.4$ and $2.8$. Based on the values of $Bi^+$ we have available, we can infer that the first regime depicted above holds for yield stress values that are below $6\%$ the wall shear stress value, while the third regime holds for yield stress values above $42\%$ the wall shear stress.

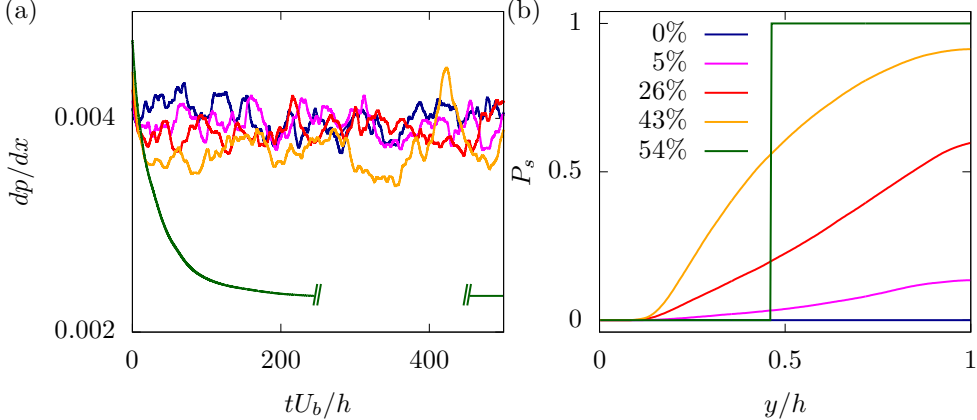
\begin{figure}
  \centering
  \input{dpdx}
  \input{volPDF} \vspace{0.4cm} \\
  \caption{(a) Time history of the streamwise pressure drop $dp/dx$ for different Bingham numbers $Bi$. (b) Probability of the fluid to be unyielded $P_s$ as a function of the wall-normal distance $y$. The percentages reported in the legend show the mean unyielded volume $Vol_s$. The blue, magenta, red, orange and green colors are used for $Bi=0$, $0.28$, $0.7$, $1.4$ and $2.8$, respectively. The viscosity ratio $\beta$ is fixed equal to $0.95$.}
  \label{fig:turbDpDx}
\end{figure}

The time history of the instantaneous pressure drop along the channel $dp/dx$ is shown in \figref[a]{fig:turbDpDx}. This quantity represents the forcing term needed to drive the flow, which in the turbulent regime oscillates around a mean value in order to maintain a constant flow rate in the domain. We observe that for the low Bingham cases ($Bi=0$, $0.28$ and $0.7$) the time histories of $dp/dx$ are very similar, with only slightly different mean values; on the other hand, for $Bi=1.4$ the mean value is further decreased while the amplitude of the oscillations increases. Finally, for $Bi=2.8$ the pressure drop smoothly decays from the turbulent value imposed as initial condition to the final laminar value. Thus, the figure clearly confirms the differences between the three regimes highlighted above.

\begin{figure}
  \centering
  \includegraphics[width=0.6\textwidth]{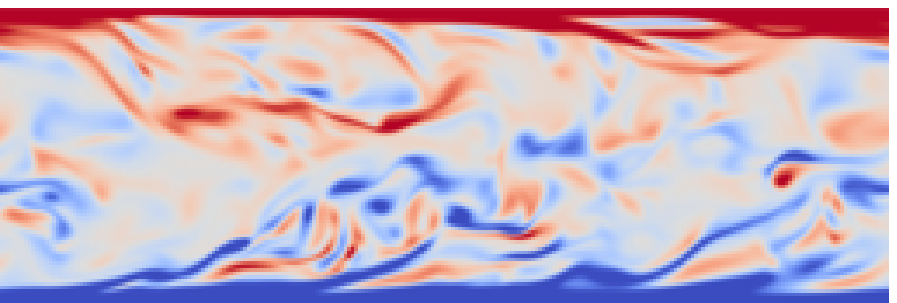}
  \includegraphics[width=0.3\textwidth]{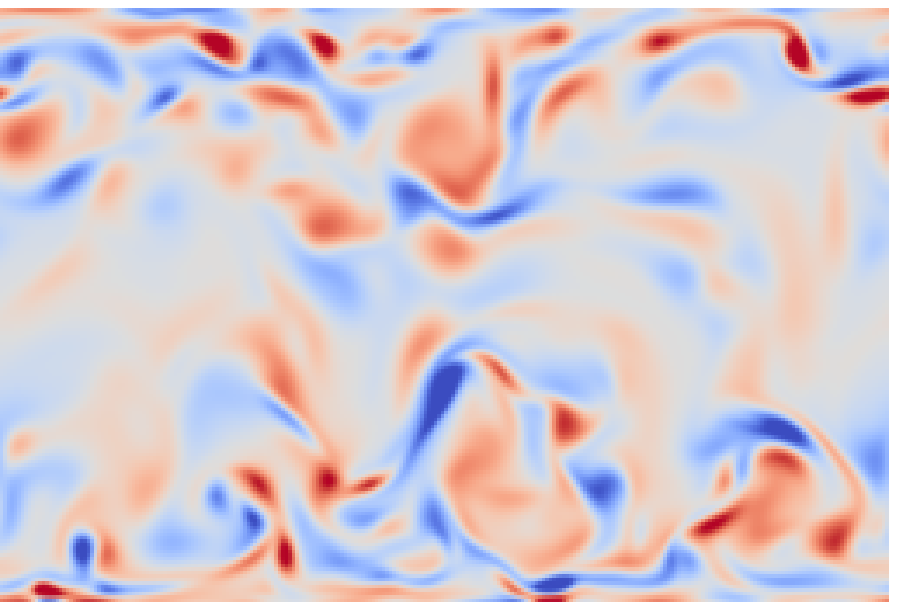} \vspace{0.08cm} \\
  \includegraphics[width=0.6\textwidth]{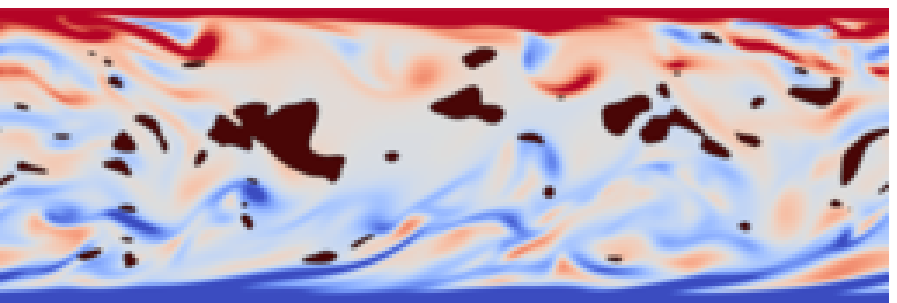}
  \includegraphics[width=0.3\textwidth]{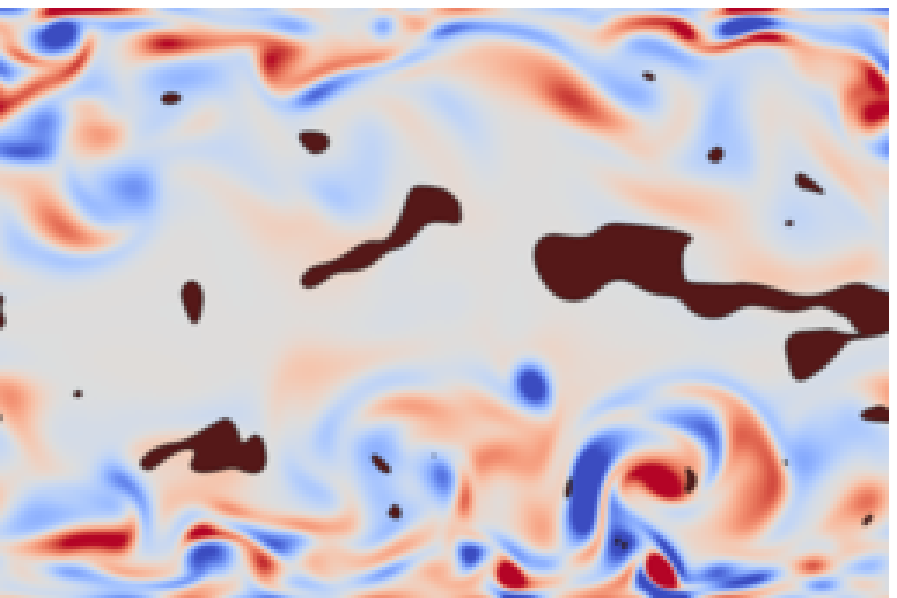} \vspace{0.08cm} \\
  \includegraphics[width=0.6\textwidth]{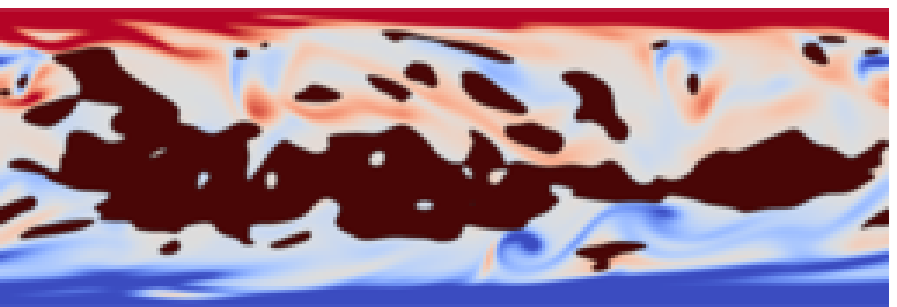}
  \includegraphics[width=0.3\textwidth]{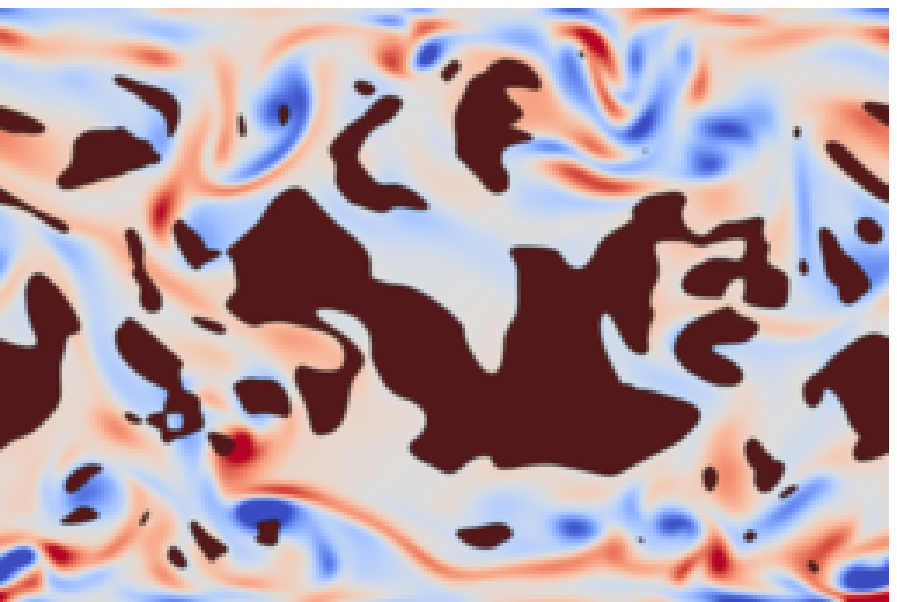} \vspace{0.08cm} \\
  \includegraphics[width=0.6\textwidth]{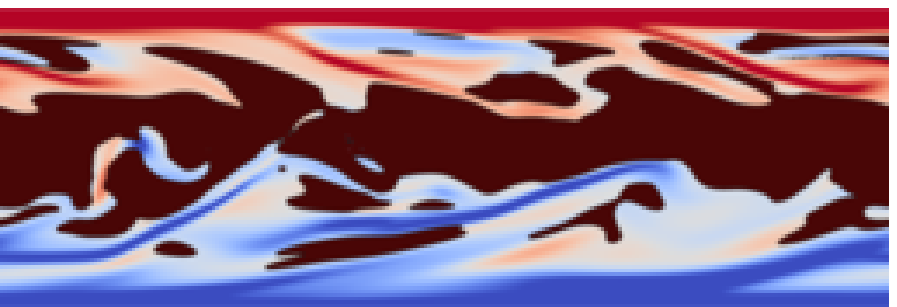}
  \includegraphics[width=0.3\textwidth]{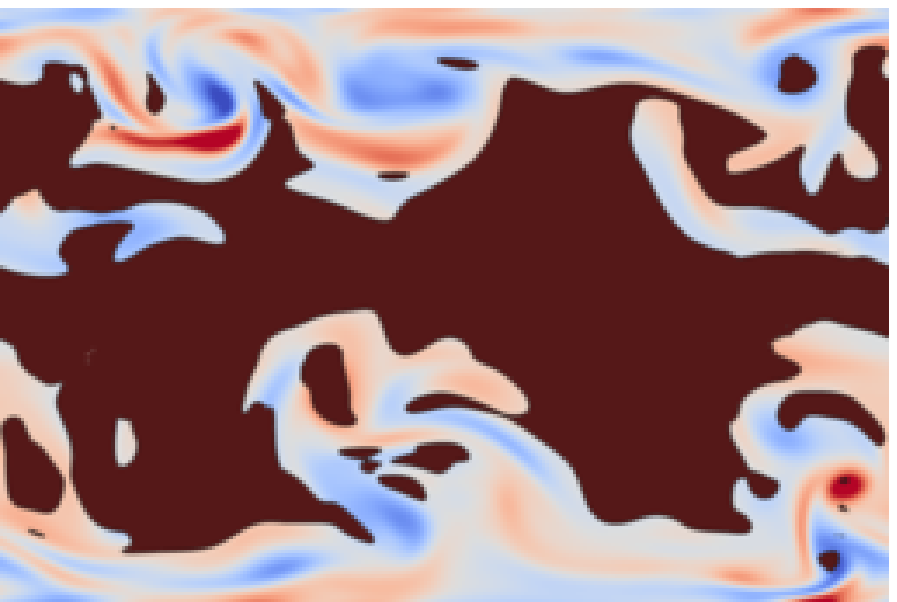} \vspace{0.08cm} \\
  \includegraphics[width=0.6\textwidth]{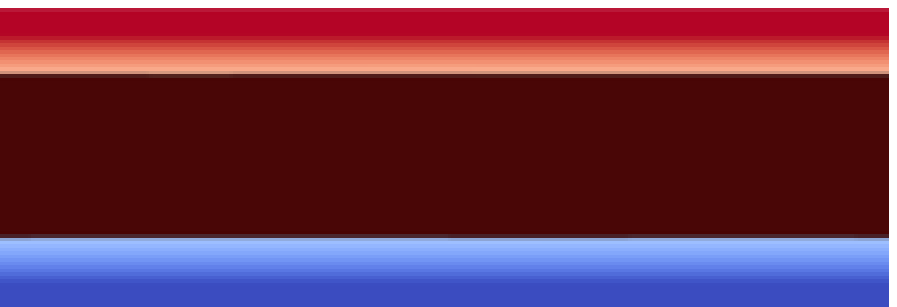}
  \includegraphics[width=0.3\textwidth]{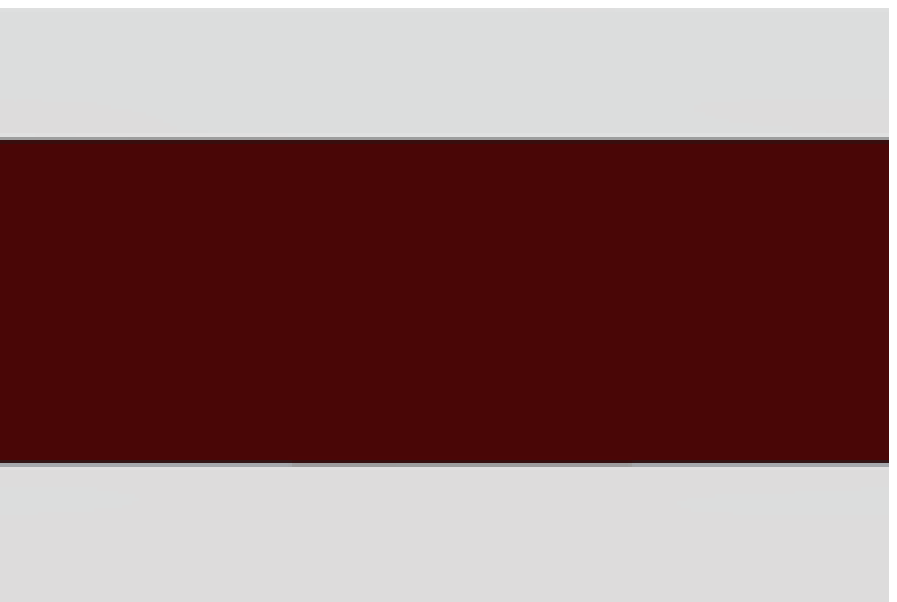} \vspace{0.4cm}
  \caption{Contours of the instantaneous spanwise vorticity $-\omega_z$ in a $x-y$ plane (left column) and of the streamwise vorticity $\omega_x$ in a $y-z$ plane (right column). Color scales ranges from $-3U_b/h$ (blue) to $3U_b/h$ (red). The brown areas represent the instantaneous regions where the flow is not yielded. The Bingham number $Bi$ increases from top to bottom ($Bi=0$, $0.28$, $0.7$, $1.4$ and $2.8$) and the viscosity ratio $\beta$ is fixed equal to $0.95$.}
  \label{fig:turbYielded2d}
\end{figure}

In order to understand the physical origin of the three regimes, we start by showing visualizations of the  instantaneous distributions of the regions where the flow is yielded and not yielded, see \figrefS{fig:turbYielded2d}. In the wall-normal and cross-stream planes in the figure we also report  color contours of the spanwise (left column corresponding to a $x-y$ plane) and  streamwise vorticity (right column displaying $y-z$ planes), $\omega_z$ and $\omega_x$. At $Bi=0$ we recognize the classic vorticity field of turbulent channel flows, with high vorticity levels at the walls, and the footprints of the classical turbulent streaky structures. For nonzero Bingham numbers, we see the appearance of unyielded regions - shown in brown - around the center of the channel and far from the walls. These regions are mostly disconnected and with a limited spanwise length for the two lowest $Bi$ (corresponding to the first regime), while for $Bi=1.4$ and $2.8$ the unyielded region extends over the full streamwise and spanwise directions. The bottom row of the figure clearly shows that the flow is fully laminar.

\begin{figure}
  \centering
  \input{inter071}
  \input{inter141}
  \input{inter211}
  \input{inter281} \vspace{0.4cm} \\
  \caption{Intermittency of the yield/unyield process (F/S) as a function of time. The four panels correspond to probes located at $x=3h$, $z=1.5h$ and different wall-normal distances: (a) $y\approx0.25h$, (b) $0.49h$, (c) $0.74$ and (d) $0.98h$. The lines in every panel are the results with different Bingham number, with the color scheme being the same as in \figrefS{fig:turbDpDx}.}
  \label{fig:inter}
\end{figure}
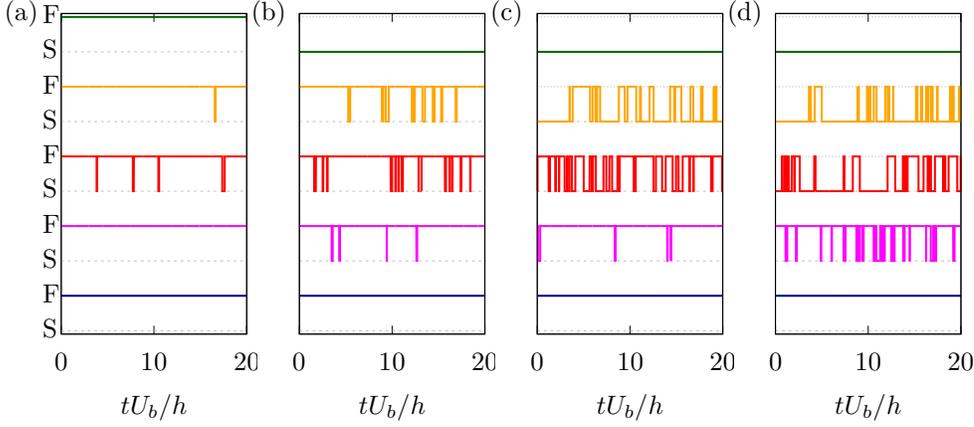

\figrefSC{fig:inter} shows the instantaneous state of the fluid: S indicates an unyielded fluid and F a yielded one. This information is extracted by various numerical probes at different wall-normal locations. In particular, we show in the figure the results for four different wall-normal distances, \ie $y\approx0.25h$, $0.49h$, $0.74$ and $0.98h$. The intermittent nature of the flow is evident; also it is clear that the fluid close to the wall is preferentially yielded, while close to the centerline it is mostly not yielded, even at low Bingham numbers.

As clearly indicated by the previous pictures, the flow and the yield/unyield process are inherently unsteady, thus we need to analyze the phenomenon in statistical terms. Going back to \figref[b]{fig:turbDpDx}, we display the probability to have an unyielded region $P_s$ as a function of the wall-normal distance $y$, together with the mean percentage of the unyielded volume $Vol_s$ reported in the legend. The value $P_s=1$ indicates a location where the material behaves as an elastic solid throughout the computational time, whereas the material behaves uniquely as a viscoelastic (Oldroyd-B) fluid when $P_s=0$. For $Bi=0$, we clearly have no solid anywhere, while as $Bi$ increases, the probability of the fluid to be not yielded increases around the centerline, while still remains null in the near wall region. Finally, for $Bi=2.8$ when the flow is fully laminar, the probability of being yielded or unyielded is either $0$ or $1$, with $54\%$ of the total volume being not yielded. Note that, even for the Bingham number $Bi=1.4$, representative of the second regime, the probability to be unyielded in the middle of the channel is not exactly equal to unity. Indeed, as also shown in the third row of \figrefS{fig:turbYielded2d}, instantaneous region where the fluid is yielded can appear around the centerline, thus decreasing the overall percentage. In particular, for $Bi=0.28$ the probability of the fluid to be yielded at the centerline is around $85\%$, for $Bi=0.7$ is $40\%$ and for $Bi=1.4$ is $8\%$. This effect contributes to the  highly unsteady and intermittent behavior discussed in relation to the pressure drop and friction factor.

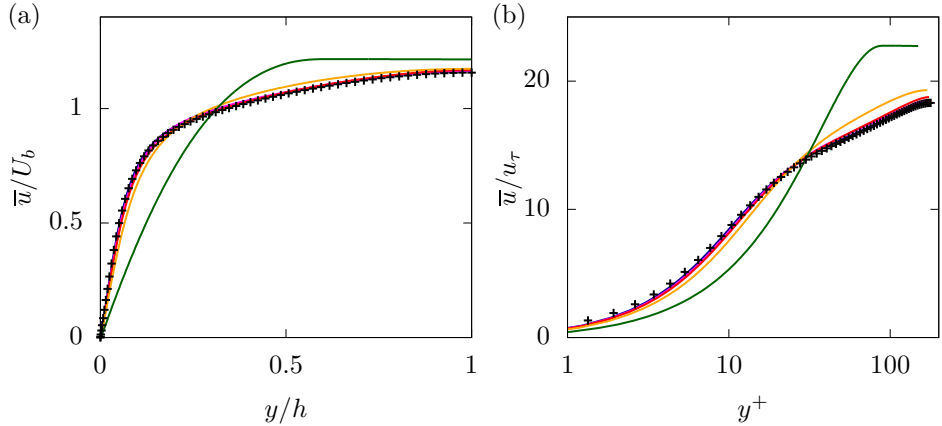
\begin{figure}
  \centering
  \input{mean}
  \input{meanLOG} \vspace{0.4cm} \\
  \caption{Mean streamwise velocity profile $\overline{u}$ as a function of the wall-normal distance $y$ in bulk (a) and wall units (b). The color scheme is the same as in \figrefS{fig:turbDpDx}, with the addition of the black symbols used for the Newtonain case, taken from the results by \citet{kim_moin_moser_1987a}. The viscosity ratio $\beta$ is fixed equal to $0.95$.}
  \label{fig:turbU}
\end{figure}

We now proceed by presenting the main flow statistics. \figrefSC{fig:turbU} shows the mean streamwise velocity component $\overline{u}$ as a function of the wall normal distance $y$. In the left panel of the figure (in bulk units) we can again find the three different behaviors described above. Up to $Bi \approx 1$, the profiles are quite similar, with only little reductions of the wall shear and an increase of the centerline velocity as $Bi$ grows. The difference with the Newtonian case becomes more noticeable for $Bi=1.4$, where a region with zero shear appears at the centerline. Finally the profile for $Bi=2.8$ clearly differs from the other ones, with a large zero-shear region occupying more than $50\%$ of the channel. In the right panel of the figure, the same velocity profiles are shown in wall units. In most of the turbulent cases we can identify three regions in the velocity profile, similarly to those found for a Newtonian turbulent channel flow (black symbols): first, the viscous sublayer for $y^+<5$ where the variation of $\overline{u}^+$ with $y^+$ is linear; then, the so-called log-law region, $y^+>30$, where the variation of $\overline{u}^+$ versus $y^+$ is logarithmic; finally, the region between $5$ and $30$ wall units is called buffer layer and neither laws hold. As $Bi$ increases, the extension of the inertial ranges reduces eventually disappearing, thus indicating the absence of an equilibrium range. By comparing the three regions discussed above present in the mean velocity profile and the extension of the unyielded region shown in \figrefS{fig:turbDpDx}, we can observe that the flow remains unyielded mostly in the logarithmic and outer layer, while it is always yielded in the viscous sublayer.

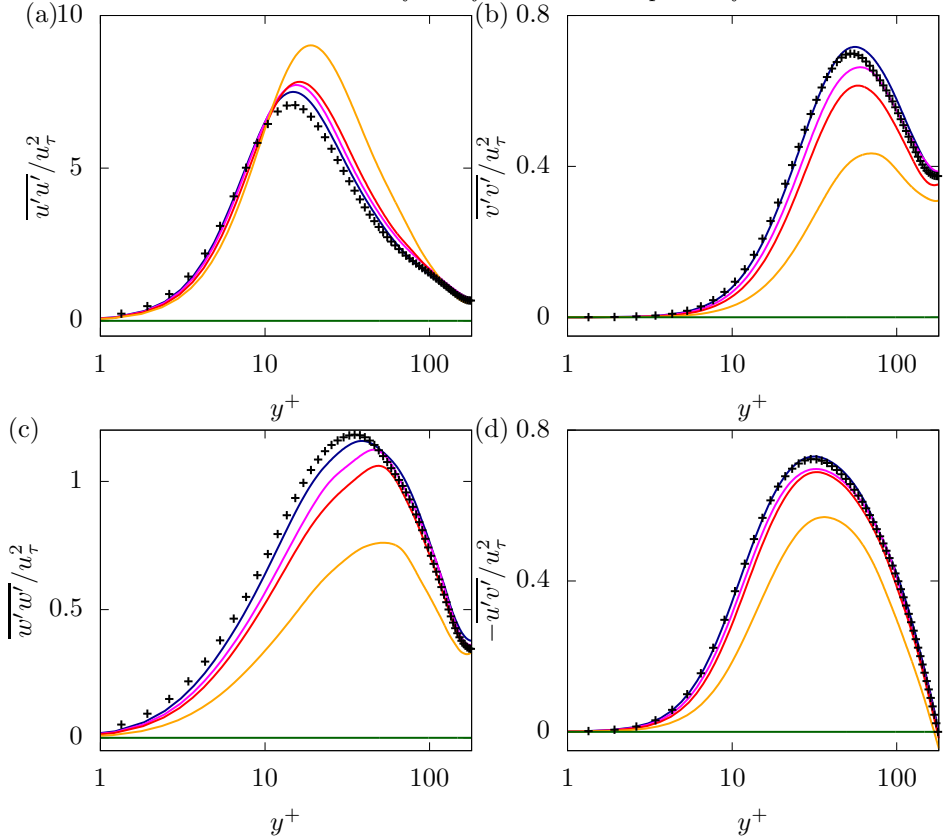
\begin{figure}
  \centering
  \input{reyU}
  \input{reyV} \vspace{0.4cm} \\
  \input{reyW}
  \input{reyUV} \vspace{0.4cm}
  \caption{Wall-normal profiles of the different components of the Reynolds stress tensor, normalized with $u_\tau^2$. Panels (a), (b), and (c) show the diagonal components $u'u'$, $v'v'$, and $w'w'$, while the panel (d) the cross term $u'v'$. The color scheme is the same as in \figrefS{fig:turbDpDx}.}
  \label{fig:turbRey}
\end{figure}
We continue our comparison between the turbulent channel of a Newtonian and elastoviscoplastic fluid by analyzing the wall-normal distribution of the diagonal component of the Reynolds stress tensor; these are shown in \figrefS{fig:turbRey} together with the data from \citet{kim_moin_moser_1987a} for the Newtonian case represented with the black symbols $+$. Also in the Reynolds stress profiles, we find the distinction between the three regimes previously mentioned. First, for low $Bi$, the fluctuations are only slightly affected, with the differences being noticeable only in the buffer layer, while the profiles in the viscous sublayer and in the inertial range still show a good collapse in wall units. Then, for high Bingham numbers ($Bi=1.4$) the profiles undergo strong modifications, which are not limited to the buffer layer, but extend to the inertial range and viscous sublayer as well. Finally, the full laminarisation of the flow for $Bi=2.8$ is proved again by showing the null values assumed by all the Reynolds stress components, in the whole domain. 

We also observe a clear trend, although not linear, of the Reynolds stress components with the Bingham number: the streamwise component $\overline{u'u'}$ increases, while the wall-normal $\overline{v'v'}$ and spanwise $\overline{w'w'}$ decrease. Also, all the peaks are displaced away from the wall, towards the centerline. In relative terms, the wall-normal and spanwise components are the most affected ones, decreasing by almost $40\%$. Finally, \figref[d]{fig:turbRey} depicts the wall-normal profile of the off-diagonal component of the Reynolds stress tensor $\overline{u'v'}$. Also this cross component is affected by the elastoviscoplasticity of the fluid in a similar fashion as the diagonal components. In particular, the maximum value decreases and moves away from the wall as the Bingham number increases. Nevertheless, the stress profiles still vary linearly between the two peaks of opposite sign close to each wall, but with different slopes (not shown here). The Reynolds stress modifications due to the elastoviscoplastic property of the fluid are similar to what observed in other drag reducing flows, such as the turbulent flow over riblets \citep{garcia-mayoral_jimenez_2011b}, the turbulent flow over anisotropic porous walls \citep{rosti_brandt_2018a}, and turbulent flows with polymers \citep{dubief_white_terrapon_shaqfeh_moin_lele_2004a, shahmardi_zade_ardekani_poole_lundell_rosti_brandt_2018a}. In general, the increased amplitude of the streamwise fluctuations, and the reduction of the other components, is usually associated with the strengthening of streaky structures above the wall, which is true also in the present case, as shown in the next paragraphs.

\begin{figure}
  \centering
  \input{reyK}
  \input{prod} \vspace{0.4cm}
  \caption{Wall-normal profiles of (a) the turbulent kinetic energy $\mathcal{K}=\left( u'^2 + v'^2 + w'^2 \right)/2$ and of (b) the turbulent production $\mathcal{P}=-\overline{u'v'} d\overline{u}/dy$, both normalized with the friction velocity $u_\tau$. The color scheme is the same as in \figrefS{fig:turbDpDx}, with the blue, magenta, red, orange and green colors are used for the turbulent cases with $Bi=0$, $0.28$, $0.7$, $1.4$ and $2.8$, respectively.}
  \label{fig:turbK}
\end{figure}
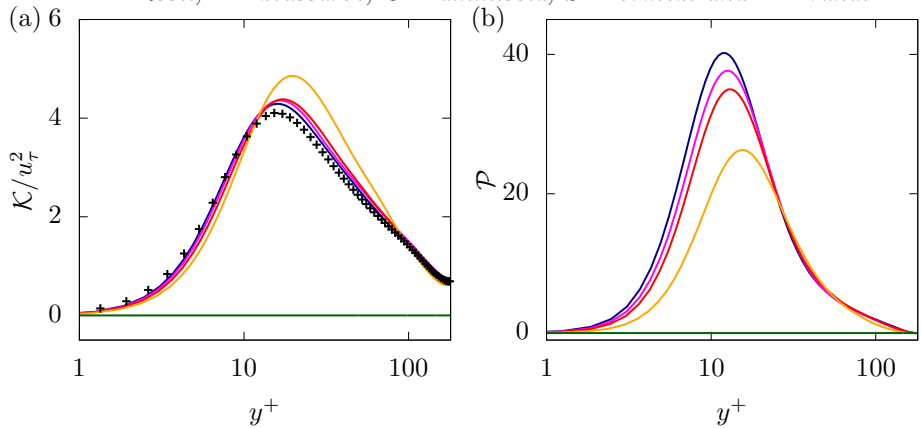
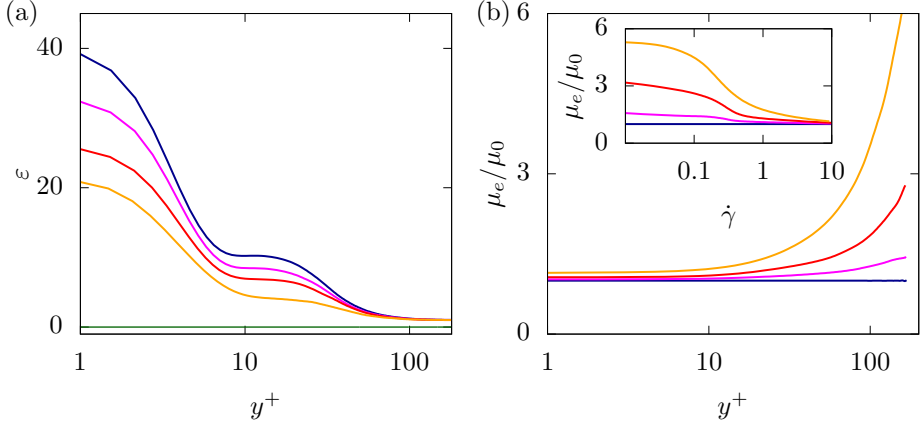
\begin{figure}
  \centering
  \input{diss}
  \input{stressVisc} \vspace{0.4cm}
  \caption{Wall-normal profiles of (a) the turbulent dissipation $\varepsilon=\mu \overline{\partial u'_i/\partial x_j \partial u'_i/\partial x_j}$ in wall units and of (b) the shear effective viscosity $\mu_e$. The color scheme is the same as in \figrefS{fig:turbDpDx}. The inset figure in the right panel shows $\mu_e$ as a function of the shear rate $\dot{\gamma}$.}
  \label{fig:diss}
\end{figure}
An overall view of the velocity fluctuations can be inferred by considering the turbulent kinetic energy $\mathcal{K}=\left( u'^2 + v'^2 + w'^2 \right)/2$, shown in \figref[a]{fig:turbK}, normalized by the friction velocity $u_\tau$. As usual, the symbols represent the profiles from the DNS of \citet{kim_moin_moser_1987a} of a turbulent channel flow. Close to the wall and close to the centerline, all the profiles coincide. On the contrary, in the region where the maximum of $\mathcal{K}$ is located, \ie the buffer layer, we observe a strong increase for $Bi=1.4$, and only a moderate one for the other values of Bingham number. Also, the peak is displaced to higher wall-normal distances $y^+$ than its Newtonian counterpart. The increased value of the peak is mainly due to the increase of the streamwise component of the velocity fluctuations discussed above. An opposite behavior is evident in \figref[b]{fig:turbK}, where the turbulent production $\mathcal{P}=-\overline{u'v'} d\overline{u}/dy$ is displayed. Indeed, although all the profiles of $\mathcal{P}$ still collapse at the wall and at the centerline, in the buffer layer the turbulent production decreases with the Bingham number, with differences noticeable in the viscous sublayer as well. \figrefC[a]{fig:diss} shows the turbulent dissipation $\varepsilon=\mu \overline{\partial u'_i/\partial x_j \partial u'_i/\partial x_j}$ of the fluctuating velocity field $u_i'$. We observe that $\varepsilon$ has a maximum at the wall and then decreases moving towards the centerline where it reaches its minimum value which is the approximately same for the considered turbulent cases. The case with $Bi=0$ shows a dissipation profile similar to the one of a Newtonian fluid \citep{rosti_cortelezzi_quadrio_2015a}, while increasing $Bi$ the dissipation decreases monotonically, being null in the laminar case when $Bi=2.8$. The decrease of dissipation with $Bi$ is consistent with the progressive decrease of $Re_\tau$ previously observed.

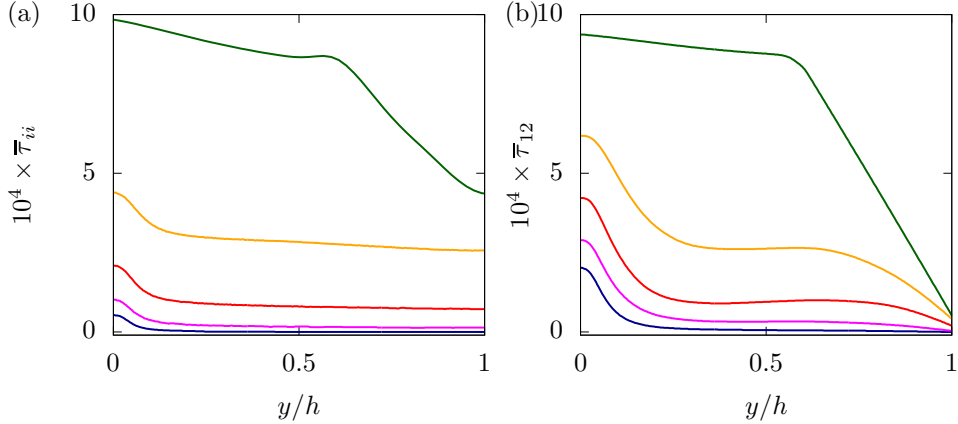
\begin{figure}
  \centering
  \input{meanSii}
  \input{meanS12} \vspace{0.4cm} \\
  \caption{(a)Trace and (b) shear component of the mean elastoviscoplastic stress tensor $\overline{\tau}_{ij}$ as a function of the wall-normal distance $y$. The color scheme is the same as in \figrefS{fig:turbDpDx}.}
  \label{fig:turbStressEVP}
\end{figure}
We now discuss in more details the elastoviscoplastic stress tensor $\tau_{ij}$. \figrefSC{fig:turbStressEVP} shows the mean profile of the microstructure stress tensor trace $\overline{\tau}_{ii}$ (a) and the shear component $\overline{\tau}_{12}$ (b) as functions of the Bingham number. All the stress profiles have their maximum values at the wall ($y=0$) and their minimum absolute value at the centerline ($y=h$), with the trace being symmetric with respect to $y=h$ and the shear component anti-symmetric. In the turbulent flows ($Bi \lesssim 2$), the normal stresses vary only slightly across the channel, except in the near wall region where they rapidly grow. On the other hand, this trend is almost inverted in the laminar flow ($Bi=2.8$). A similar behavior is shown by the shear stress component $\overline{\tau}_{12}$, except that the almost uniform region is less wide, since in the middle of the channel the stress needs to vanish. Moreover, this region with an almost uniform elastoviscoplastic shear stress further reduces with the Bingham number $Bi$. We observe that the shear stress component and the trace of the stress tensor are approximately of the same magnitude, and that the values of the stress components approximately scale with $Bi$, but only when the flow is turbulent.

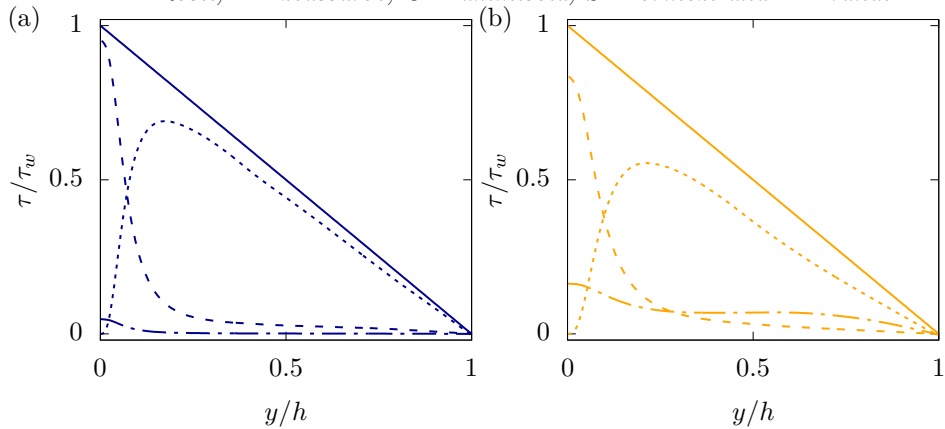
\begin{figure}
  \centering
  \input{stress01}
  \input{stress04} \vspace{0.4cm} \\
  \caption{Normalised shear stress balance across the channel, for (a) $Bi=0$ and (b) $Bi=1.4$. The dashed, dotted and dash-dotted lines are used for the viscous, Reynolds and elastoviscoplastic shear stress, respectively, while the solid line is the total shear stress which varies linearly across the channel height.}
  \label{fig:turbStress}
\end{figure}
To gain further understanding, we report in \figrefS{fig:turbStress} the shear stress budget for the cases with $Bi=0$ and $Bi=1.4$, normalized with the corresponding wall stress. For $Bi=0$, the additional elastoviscoplastic stress is very small and the behavior is therefore similar to that of a standard Newtonian turbulent channel flow, with the viscous stress dominating at the wall and then rapidly decreasing towards the channel core; the Reynolds stresses, on the other hand, are zero at the wall and at the centerline and attains a maximum relatively close to the wall. 
Note that, although very small, the elastoviscoplastic stress is not null at the wall, thus the total wall shear stress is the sum of the two contributions, the elastoviscoplastic and viscous stresses. 
The situation differs in the flow at the Bingham number $Bi=1.4$. Here, the elastoviscoplastic stress increases across the whole channel, reaching approximately $15\%$ of the total stress at the wall. Its increase is compensated by a changes of the other two stress components; in particular, the Reynolds stress peak reduces from $70\%$ to $55\%$ of the total, while the viscous stress peak from $95\%$ to $85\%$. An overall picture of the total shear stress profile can be gained by studying the effective shear viscosity $\mu_e$ normalised with the total viscosity $\mu_0$, reported in \figref[b]{fig:diss}. This is defined as follows:
\begin{equation} \label{eq:shearvisc}
\frac{\mu_e}{\mu_0} = \frac{\mu_f \dfrac{d\overline{u}}{dy} + \overline{\tau}_{12}}{\mu_0 \dfrac{d\overline{u}}{dy}}.
\end{equation}
The effective shear viscosity $\mu_e$ grows with the Bingham number $Bi$ and moving from the wall towards the center of the channel. The inset of the figure shows the same quantity $\mu_e$ as a function of the shear rate $\dot{\gamma}$ here defined as $\mu_0 d\overline{u}/dy$, \ie the denominator of \equref{eq:shearvisc}; from the figure we can appreciate the shear thinning behavior of the elastoviscoplastic fluid described by the Saramito model \citep{saramito_2007a}.

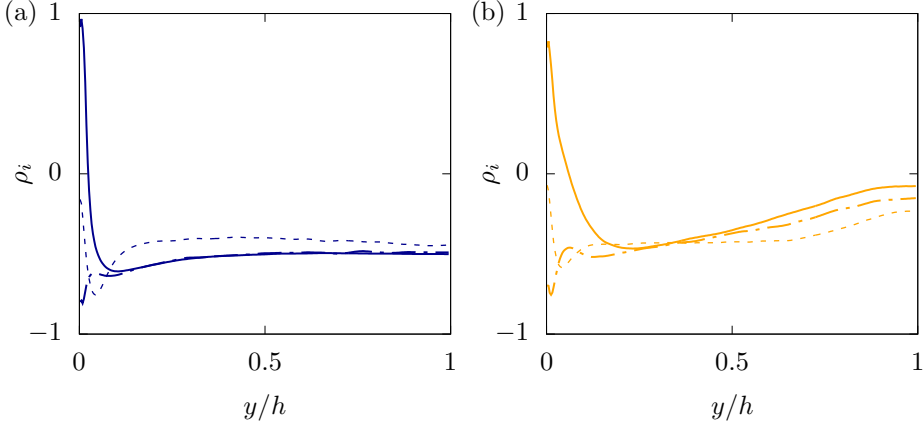
\begin{figure}
  \centering
  \input{dubi01}
  \input{dubi04} \vspace{0.4cm} \\
  \caption{Wall-normal profiles of the cross-correlations $\rho_i$ of the streamwise (dashed line), wall-normal (solid line) and spanwise (dash-dotted line) velocity component $u_i$ and elastoviscoplastic contribution $f_i$, defined in \equref{eq:crosscorr}, for (a) $Bi=0$ and (b) $Bi=1.4$.}
  \label{fig:evpcrosscorr}
\end{figure}
Finally, \figrefS{fig:evpcrosscorr} shows the cross-correlation $\rho_i$ defined as
\begin{equation} \label{eq:crosscorr}
\rho_i= \dfrac{\overline{u'_i f'_i}}{u'_i f'_i},
\end{equation}
where $f_i$ is the elastoviscoplastic volume force, \ie the contribution to the Navier-Stokes equation of the elastoviscoplastic stress tensor $\tau_{ij}$ defined as $f_i=\partial \tau_{ij}/\partial x_j$. Note that, there is no summation over the repeated indices in the previous relation. The left and right panels of the plot show the cases with $Bi=0$ (a) and $Bi=1.4$ (b), with the dashed, solid and dash-dotted lines in each plot corresponding to the streamwise, wall-normal and spanwise cross-correlations: $\rho_1$, $\rho_2$ and $\rho_3$. In general, when $\rho_i$ equals $1$ or $-1$ the flow velocity and elastoviscoplastic force are perfectly correlated or anti-correlated, while when $\rho_i=0$ they are not correlated. We observe that for the case $Bi=0$ all the cross-correlation components $\rho_i$ are negative in most of the channel, except in the near-wall region where $\rho_1$ equals $1$ and $\rho_2$ equals $0$. In this case, the cross-correlation is almost uniform across the whole channel-height, with a negative value equal to $\rho_i \approx -0.5$: the elastoviscoplastic body force and velocity are anti-correlated in the bulk of the flow away from the walls, thus indicating that the elastoviscoplastic contribution is opposing to the turbulent fluctuations. Close to the wall, however, the high positive values attained by $\rho_1$ suggest a role played by the viscoelastic stresses on the increase of the streamwise velocity fluctuations. Note that, this is similar to what found by \citet{dubief_terrapon_white_shaqfeh_moin_lele_2005a} for a turbulent channel flow with a polymer solution. On the other hand, the high Bingham case ($Bi=1.4$) shows a similar trend only close to the wall ($y \lesssim 0.5h$), while in the bulk the cross-correlations $\rho_i$ interestingly go to zero. The fact that the flow velocity and the elastoviscoplastic stress tensor are not correlated around the centerline can be associated to the continuous cycle of yielding and unyielding process.

\begin{figure}
  \centering
  \includegraphics[width=0.33\textwidth]{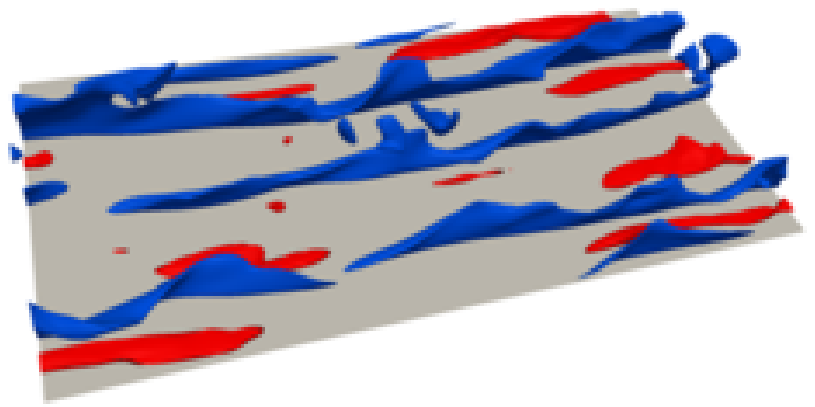} \includegraphics[width=0.31\textwidth]{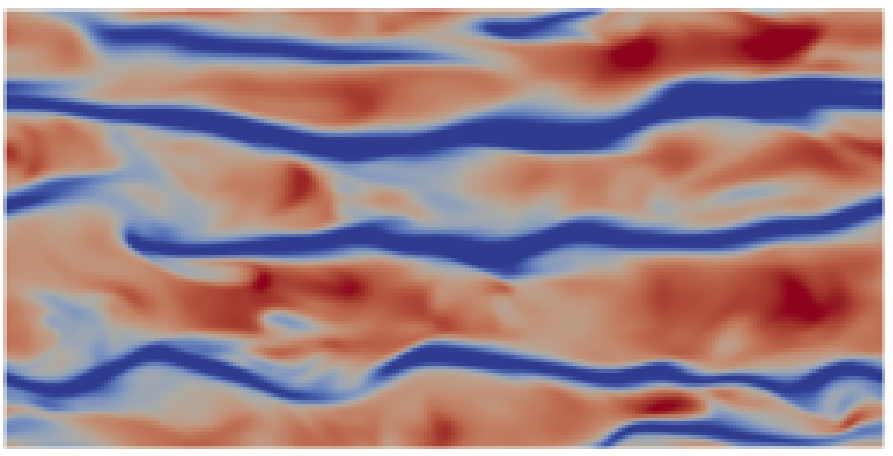} \includegraphics[width=0.31\textwidth]{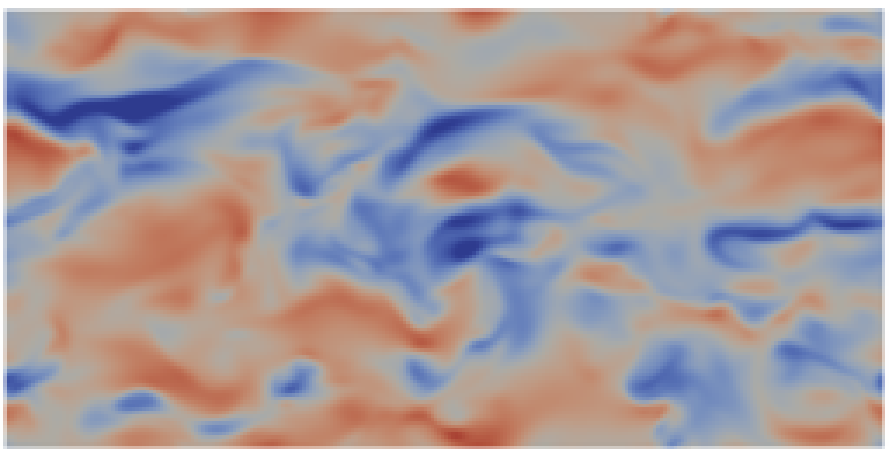} \vspace{0.1cm} \\
  \includegraphics[width=0.33\textwidth]{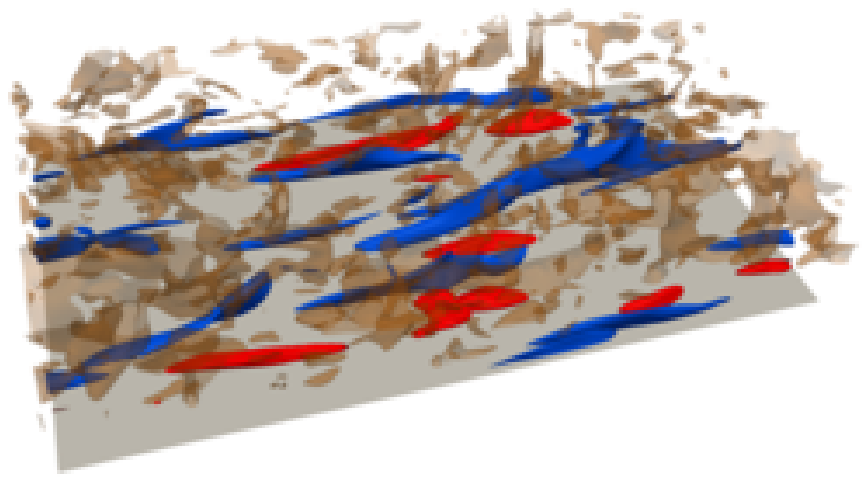} \includegraphics[width=0.31\textwidth]{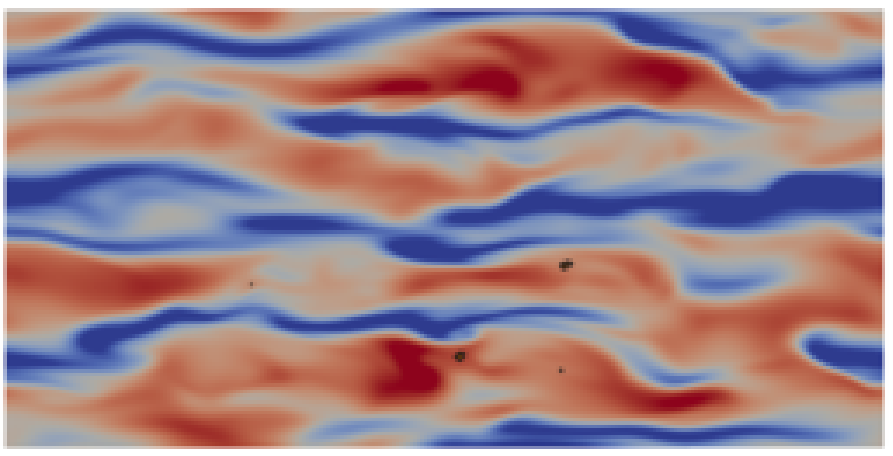} \includegraphics[width=0.31\textwidth]{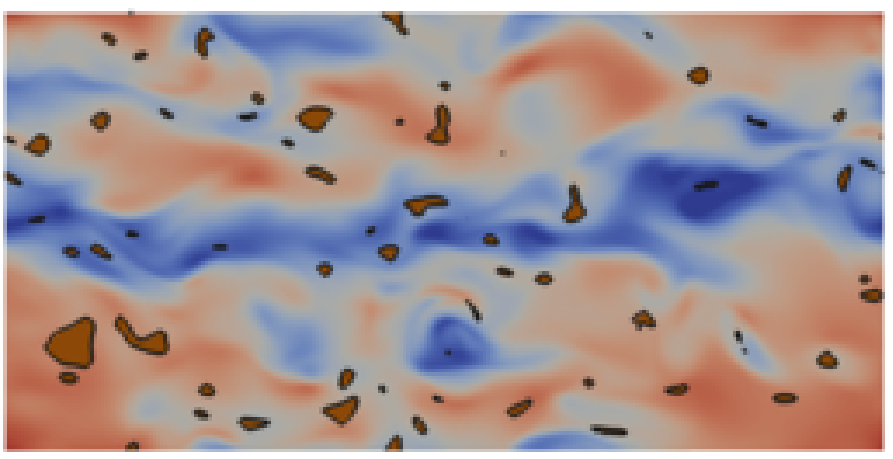} \vspace{0.1cm} \\
  \includegraphics[width=0.33\textwidth]{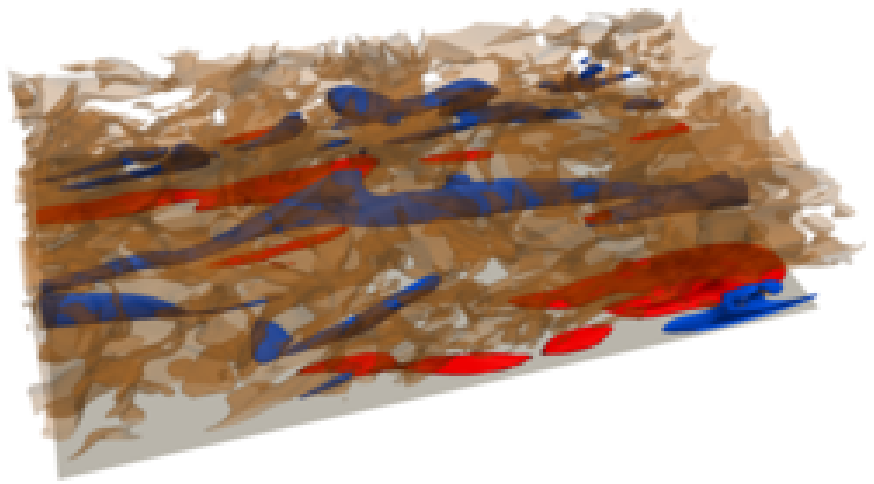} \includegraphics[width=0.31\textwidth]{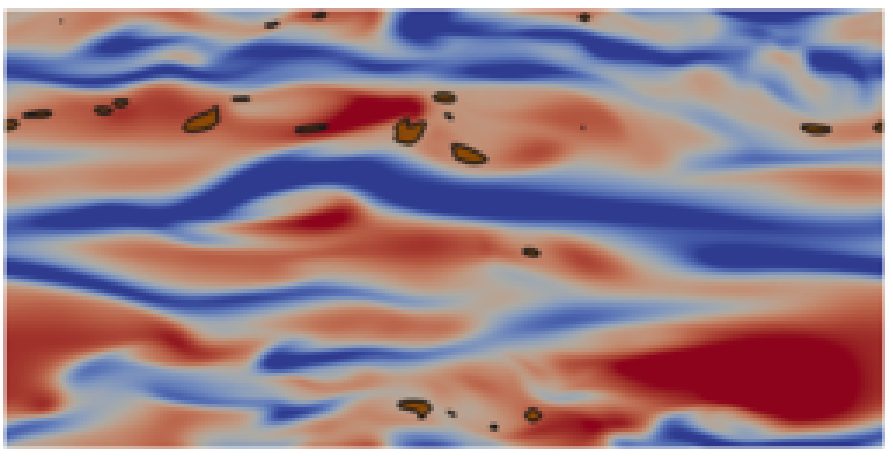} \includegraphics[width=0.31\textwidth]{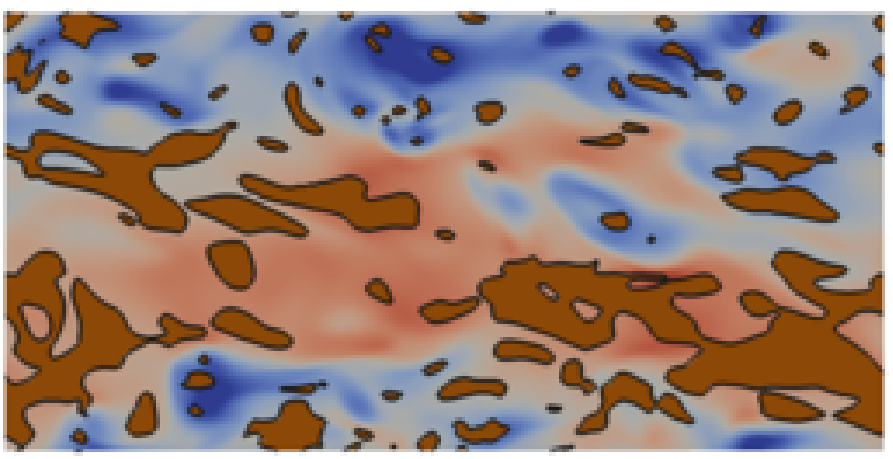} \vspace{0.1cm} \\
  \includegraphics[width=0.33\textwidth]{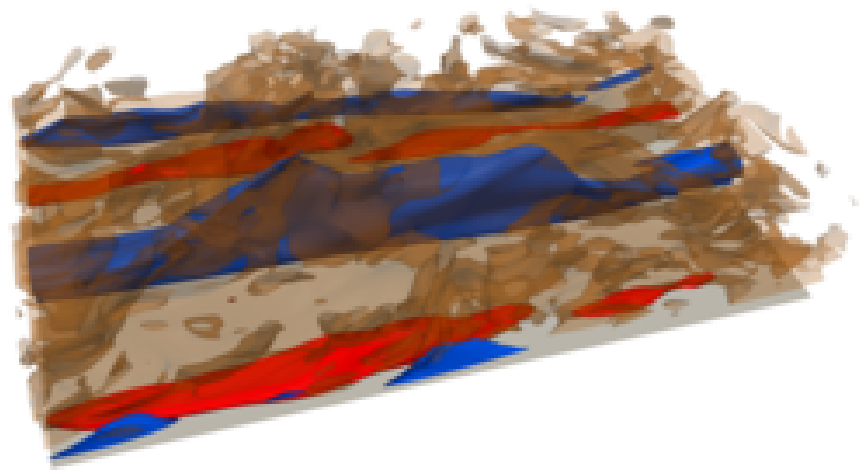} \includegraphics[width=0.31\textwidth]{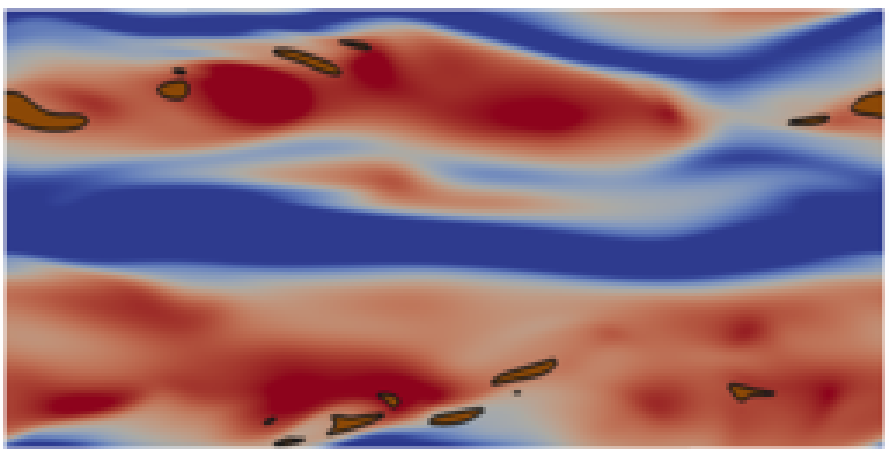} \includegraphics[width=0.31\textwidth]{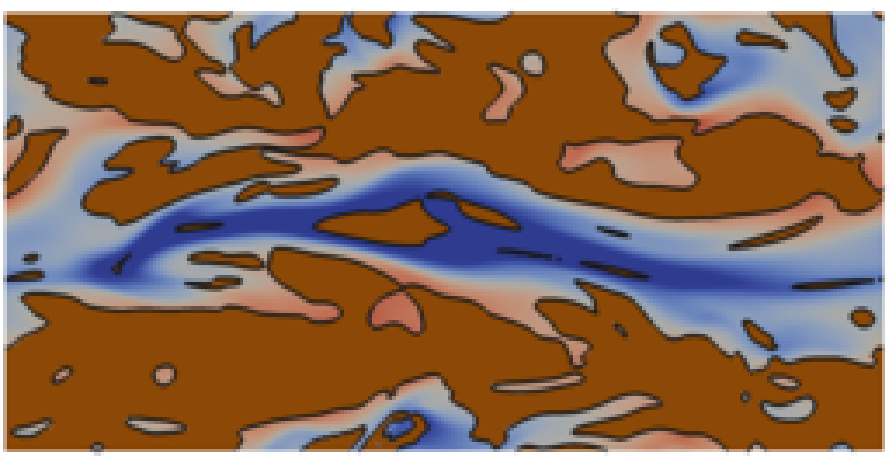} \vspace{0.1cm} \\
  \includegraphics[width=0.33\textwidth]{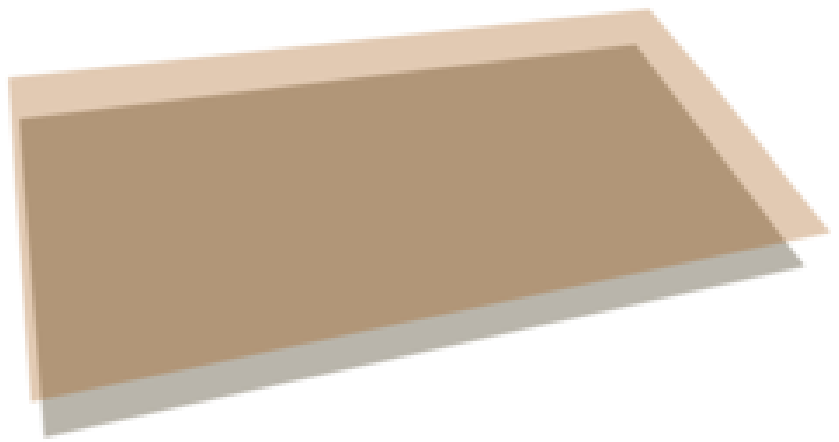} \includegraphics[width=0.31\textwidth]{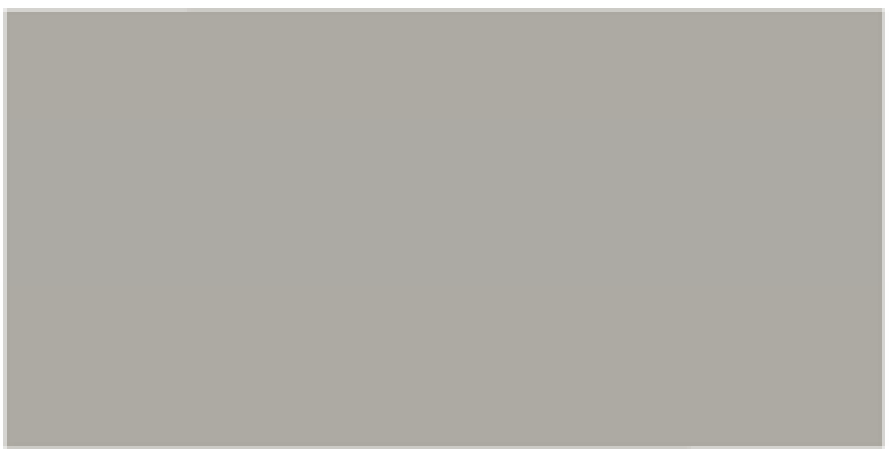} \includegraphics[width=0.31\textwidth]{up05}     \vspace{0.4cm}
  \caption{(left) Isosurfaces and (middle and right) contours of instantaneous streamwise velocity fluctuation $u'$. The flow goes from left to right and the color scale ranges from $-0.25U_b$ (blue) to $0.25U_b$ (red). The brown regions in all the figures represent the unyielded fluid. The two slices on the right are $x-z$ planes at $y = 0.15h$ and $y=0.44h$. The Bingham number $Bi$ increases from top to bottom ($Bi=0$, $0.28$, $0.7$, $1.4$ and $2.8$), and the viscosity ratio $\beta$ is fixed equal to $0.95$.}
  \label{fig:turbYielded3d}
\end{figure}

The elastoviscoplastic character of the flow affects the near-wall turbulent structures, and this is visually confirmed in \figrefS{fig:turbYielded3d}. The left panels identify the low- (blue) and high-speed (red) near-wall streaks with isosurfaces of the streamwise velocity fluctuations $u'$ corresponding to the levels $u'^+ = \pm 0.25 U_b$, while the pictures in the middle and right columns show the footprints of these structures on the wall-parallel planes at $y=0.15h$ and $0.44h$. It is evident that the structures in the buffer layer are less fragmented and more elongated in the streamwise direction as the Bingham number increases. Also, their spanwise extension increases, and consequently the number of streaks reduces. Indeed, the attenuation of the small-scale features is consistent with a picture where the larger coherent structures grow in size due to the reduction of the friction Reynolds number $Re_\tau$, \ie drag reduction. This effect - decreasing drag and  wider and more coherent structures - is similar to what found in other drag reducing flows, such as riblets, polymer suspensions and anisotropic porous walls as already discussed previously. From the $3D$ visualizations, we observe that the low-speed streaks penetrate to higher wall-normal distances than the high-speed ones; the former are usually associated with wall-normal velocity fluctuations $v'$ away from the wall, while their high-speed counterparts with wall-normal velocities towards the wall \citep{kim_moin_moser_1987a}. This tendency interacts with the yield/unyield process; indeed, from the rightmost panels in the figure we note that the regions where the fluid is not yielded are mostly located in the positions above an high-speed streak, while almost all the fluid above the low-speed streaks remains fully yielded. These visual observations will be now quantified statistically by analyzing the autocorrelation functions.

\begin{figure}
  \centering
  \includegraphics[width=0.24\textwidth]{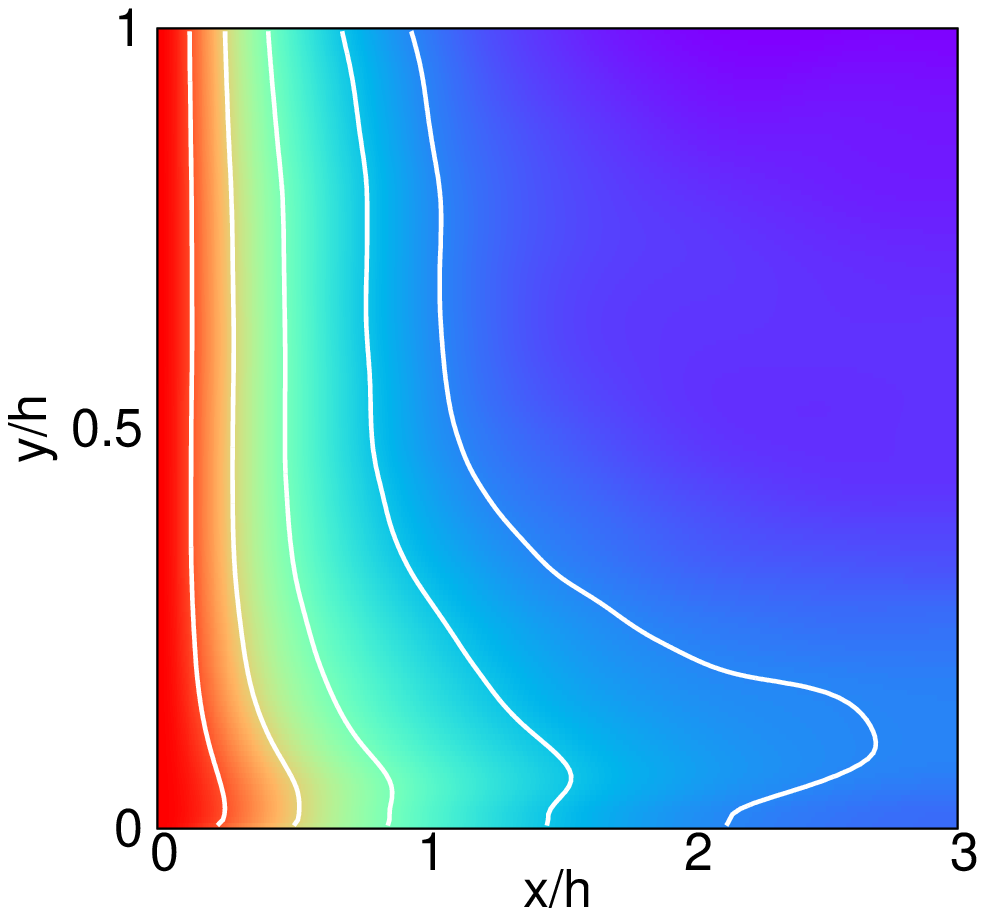}
  \includegraphics[width=0.24\textwidth]{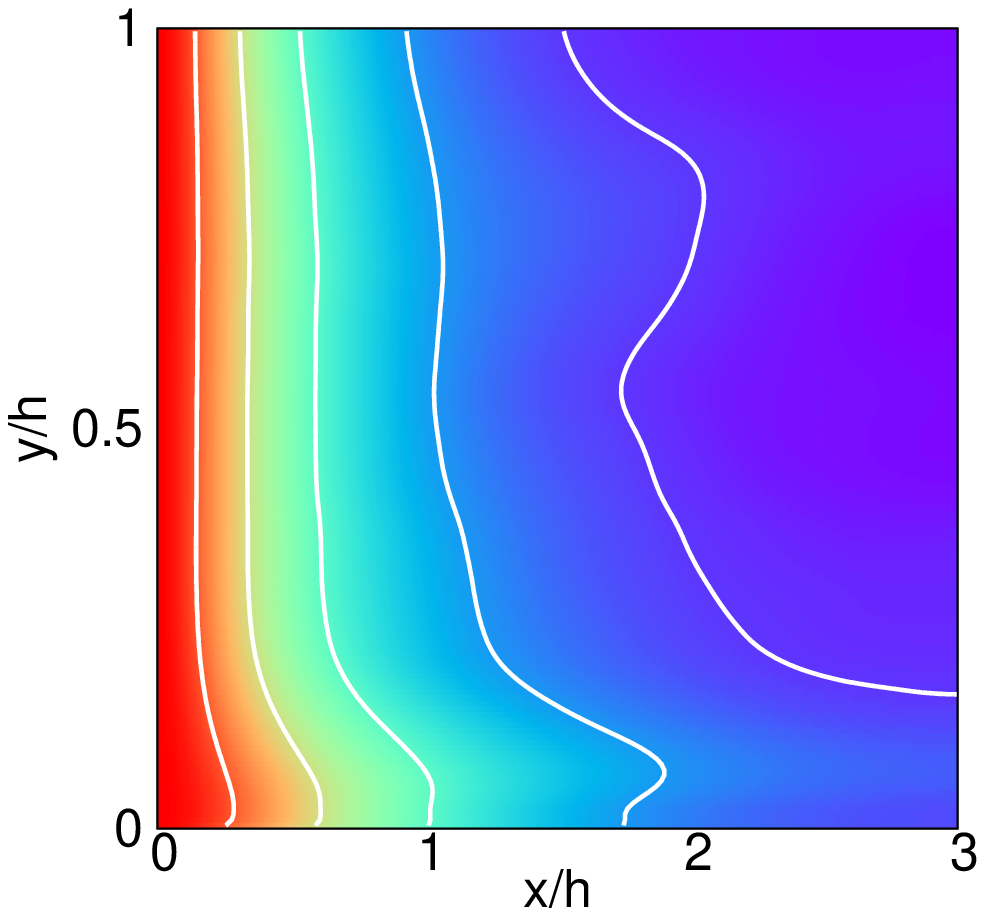}
  \includegraphics[width=0.24\textwidth]{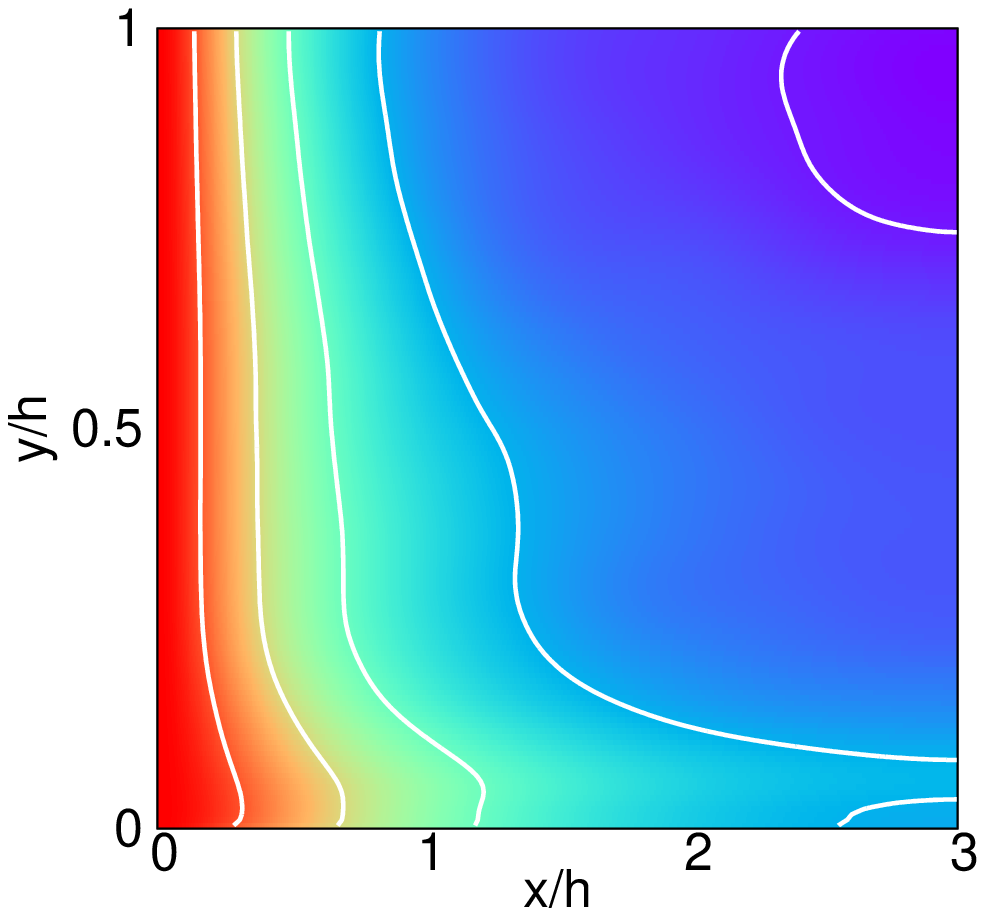}
  \includegraphics[width=0.24\textwidth]{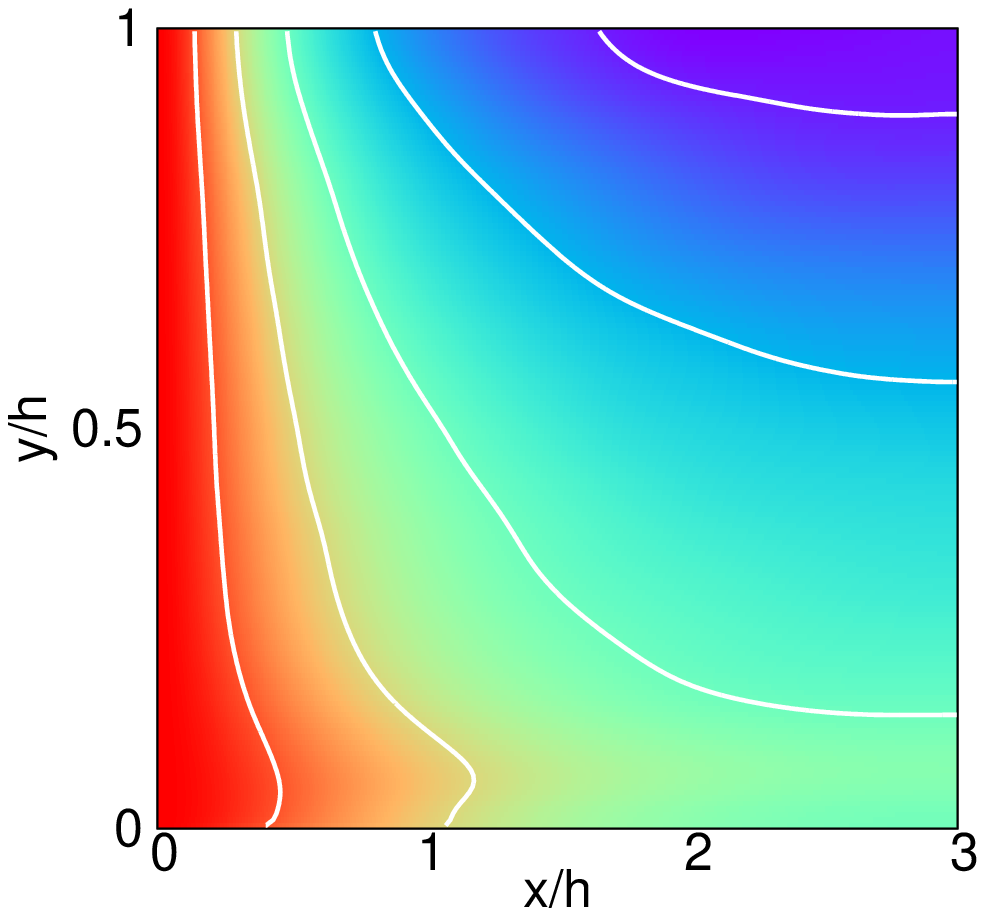} \vspace{0.2cm} \\
  \includegraphics[width=0.24\textwidth]{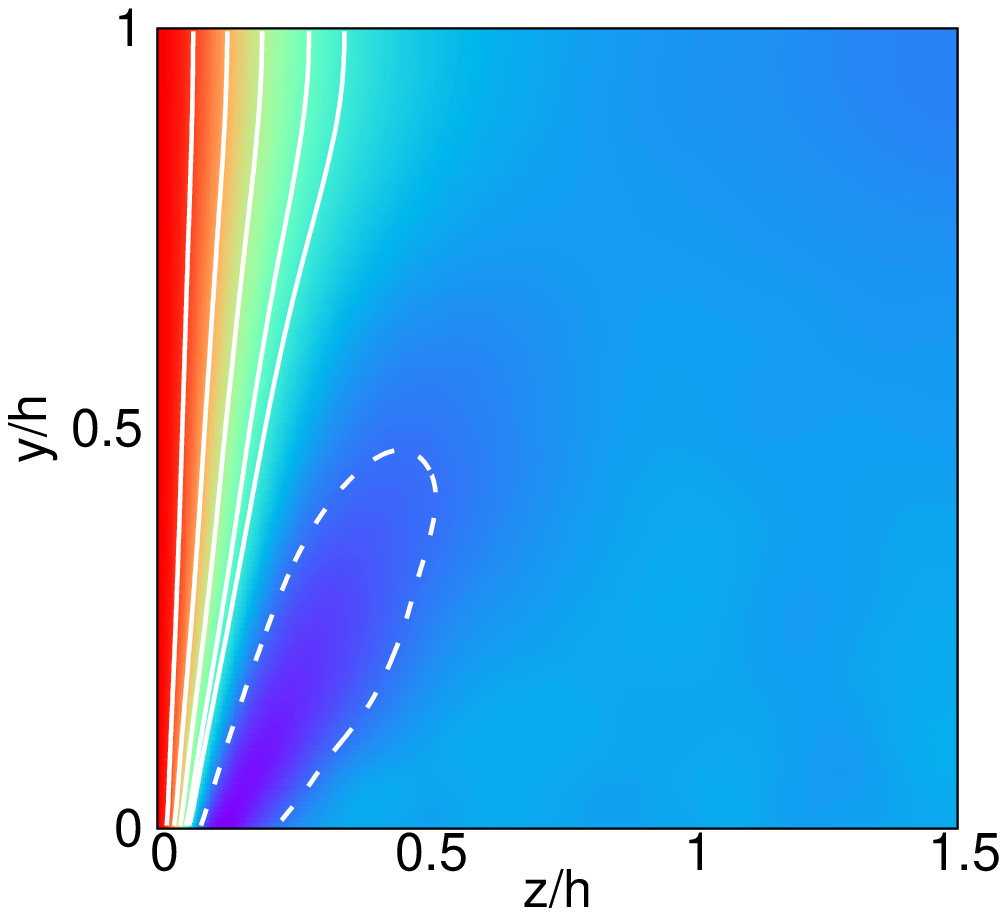}
  \includegraphics[width=0.24\textwidth]{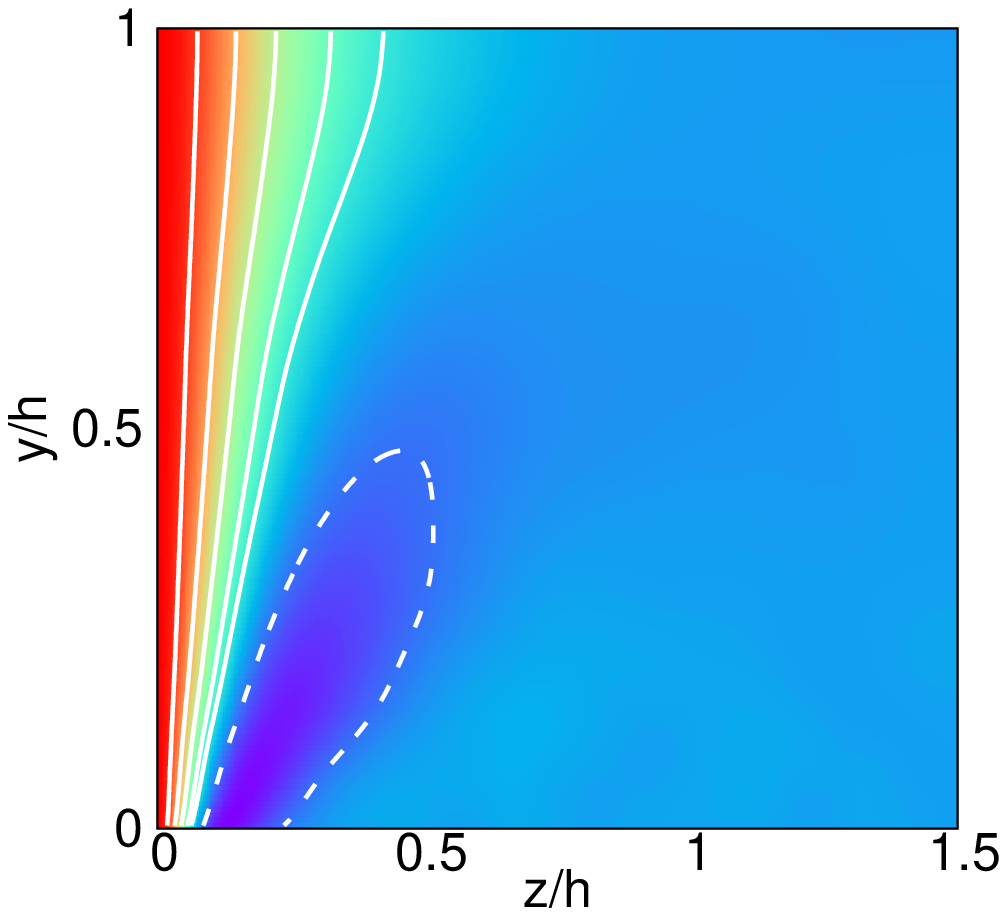}
  \includegraphics[width=0.24\textwidth]{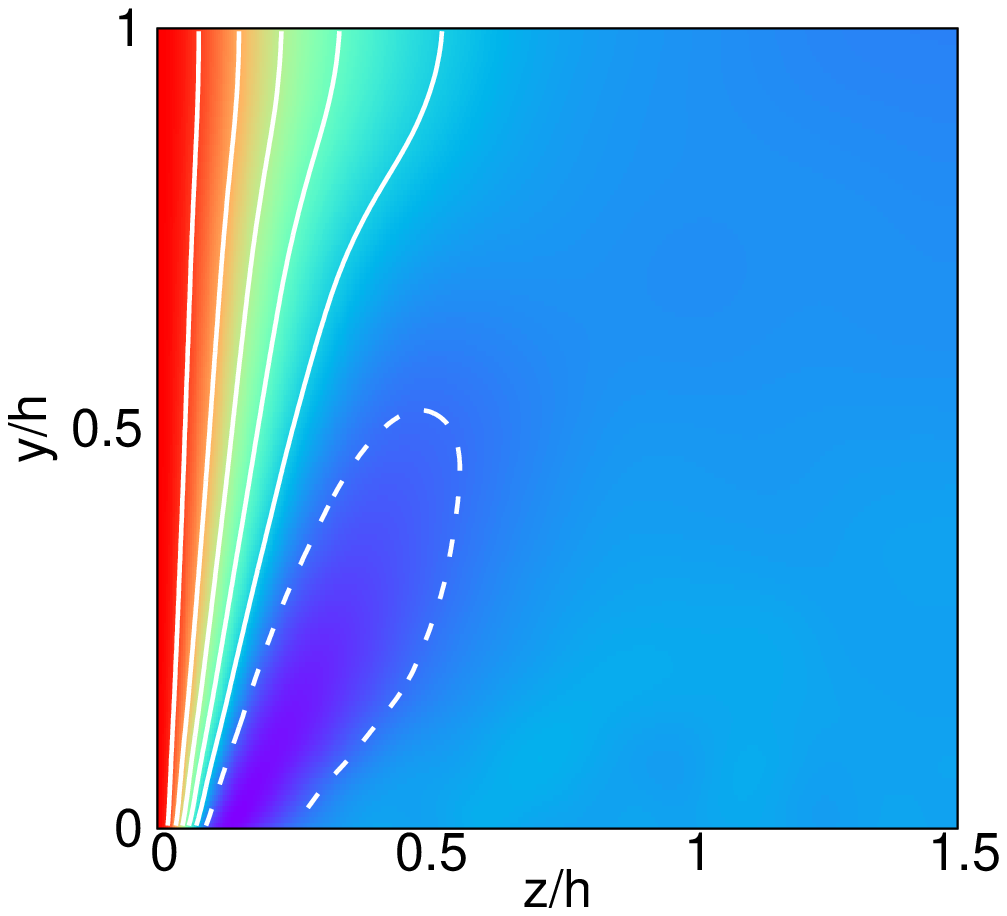}
  \includegraphics[width=0.24\textwidth]{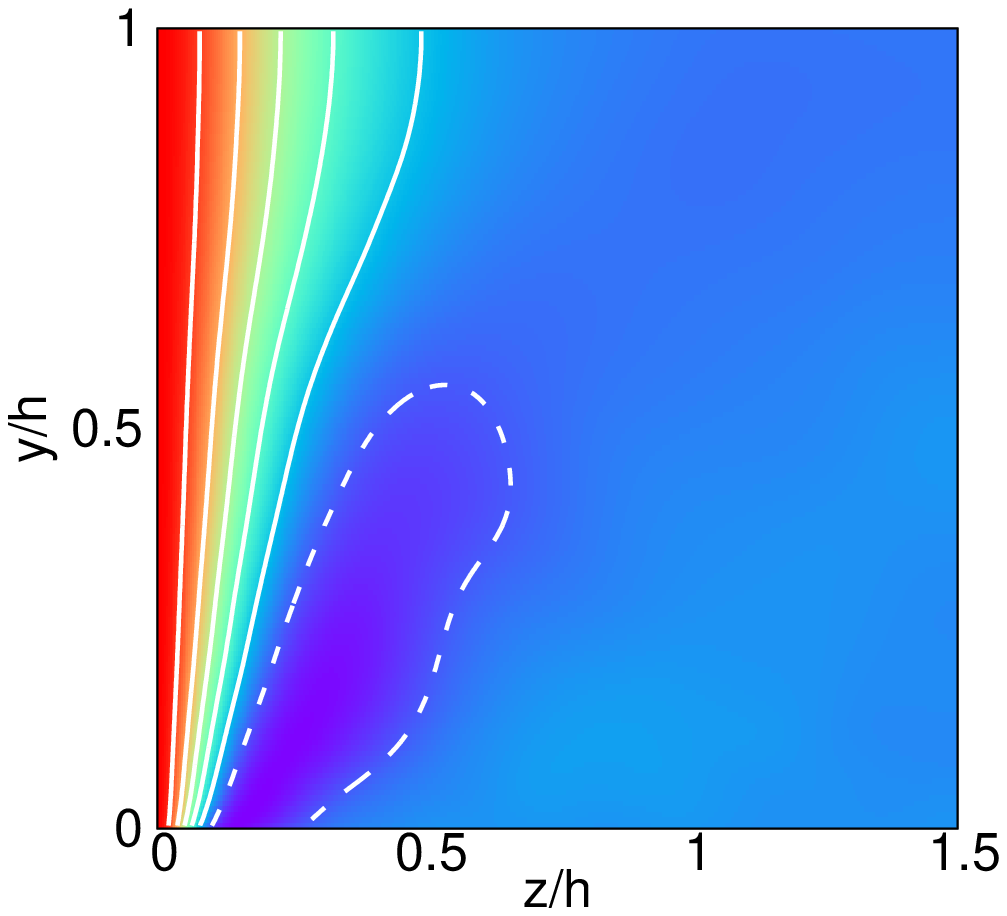}
  \caption{Stack of two-point velocity auto-correlation functions across the channel $\mathcal{R}_{ii}(y)$. The top row shows the streamwise autocorrelation function of the streamwise velocity $\mathcal{R}_{11}$, while the bottom one the spanwise autocorrelation function of the wall-normal velocity $\mathcal{R}_{22}$. The Bingham numbers increases from left to right ($Bi=0$, $0.28$, $0.7$ and $1.4$). The solid and dashed lines correspond to positive and negative values of auto-correlation, ranging from $-0.1$ to $0.9$ with a step of $0.2$ between two neighboring lines. The colour scale ranges from $-0.1$ (purple) to $1.0$ (red).}
  \label{fig:turbCorr}
\end{figure}

The effect of the Bingham number $Bi$ on the flow coherence is quantified by the two-point velocity auto-correlation functions, reported in \figrefS{fig:turbCorr} for the cases with $Bi=0$, $0.28$, $0.7$ and $1.4$. The two-point auto-correlation function $\mathcal{R}_{ii}$ is defined here as
\begin{equation} \label{eq:correlation}
\mathcal{R}_{ii}(\bm{x},\bm{r})=\frac{\overline{u_i'(\bm{x})u_i'(\bm{x}+\bm{r})}}{\overline{u_i'^2(\bm{x})}},
\end{equation}
where the bar denotes average over time and the two homogeneous directions, and the prime the velocity fluctuation. Note that, there is no summation over the repeated indices in the previous relation. The top row in the figure shows the distribution in the $x-y$ plane of the streamwise velocity component auto-correlation along the streamwise direction $x$, while the bottom row the distribution in the $z-y$ plane of the wall-normal velocity component auto-correlation along the spanwise direction $z$. In the case $Bi=0$ (shown in the leftmost column), the correlations appear to be very similar to the baseline Newtonian case with highly elongated streaky structures that alternate at the canonical spanwise distance  (\ie $\Delta z^+\simeq 100^+$) \citep{kim_moin_moser_1987a}. As the Bingham number $Bi$ is increased (panels from left to right), the streamwise correlation length monotonically increases, thus indicating a higher level of coherency of the flow structures; in particular, this reveals that the velocity streaks are more elongated in the streamwise direction. The spanwise correlation also increases when increasing $Bi$ as the velocity streaks become wider than in a Newtonian fluid, despite the existence of yielded regions in the channel core (see \figrefS{fig:turbYielded3d}). Note also that, the increased correlation lengths in both the streamwise and spanwise directions, are not limited to the near wall-regions occupied by the streaky structures, as in the other drag reducing flows cited above \citep{garcia-mayoral_jimenez_2011b, rosti_brandt_2018a, dubief_white_terrapon_shaqfeh_moin_lele_2004a, shahmardi_zade_ardekani_poole_lundell_rosti_brandt_2018a}, but extends up to the centerline. However, this increased spanwise correlation does not extend towards the centreline as much as the streamwise coherence. This difference originates from the fact that the flow at a high wall-normal distance $y$ is still not yielded, thus behaving as a viscoelastic solid.

\section{Conclusion} \label{sec:conclusion}
We present numerical simulations of laminar and turbulent channel flow of a non-Newtonian elastoviscoplastic fluid. The  elastoviscoplastic flow is simulated with the model proposed by \citet{saramito_2007a}, where the incompressible Navier-Stokes equations are coupled with an additional equation for the evolution of the elastoviscoplastic stress tensor. In particular, the model predicts only recoverable Kelvin-Voigt viscoelastic deformation for stress below the yield stress value, while  when the stress exceeds the yield value, the fluid behaves as an Oldroyd-B viscoelastic fluid. For both the laminar and turbulent regime, we examine the flow behavior when changing the values of the material plasticity (Bingham number) and viscosity ratio ($\beta$), while keeping the elasticity constant (Weissenberg number) to a small value, to more clearly identify the role of the plasticity.

For the laminar channel flow, we carried out a full parametric study and find that the friction factor  increases with the Bingham number and decreases with the Reynolds number and the viscosity ratio. The drag increase due to the Bingham number originates from the increase of the portion of the channel where the stress is below the yield stress value and thus the fluid is not yielded. In these regions, forming initially at the centerline and then growing towards the wall as the Bingham number increases, the flow presents a flat velocity profile with zero shear. We propose an empirical correlation for the friction factor in a laminar channel flow, which is function of the Reynolds number, Bingham number, and viscosity ratio. We show that the Fanning friction factor is inversely proportional to the Reynolds number and proportional to the square root of the Bingham number (a results interestingly found also for the flow of the same kind of fluid in a porous media).

In the turbulent flows, the bulk Reynolds number is fixed to $2800$ due to computational costs and the effect of different yield stress values and viscosity ratios is studied via both statistical data and instantaneous visualizations. Unlike the case of laminar flows, the Fanning friction factor is almost independent of the viscosity ratio and decreases with the Bingham number. We show that all the elastoviscoplastic flow configurations analyzed are drag reducing, and since the Weissenberg number considered is very low, we have demonstrated that this is not an effect of the elasticity. We identify three different regimes  depending on the value of the Bingham number: for low Bingham numbers, the turbulence is only slightly modified, except for a slowly progressive reduction of the friction Reynolds number. Next, for intermediate Bingham numbers, the flow becomes highly intermittent, with a continuous cycle of yielding and unyielding process in the center of the channel which is the main responsible for the increased fluid oscillations. We also document strong streamwise velocity fluctuations, with the mean velocity profile departing from the usual log-law and with the loss of the inertial equilibrium range; all the flow statistics are affected both in the buffer and logarithmic layers. Finally, for high values of the Bingham number, the flow fully laminarises. 

We show that the progressive increase of the amount of fluid which is not yielded with the Bingham number has a strong influence on the flow. Indeed, these regions grow from the centerline towards the walls as the Bingham number increases, similarly to the laminar regime, but introduce a strong unsteadiness in the flow when they extend over the full streamwise and spanwise dimensions.

In the elastoviscoplsatic flow we observed an enhancement of the near-wall streak intensity and of the associated quasi-streamwise vortices, regions with localized high stress values. The low-speed streaks, usually associated to positive wall-normal fluctuations, reach higher wall-normal distances than the high-speed streaks, thus inducing the flow to yield at higher wall-normal distances if the local stress reaches the yield stress threshold. Indeed, the unyielded regions preferentially form above high speed streaks. Overall, the flow becomes more and more correlated in the streamwise direction when increasing the Bingham number, with high levels of flow anisotropy close to the wall, similarly to what observed in other drag reducing flows. Differently from the other flows, however, both the streamwise and spanwise correlations grow with the Bingham number also away from from the wall, due to the growth of the unyielded region.

The analysis performed here assumed a very low level of elasticity of the flow. The present results can therefore be extended by introducing this additional effect and investigating how the dynamics described here changes. Furthermore, more complex flow configurations, \eg separating and fully inhomogeneous flows, as well as the addition of a dispersed solid phase in this complex matrix deserve further consideration. Another interesting extension of the present work is the analysis of these flows at higher Reynolds numbers, investigating how the friction factor depends on the Bingham number and the absence of unyielded regions in the viscous sublayer.

\section*{Acknowledgment}
M.R. and L.B. were supported by the European Research Council Grant no. ERC-2013-CoG-616186, TRITOS and by the Swedish Research Council Grant no. VR 2014-5001. D.I. and O.T. acknowledge financial support  by the Swedish Research Council through grants No. VR2013-5789 and No. VR 2014-5001. S.H. acknowledges financial support by NSF (Grant No. CBET-1554044-CAREER), NSF-ERC (Grant No. CBET-1641152 Supplementary CAREER). The authors acknowledge computer time provided by SNIC (Swedish National Infrastructure for Computing). 

\bibliographystyle{jfm}
\bibliography{bibliography.bib}
\end{document}

%% file: saramitoS.tex
% GNUPLOT: LaTeX picture with Postscript
\begingroup
  \makeatletter
  \providecommand\color[2][]{%
    \GenericError{(gnuplot) \space\space\space\@spaces}{%
      Package color not loaded in conjunction with
      terminal option `colourtext'%
    }{See the gnuplot documentation for explanation.%
    }{Either use 'blacktext' in gnuplot or load the package
      color.sty in LaTeX.}%
    \renewcommand\color[2][]{}%
  }%
  \providecommand\includegraphics[2][]{%
    \GenericError{(gnuplot) \space\space\space\@spaces}{%
      Package graphicx or graphics not loaded%
    }{See the gnuplot documentation for explanation.%
    }{The gnuplot epslatex terminal needs graphicx.sty or graphics.sty.}%
    \renewcommand\includegraphics[2][]{}%
  }%
  \providecommand\rotatebox[2]{#2}%
  \@ifundefined{ifGPcolor}{%
    \newif\ifGPcolor
    \GPcolortrue
  }{}%
  \@ifundefined{ifGPblacktext}{%
    \newif\ifGPblacktext
    \GPblacktexttrue
  }{}%
  % define a \g@addto@macro without @ in the name:
  \let\gplgaddtomacro\g@addto@macro
  % define empty templates for all commands taking text:
  \gdef\gplbacktext{}%
  \gdef\gplfronttext{}%
  \makeatother
  \ifGPblacktext
    % no textcolor at all
    \def\colorrgb#1{}%
    \def\colorgray#1{}%
  \else
    % gray or color?
    \ifGPcolor
      \def\colorrgb#1{\color[rgb]{#1}}%
      \def\colorgray#1{\color[gray]{#1}}%
      \expandafter\def\csname LTw\endcsname{\color{white}}%
      \expandafter\def\csname LTb\endcsname{\color{black}}%
      \expandafter\def\csname LTa\endcsname{\color{black}}%
      \expandafter\def\csname LT0\endcsname{\color[rgb]{1,0,0}}%
      \expandafter\def\csname LT1\endcsname{\color[rgb]{0,1,0}}%
      \expandafter\def\csname LT2\endcsname{\color[rgb]{0,0,1}}%
      \expandafter\def\csname LT3\endcsname{\color[rgb]{1,0,1}}%
      \expandafter\def\csname LT4\endcsname{\color[rgb]{0,1,1}}%
      \expandafter\def\csname LT5\endcsname{\color[rgb]{1,1,0}}%
      \expandafter\def\csname LT6\endcsname{\color[rgb]{0,0,0}}%
      \expandafter\def\csname LT7\endcsname{\color[rgb]{1,0.3,0}}%
      \expandafter\def\csname LT8\endcsname{\color[rgb]{0.5,0.5,0.5}}%
    \else
      % gray
      \def\colorrgb#1{\color{black}}%
      \def\colorgray#1{\color[gray]{#1}}%
      \expandafter\def\csname LTw\endcsname{\color{white}}%
      \expandafter\def\csname LTb\endcsname{\color{black}}%
      \expandafter\def\csname LTa\endcsname{\color{black}}%
      \expandafter\def\csname LT0\endcsname{\color{black}}%
      \expandafter\def\csname LT1\endcsname{\color{black}}%
      \expandafter\def\csname LT2\endcsname{\color{black}}%
      \expandafter\def\csname LT3\endcsname{\color{black}}%
      \expandafter\def\csname LT4\endcsname{\color{black}}%
      \expandafter\def\csname LT5\endcsname{\color{black}}%
      \expandafter\def\csname LT6\endcsname{\color{black}}%
      \expandafter\def\csname LT7\endcsname{\color{black}}%
      \expandafter\def\csname LT8\endcsname{\color{black}}%
    \fi
  \fi
    \setlength{\unitlength}{0.0500bp}%
    \ifx\gptboxheight\undefined%
      \newlength{\gptboxheight}%
      \newlength{\gptboxwidth}%
      \newsavebox{\gptboxtext}%
    \fi%
    \setlength{\fboxrule}{0.5pt}%
    \setlength{\fboxsep}{1pt}%
\begin{picture}(3456.00,2880.00)%
\definecolor{gpBackground}{rgb}{1.000, 1.000, 1.000}%
\put(0,0){\colorbox{gpBackground}{\makebox(3456.00,2880.00)[]{}}}%
    \gplgaddtomacro\gplbacktext{%
      \csname LTb\endcsname%
      \put(490,432){\makebox(0,0)[r]{\strut{}$0.01$}}%
      \put(490,1238){\makebox(0,0)[r]{\strut{}$0.1$}}%
      \put(490,2044){\makebox(0,0)[r]{\strut{}$1$}}%
      \put(490,2850){\makebox(0,0)[r]{\strut{}$10$}}%
      \put(622,212){\makebox(0,0){\strut{}$0.1$}}%
      \put(1659,212){\makebox(0,0){\strut{}$1$}}%
      \put(2695,212){\makebox(0,0){\strut{}$10$}}%
    }%
    \gplgaddtomacro\gplfronttext{%
      \csname LTb\endcsname%
      \put(-82,2850){\rotatebox{0}{\makebox(0,0){\strut{}(a)}}}%
      \put(-82,1641){\rotatebox{-270}{\makebox(0,0){\strut{}$\tau_{11}-\tau_{22}, \tau_{12}$}}}%
      \put(2021,-118){\makebox(0,0){\strut{}$t \dot{\gamma}_0$}}%
      \put(2021,2740){\makebox(0,0){\strut{}}}%
    }%
    \gplbacktext
    \put(0,0){\includegraphics{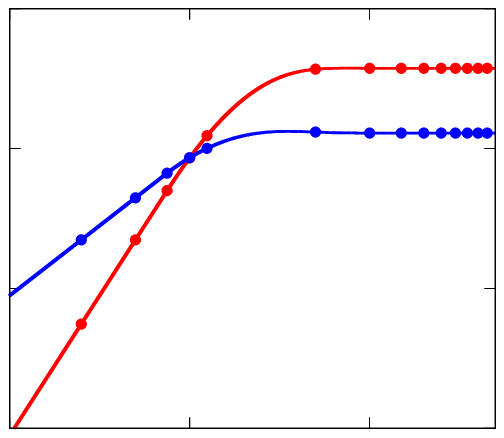}}%
    \gplfronttext
  \end{picture}%
\endgroup

%% file: saramitoO.tex
% GNUPLOT: LaTeX picture with Postscript
\begingroup
  \makeatletter
  \providecommand\color[2][]{%
    \GenericError{(gnuplot) \space\space\space\@spaces}{%
      Package color not loaded in conjunction with
      terminal option `colourtext'%
    }{See the gnuplot documentation for explanation.%
    }{Either use 'blacktext' in gnuplot or load the package
      color.sty in LaTeX.}%
    \renewcommand\color[2][]{}%
  }%
  \providecommand\includegraphics[2][]{%
    \GenericError{(gnuplot) \space\space\space\@spaces}{%
      Package graphicx or graphics not loaded%
    }{See the gnuplot documentation for explanation.%
    }{The gnuplot epslatex terminal needs graphicx.sty or graphics.sty.}%
    \renewcommand\includegraphics[2][]{}%
  }%
  \providecommand\rotatebox[2]{#2}%
  \@ifundefined{ifGPcolor}{%
    \newif\ifGPcolor
    \GPcolortrue
  }{}%
  \@ifundefined{ifGPblacktext}{%
    \newif\ifGPblacktext
    \GPblacktexttrue
  }{}%
  % define a \g@addto@macro without @ in the name:
  \let\gplgaddtomacro\g@addto@macro
  % define empty templates for all commands taking text:
  \gdef\gplbacktext{}%
  \gdef\gplfronttext{}%
  \makeatother
  \ifGPblacktext
    % no textcolor at all
    \def\colorrgb#1{}%
    \def\colorgray#1{}%
  \else
    % gray or color?
    \ifGPcolor
      \def\colorrgb#1{\color[rgb]{#1}}%
      \def\colorgray#1{\color[gray]{#1}}%
      \expandafter\def\csname LTw\endcsname{\color{white}}%
      \expandafter\def\csname LTb\endcsname{\color{black}}%
      \expandafter\def\csname LTa\endcsname{\color{black}}%
      \expandafter\def\csname LT0\endcsname{\color[rgb]{1,0,0}}%
      \expandafter\def\csname LT1\endcsname{\color[rgb]{0,1,0}}%
      \expandafter\def\csname LT2\endcsname{\color[rgb]{0,0,1}}%
      \expandafter\def\csname LT3\endcsname{\color[rgb]{1,0,1}}%
      \expandafter\def\csname LT4\endcsname{\color[rgb]{0,1,1}}%
      \expandafter\def\csname LT5\endcsname{\color[rgb]{1,1,0}}%
      \expandafter\def\csname LT6\endcsname{\color[rgb]{0,0,0}}%
      \expandafter\def\csname LT7\endcsname{\color[rgb]{1,0.3,0}}%
      \expandafter\def\csname LT8\endcsname{\color[rgb]{0.5,0.5,0.5}}%
    \else
      % gray
      \def\colorrgb#1{\color{black}}%
      \def\colorgray#1{\color[gray]{#1}}%
      \expandafter\def\csname LTw\endcsname{\color{white}}%
      \expandafter\def\csname LTb\endcsname{\color{black}}%
      \expandafter\def\csname LTa\endcsname{\color{black}}%
      \expandafter\def\csname LT0\endcsname{\color{black}}%
      \expandafter\def\csname LT1\endcsname{\color{black}}%
      \expandafter\def\csname LT2\endcsname{\color{black}}%
      \expandafter\def\csname LT3\endcsname{\color{black}}%
      \expandafter\def\csname LT4\endcsname{\color{black}}%
      \expandafter\def\csname LT5\endcsname{\color{black}}%
      \expandafter\def\csname LT6\endcsname{\color{black}}%
      \expandafter\def\csname LT7\endcsname{\color{black}}%
      \expandafter\def\csname LT8\endcsname{\color{black}}%
    \fi
  \fi
    \setlength{\unitlength}{0.0500bp}%
    \ifx\gptboxheight\undefined%
      \newlength{\gptboxheight}%
      \newlength{\gptboxwidth}%
      \newsavebox{\gptboxtext}%
    \fi%
    \setlength{\fboxrule}{0.5pt}%
    \setlength{\fboxsep}{1pt}%
\begin{picture}(3456.00,2880.00)%
\definecolor{gpBackground}{rgb}{1.000, 1.000, 1.000}%
\put(0,0){\colorbox{gpBackground}{\makebox(3456.00,2880.00)[]{}}}%
    \gplgaddtomacro\gplbacktext{%
      \csname LTb\endcsname%
      \put(490,835){\makebox(0,0)[r]{\strut{}$-10$}}%
      \put(490,1641){\makebox(0,0)[r]{\strut{}$0$}}%
      \put(490,2447){\makebox(0,0)[r]{\strut{}$10$}}%
      \put(622,212){\makebox(0,0){\strut{}$0$}}%
      \put(1513,212){\makebox(0,0){\strut{}$2$}}%
      \put(2403,212){\makebox(0,0){\strut{}$4$}}%
      \put(3294,212){\makebox(0,0){\strut{}$6$}}%
    }%
    \gplgaddtomacro\gplfronttext{%
      \csname LTb\endcsname%
      \put(100,2850){\rotatebox{0}{\makebox(0,0){\strut{}(b)}}}%
      \put(100,1641){\rotatebox{-270}{\makebox(0,0){\strut{}$\tau_{12}$}}}%
      \put(2021,-118){\makebox(0,0){\strut{}$t \dot{\gamma}_0$}}%
      \put(2021,2740){\makebox(0,0){\strut{}}}%
    }%
    \gplbacktext
    \put(0,0){\includegraphics{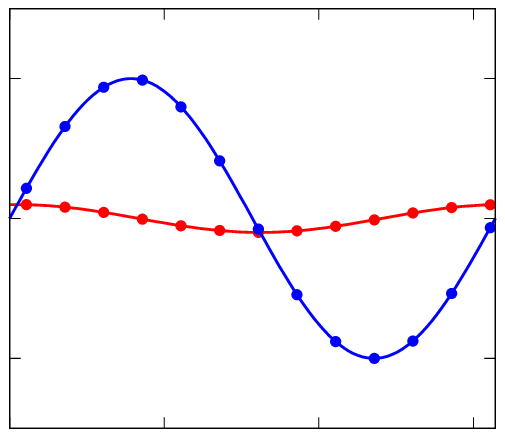}}%
    \gplfronttext
  \end{picture}%
\endgroup

%% file: fan.tex
% GNUPLOT: LaTeX picture with Postscript
\begingroup
  \makeatletter
  \providecommand\color[2][]{%
    \GenericError{(gnuplot) \space\space\space\@spaces}{%
      Package color not loaded in conjunction with
      terminal option `colourtext'%
    }{See the gnuplot documentation for explanation.%
    }{Either use 'blacktext' in gnuplot or load the package
      color.sty in LaTeX.}%
    \renewcommand\color[2][]{}%
  }%
  \providecommand\includegraphics[2][]{%
    \GenericError{(gnuplot) \space\space\space\@spaces}{%
      Package graphicx or graphics not loaded%
    }{See the gnuplot documentation for explanation.%
    }{The gnuplot epslatex terminal needs graphicx.sty or graphics.sty.}%
    \renewcommand\includegraphics[2][]{}%
  }%
  \providecommand\rotatebox[2]{#2}%
  \@ifundefined{ifGPcolor}{%
    \newif\ifGPcolor
    \GPcolortrue
  }{}%
  \@ifundefined{ifGPblacktext}{%
    \newif\ifGPblacktext
    \GPblacktexttrue
  }{}%
  % define a \g@addto@macro without @ in the name:
  \let\gplgaddtomacro\g@addto@macro
  % define empty templates for all commands taking text:
  \gdef\gplbacktext{}%
  \gdef\gplfronttext{}%
  \makeatother
  \ifGPblacktext
    % no textcolor at all
    \def\colorrgb#1{}%
    \def\colorgray#1{}%
  \else
    % gray or color?
    \ifGPcolor
      \def\colorrgb#1{\color[rgb]{#1}}%
      \def\colorgray#1{\color[gray]{#1}}%
      \expandafter\def\csname LTw\endcsname{\color{white}}%
      \expandafter\def\csname LTb\endcsname{\color{black}}%
      \expandafter\def\csname LTa\endcsname{\color{black}}%
      \expandafter\def\csname LT0\endcsname{\color[rgb]{1,0,0}}%
      \expandafter\def\csname LT1\endcsname{\color[rgb]{0,1,0}}%
      \expandafter\def\csname LT2\endcsname{\color[rgb]{0,0,1}}%
      \expandafter\def\csname LT3\endcsname{\color[rgb]{1,0,1}}%
      \expandafter\def\csname LT4\endcsname{\color[rgb]{0,1,1}}%
      \expandafter\def\csname LT5\endcsname{\color[rgb]{1,1,0}}%
      \expandafter\def\csname LT6\endcsname{\color[rgb]{0,0,0}}%
      \expandafter\def\csname LT7\endcsname{\color[rgb]{1,0.3,0}}%
      \expandafter\def\csname LT8\endcsname{\color[rgb]{0.5,0.5,0.5}}%
    \else
      % gray
      \def\colorrgb#1{\color{black}}%
      \def\colorgray#1{\color[gray]{#1}}%
      \expandafter\def\csname LTw\endcsname{\color{white}}%
      \expandafter\def\csname LTb\endcsname{\color{black}}%
      \expandafter\def\csname LTa\endcsname{\color{black}}%
      \expandafter\def\csname LT0\endcsname{\color{black}}%
      \expandafter\def\csname LT1\endcsname{\color{black}}%
      \expandafter\def\csname LT2\endcsname{\color{black}}%
      \expandafter\def\csname LT3\endcsname{\color{black}}%
      \expandafter\def\csname LT4\endcsname{\color{black}}%
      \expandafter\def\csname LT5\endcsname{\color{black}}%
      \expandafter\def\csname LT6\endcsname{\color{black}}%
      \expandafter\def\csname LT7\endcsname{\color{black}}%
      \expandafter\def\csname LT8\endcsname{\color{black}}%
    \fi
  \fi
    \setlength{\unitlength}{0.0500bp}%
    \ifx\gptboxheight\undefined%
      \newlength{\gptboxheight}%
      \newlength{\gptboxwidth}%
      \newsavebox{\gptboxtext}%
    \fi%
    \setlength{\fboxrule}{0.5pt}%
    \setlength{\fboxsep}{1pt}%
\begin{picture}(3456.00,2880.00)%
\definecolor{gpBackground}{rgb}{1.000, 1.000, 1.000}%
\put(0,0){\colorbox{gpBackground}{\makebox(3456.00,2880.00)[]{}}}%
    \gplgaddtomacro\gplbacktext{%
      \csname LTb\endcsname%
      \put(490,432){\makebox(0,0)[r]{\strut{}$10^{-3}$}}%
      \put(490,777){\makebox(0,0)[r]{\strut{}$10^{-2}$}}%
      \put(490,1123){\makebox(0,0)[r]{\strut{}$10^{-1}$}}%
      \put(490,1468){\makebox(0,0)[r]{\strut{}$10^{0}$}}%
      \put(490,1814){\makebox(0,0)[r]{\strut{}$10^{1}$}}%
      \put(490,2159){\makebox(0,0)[r]{\strut{}$10^{2}$}}%
      \put(490,2505){\makebox(0,0)[r]{\strut{}$10^{3}$}}%
      \put(490,2850){\makebox(0,0)[r]{\strut{}$10^{4}$}}%
      \put(622,212){\makebox(0,0){\strut{}$10^{-1}$}}%
      \put(1247,212){\makebox(0,0){\strut{}$10^{0}$}}%
      \put(1872,212){\makebox(0,0){\strut{}$10^{1}$}}%
      \put(2497,212){\makebox(0,0){\strut{}$10^{2}$}}%
      \put(3122,212){\makebox(0,0){\strut{}$10^{3}$}}%
      \put(2497,1814){\makebox(0,0)[l]{\strut{}$Bi$}}%
    }%
    \gplgaddtomacro\gplfronttext{%
      \csname LTb\endcsname%
      \put(-82,2850){\rotatebox{0}{\makebox(0,0){\strut{}(a)}}}%
      \put(-82,1641){\rotatebox{-270}{\makebox(0,0){\strut{}$f$}}}%
      \put(2021,-118){\makebox(0,0){\strut{}$Re_b$}}%
      \put(2021,2740){\makebox(0,0){\strut{}}}%
    }%
    \gplbacktext
    \put(0,0){\includegraphics{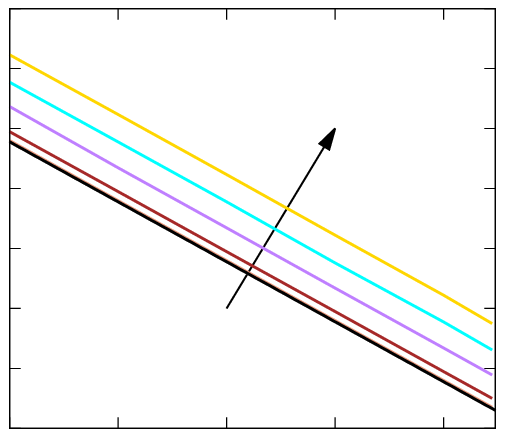}}%
    \gplfronttext
  \end{picture}%
\endgroup

%% file: fanB.tex
% GNUPLOT: LaTeX picture with Postscript
\begingroup
  \makeatletter
  \providecommand\color[2][]{%
    \GenericError{(gnuplot) \space\space\space\@spaces}{%
      Package color not loaded in conjunction with
      terminal option `colourtext'%
    }{See the gnuplot documentation for explanation.%
    }{Either use 'blacktext' in gnuplot or load the package
      color.sty in LaTeX.}%
    \renewcommand\color[2][]{}%
  }%
  \providecommand\includegraphics[2][]{%
    \GenericError{(gnuplot) \space\space\space\@spaces}{%
      Package graphicx or graphics not loaded%
    }{See the gnuplot documentation for explanation.%
    }{The gnuplot epslatex terminal needs graphicx.sty or graphics.sty.}%
    \renewcommand\includegraphics[2][]{}%
  }%
  \providecommand\rotatebox[2]{#2}%
  \@ifundefined{ifGPcolor}{%
    \newif\ifGPcolor
    \GPcolortrue
  }{}%
  \@ifundefined{ifGPblacktext}{%
    \newif\ifGPblacktext
    \GPblacktexttrue
  }{}%
  % define a \g@addto@macro without @ in the name:
  \let\gplgaddtomacro\g@addto@macro
  % define empty templates for all commands taking text:
  \gdef\gplbacktext{}%
  \gdef\gplfronttext{}%
  \makeatother
  \ifGPblacktext
    % no textcolor at all
    \def\colorrgb#1{}%
    \def\colorgray#1{}%
  \else
    % gray or color?
    \ifGPcolor
      \def\colorrgb#1{\color[rgb]{#1}}%
      \def\colorgray#1{\color[gray]{#1}}%
      \expandafter\def\csname LTw\endcsname{\color{white}}%
      \expandafter\def\csname LTb\endcsname{\color{black}}%
      \expandafter\def\csname LTa\endcsname{\color{black}}%
      \expandafter\def\csname LT0\endcsname{\color[rgb]{1,0,0}}%
      \expandafter\def\csname LT1\endcsname{\color[rgb]{0,1,0}}%
      \expandafter\def\csname LT2\endcsname{\color[rgb]{0,0,1}}%
      \expandafter\def\csname LT3\endcsname{\color[rgb]{1,0,1}}%
      \expandafter\def\csname LT4\endcsname{\color[rgb]{0,1,1}}%
      \expandafter\def\csname LT5\endcsname{\color[rgb]{1,1,0}}%
      \expandafter\def\csname LT6\endcsname{\color[rgb]{0,0,0}}%
      \expandafter\def\csname LT7\endcsname{\color[rgb]{1,0.3,0}}%
      \expandafter\def\csname LT8\endcsname{\color[rgb]{0.5,0.5,0.5}}%
    \else
      % gray
      \def\colorrgb#1{\color{black}}%
      \def\colorgray#1{\color[gray]{#1}}%
      \expandafter\def\csname LTw\endcsname{\color{white}}%
      \expandafter\def\csname LTb\endcsname{\color{black}}%
      \expandafter\def\csname LTa\endcsname{\color{black}}%
      \expandafter\def\csname LT0\endcsname{\color{black}}%
      \expandafter\def\csname LT1\endcsname{\color{black}}%
      \expandafter\def\csname LT2\endcsname{\color{black}}%
      \expandafter\def\csname LT3\endcsname{\color{black}}%
      \expandafter\def\csname LT4\endcsname{\color{black}}%
      \expandafter\def\csname LT5\endcsname{\color{black}}%
      \expandafter\def\csname LT6\endcsname{\color{black}}%
      \expandafter\def\csname LT7\endcsname{\color{black}}%
      \expandafter\def\csname LT8\endcsname{\color{black}}%
    \fi
  \fi
    \setlength{\unitlength}{0.0500bp}%
    \ifx\gptboxheight\undefined%
      \newlength{\gptboxheight}%
      \newlength{\gptboxwidth}%
      \newsavebox{\gptboxtext}%
    \fi%
    \setlength{\fboxrule}{0.5pt}%
    \setlength{\fboxsep}{1pt}%
\begin{picture}(3456.00,2880.00)%
\definecolor{gpBackground}{rgb}{1.000, 1.000, 1.000}%
\put(0,0){\colorbox{gpBackground}{\makebox(3456.00,2880.00)[]{}}}%
    \gplgaddtomacro\gplbacktext{%
      \csname LTb\endcsname%
      \put(490,833){\makebox(0,0)[r]{\strut{}$10^{1}$}}%
      \put(490,1842){\makebox(0,0)[r]{\strut{}$10^{2}$}}%
      \put(490,2850){\makebox(0,0)[r]{\strut{}$10^{3}$}}%
      \put(622,212){\makebox(0,0){\strut{}$0.2$}}%
      \put(1322,212){\makebox(0,0){\strut{}$0.4$}}%
      \put(2021,212){\makebox(0,0){\strut{}$0.6$}}%
      \put(2721,212){\makebox(0,0){\strut{}$0.8$}}%
      \put(3420,212){\makebox(0,0){\strut{}$1$}}%
    }%
    \gplgaddtomacro\gplfronttext{%
      \csname LTb\endcsname%
      \put(50,2850){\rotatebox{0}{\makebox(0,0){\strut{}(b)}}}%
      \put(50,1641){\rotatebox{-270}{\makebox(0,0){\strut{}$f$}}}%
      \put(2021,-118){\makebox(0,0){\strut{}$\beta$}}%
      \put(2021,2740){\makebox(0,0){\strut{}}}%
    }%
    \gplbacktext
    \put(0,0){\includegraphics{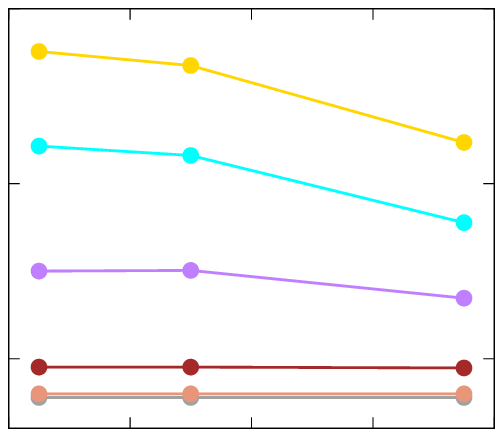}}%
    \gplfronttext
  \end{picture}%
\endgroup

%% file: meanL.tex
% GNUPLOT: LaTeX picture with Postscript
\begingroup
  \makeatletter
  \providecommand\color[2][]{%
    \GenericError{(gnuplot) \space\space\space\@spaces}{%
      Package color not loaded in conjunction with
      terminal option `colourtext'%
    }{See the gnuplot documentation for explanation.%
    }{Either use 'blacktext' in gnuplot or load the package
      color.sty in LaTeX.}%
    \renewcommand\color[2][]{}%
  }%
  \providecommand\includegraphics[2][]{%
    \GenericError{(gnuplot) \space\space\space\@spaces}{%
      Package graphicx or graphics not loaded%
    }{See the gnuplot documentation for explanation.%
    }{The gnuplot epslatex terminal needs graphicx.sty or graphics.sty.}%
    \renewcommand\includegraphics[2][]{}%
  }%
  \providecommand\rotatebox[2]{#2}%
  \@ifundefined{ifGPcolor}{%
    \newif\ifGPcolor
    \GPcolortrue
  }{}%
  \@ifundefined{ifGPblacktext}{%
    \newif\ifGPblacktext
    \GPblacktexttrue
  }{}%
  % define a \g@addto@macro without @ in the name:
  \let\gplgaddtomacro\g@addto@macro
  % define empty templates for all commands taking text:
  \gdef\gplbacktext{}%
  \gdef\gplfronttext{}%
  \makeatother
  \ifGPblacktext
    % no textcolor at all
    \def\colorrgb#1{}%
    \def\colorgray#1{}%
  \else
    % gray or color?
    \ifGPcolor
      \def\colorrgb#1{\color[rgb]{#1}}%
      \def\colorgray#1{\color[gray]{#1}}%
      \expandafter\def\csname LTw\endcsname{\color{white}}%
      \expandafter\def\csname LTb\endcsname{\color{black}}%
      \expandafter\def\csname LTa\endcsname{\color{black}}%
      \expandafter\def\csname LT0\endcsname{\color[rgb]{1,0,0}}%
      \expandafter\def\csname LT1\endcsname{\color[rgb]{0,1,0}}%
      \expandafter\def\csname LT2\endcsname{\color[rgb]{0,0,1}}%
      \expandafter\def\csname LT3\endcsname{\color[rgb]{1,0,1}}%
      \expandafter\def\csname LT4\endcsname{\color[rgb]{0,1,1}}%
      \expandafter\def\csname LT5\endcsname{\color[rgb]{1,1,0}}%
      \expandafter\def\csname LT6\endcsname{\color[rgb]{0,0,0}}%
      \expandafter\def\csname LT7\endcsname{\color[rgb]{1,0.3,0}}%
      \expandafter\def\csname LT8\endcsname{\color[rgb]{0.5,0.5,0.5}}%
    \else
      % gray
      \def\colorrgb#1{\color{black}}%
      \def\colorgray#1{\color[gray]{#1}}%
      \expandafter\def\csname LTw\endcsname{\color{white}}%
      \expandafter\def\csname LTb\endcsname{\color{black}}%
      \expandafter\def\csname LTa\endcsname{\color{black}}%
      \expandafter\def\csname LT0\endcsname{\color{black}}%
      \expandafter\def\csname LT1\endcsname{\color{black}}%
      \expandafter\def\csname LT2\endcsname{\color{black}}%
      \expandafter\def\csname LT3\endcsname{\color{black}}%
      \expandafter\def\csname LT4\endcsname{\color{black}}%
      \expandafter\def\csname LT5\endcsname{\color{black}}%
      \expandafter\def\csname LT6\endcsname{\color{black}}%
      \expandafter\def\csname LT7\endcsname{\color{black}}%
      \expandafter\def\csname LT8\endcsname{\color{black}}%
    \fi
  \fi
    \setlength{\unitlength}{0.0500bp}%
    \ifx\gptboxheight\undefined%
      \newlength{\gptboxheight}%
      \newlength{\gptboxwidth}%
      \newsavebox{\gptboxtext}%
    \fi%
    \setlength{\fboxrule}{0.5pt}%
    \setlength{\fboxsep}{1pt}%
\begin{picture}(3456.00,2880.00)%
\definecolor{gpBackground}{rgb}{1.000, 1.000, 1.000}%
\put(0,0){\colorbox{gpBackground}{\makebox(3456.00,2880.00)[]{}}}%
    \gplgaddtomacro\gplbacktext{%
      \csname LTb\endcsname%
      \put(490,432){\makebox(0,0)[r]{\strut{}$0$}}%
      \put(490,1188){\makebox(0,0)[r]{\strut{}$0.5$}}%
      \put(490,1943){\makebox(0,0)[r]{\strut{}$1$}}%
      \put(622,212){\makebox(0,0){\strut{}$0$}}%
      \put(2021,212){\makebox(0,0){\strut{}$0.5$}}%
      \put(3420,212){\makebox(0,0){\strut{}$1$}}%
      \put(3000,1565){\makebox(0,0)[l]{\strut{}$Bi$}}%
    }%
    \gplgaddtomacro\gplfronttext{%
      \csname LTb\endcsname%
      \put(50,2850){\rotatebox{0}{\makebox(0,0){\strut{}(a)}}}%
      \put(50,1641){\rotatebox{-270}{\makebox(0,0){\strut{}$u/U_b$}}}%
      \put(2021,-118){\makebox(0,0){\strut{}$y/h$}}%
      \put(2021,2740){\makebox(0,0){\strut{}}}%
    }%
    \gplbacktext
    \put(0,0){\includegraphics{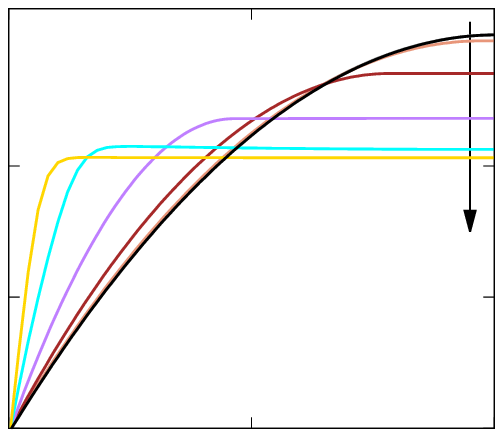}}%
    \gplfronttext
  \end{picture}%
\endgroup

%% file: volL.tex
% GNUPLOT: LaTeX picture with Postscript
\begingroup
  \makeatletter
  \providecommand\color[2][]{%
    \GenericError{(gnuplot) \space\space\space\@spaces}{%
      Package color not loaded in conjunction with
      terminal option `colourtext'%
    }{See the gnuplot documentation for explanation.%
    }{Either use 'blacktext' in gnuplot or load the package
      color.sty in LaTeX.}%
    \renewcommand\color[2][]{}%
  }%
  \providecommand\includegraphics[2][]{%
    \GenericError{(gnuplot) \space\space\space\@spaces}{%
      Package graphicx or graphics not loaded%
    }{See the gnuplot documentation for explanation.%
    }{The gnuplot epslatex terminal needs graphicx.sty or graphics.sty.}%
    \renewcommand\includegraphics[2][]{}%
  }%
  \providecommand\rotatebox[2]{#2}%
  \@ifundefined{ifGPcolor}{%
    \newif\ifGPcolor
    \GPcolortrue
  }{}%
  \@ifundefined{ifGPblacktext}{%
    \newif\ifGPblacktext
    \GPblacktexttrue
  }{}%
  % define a \g@addto@macro without @ in the name:
  \let\gplgaddtomacro\g@addto@macro
  % define empty templates for all commands taking text:
  \gdef\gplbacktext{}%
  \gdef\gplfronttext{}%
  \makeatother
  \ifGPblacktext
    % no textcolor at all
    \def\colorrgb#1{}%
    \def\colorgray#1{}%
  \else
    % gray or color?
    \ifGPcolor
      \def\colorrgb#1{\color[rgb]{#1}}%
      \def\colorgray#1{\color[gray]{#1}}%
      \expandafter\def\csname LTw\endcsname{\color{white}}%
      \expandafter\def\csname LTb\endcsname{\color{black}}%
      \expandafter\def\csname LTa\endcsname{\color{black}}%
      \expandafter\def\csname LT0\endcsname{\color[rgb]{1,0,0}}%
      \expandafter\def\csname LT1\endcsname{\color[rgb]{0,1,0}}%
      \expandafter\def\csname LT2\endcsname{\color[rgb]{0,0,1}}%
      \expandafter\def\csname LT3\endcsname{\color[rgb]{1,0,1}}%
      \expandafter\def\csname LT4\endcsname{\color[rgb]{0,1,1}}%
      \expandafter\def\csname LT5\endcsname{\color[rgb]{1,1,0}}%
      \expandafter\def\csname LT6\endcsname{\color[rgb]{0,0,0}}%
      \expandafter\def\csname LT7\endcsname{\color[rgb]{1,0.3,0}}%
      \expandafter\def\csname LT8\endcsname{\color[rgb]{0.5,0.5,0.5}}%
    \else
      % gray
      \def\colorrgb#1{\color{black}}%
      \def\colorgray#1{\color[gray]{#1}}%
      \expandafter\def\csname LTw\endcsname{\color{white}}%
      \expandafter\def\csname LTb\endcsname{\color{black}}%
      \expandafter\def\csname LTa\endcsname{\color{black}}%
      \expandafter\def\csname LT0\endcsname{\color{black}}%
      \expandafter\def\csname LT1\endcsname{\color{black}}%
      \expandafter\def\csname LT2\endcsname{\color{black}}%
      \expandafter\def\csname LT3\endcsname{\color{black}}%
      \expandafter\def\csname LT4\endcsname{\color{black}}%
      \expandafter\def\csname LT5\endcsname{\color{black}}%
      \expandafter\def\csname LT6\endcsname{\color{black}}%
      \expandafter\def\csname LT7\endcsname{\color{black}}%
      \expandafter\def\csname LT8\endcsname{\color{black}}%
    \fi
  \fi
    \setlength{\unitlength}{0.0500bp}%
    \ifx\gptboxheight\undefined%
      \newlength{\gptboxheight}%
      \newlength{\gptboxwidth}%
      \newsavebox{\gptboxtext}%
    \fi%
    \setlength{\fboxrule}{0.5pt}%
    \setlength{\fboxsep}{1pt}%
\begin{picture}(3456.00,2880.00)%
\definecolor{gpBackground}{rgb}{1.000, 1.000, 1.000}%
\put(0,0){\colorbox{gpBackground}{\makebox(3456.00,2880.00)[]{}}}%
    \gplgaddtomacro\gplbacktext{%
      \csname LTb\endcsname%
      \put(490,432){\makebox(0,0)[r]{\strut{}$0$}}%
      \put(490,1641){\makebox(0,0)[r]{\strut{}$50$}}%
      \put(490,2850){\makebox(0,0)[r]{\strut{}$100$}}%
      \put(649,212){\makebox(0,0){\strut{}$0$}}%
      \put(2021,212){\makebox(0,0){\strut{}$500$}}%
      \put(3393,212){\makebox(0,0){\strut{}$1000$}}%
    }%
    \gplgaddtomacro\gplfronttext{%
      \csname LTb\endcsname%
      \put(50,2850){\rotatebox{0}{\makebox(0,0){\strut{}(b)}}}%
      \put(50,1641){\rotatebox{-270}{\makebox(0,0){\strut{}$Vol_s \%$}}}%
      \put(2021,-118){\makebox(0,0){\strut{}$Bi$}}%
      \put(2021,2740){\makebox(0,0){\strut{}}}%
    }%
    \gplbacktext
    \put(0,0){\includegraphics{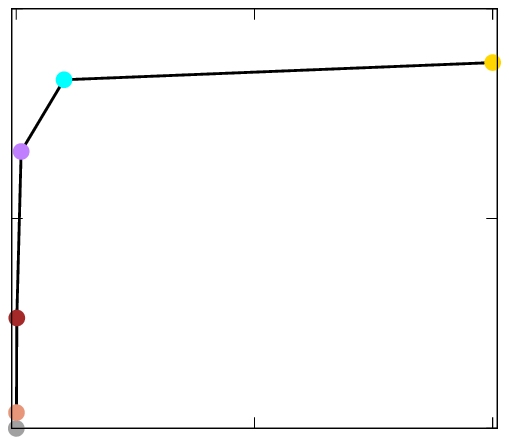}}%
    \gplfronttext
  \end{picture}%
\endgroup

%% file: fan2800.tex
% GNUPLOT: LaTeX picture with Postscript
\begingroup
  \makeatletter
  \providecommand\color[2][]{%
    \GenericError{(gnuplot) \space\space\space\@spaces}{%
      Package color not loaded in conjunction with
      terminal option `colourtext'%
    }{See the gnuplot documentation for explanation.%
    }{Either use 'blacktext' in gnuplot or load the package
      color.sty in LaTeX.}%
    \renewcommand\color[2][]{}%
  }%
  \providecommand\includegraphics[2][]{%
    \GenericError{(gnuplot) \space\space\space\@spaces}{%
      Package graphicx or graphics not loaded%
    }{See the gnuplot documentation for explanation.%
    }{The gnuplot epslatex terminal needs graphicx.sty or graphics.sty.}%
    \renewcommand\includegraphics[2][]{}%
  }%
  \providecommand\rotatebox[2]{#2}%
  \@ifundefined{ifGPcolor}{%
    \newif\ifGPcolor
    \GPcolortrue
  }{}%
  \@ifundefined{ifGPblacktext}{%
    \newif\ifGPblacktext
    \GPblacktexttrue
  }{}%
  % define a \g@addto@macro without @ in the name:
  \let\gplgaddtomacro\g@addto@macro
  % define empty templates for all commands taking text:
  \gdef\gplbacktext{}%
  \gdef\gplfronttext{}%
  \makeatother
  \ifGPblacktext
    % no textcolor at all
    \def\colorrgb#1{}%
    \def\colorgray#1{}%
  \else
    % gray or color?
    \ifGPcolor
      \def\colorrgb#1{\color[rgb]{#1}}%
      \def\colorgray#1{\color[gray]{#1}}%
      \expandafter\def\csname LTw\endcsname{\color{white}}%
      \expandafter\def\csname LTb\endcsname{\color{black}}%
      \expandafter\def\csname LTa\endcsname{\color{black}}%
      \expandafter\def\csname LT0\endcsname{\color[rgb]{1,0,0}}%
      \expandafter\def\csname LT1\endcsname{\color[rgb]{0,1,0}}%
      \expandafter\def\csname LT2\endcsname{\color[rgb]{0,0,1}}%
      \expandafter\def\csname LT3\endcsname{\color[rgb]{1,0,1}}%
      \expandafter\def\csname LT4\endcsname{\color[rgb]{0,1,1}}%
      \expandafter\def\csname LT5\endcsname{\color[rgb]{1,1,0}}%
      \expandafter\def\csname LT6\endcsname{\color[rgb]{0,0,0}}%
      \expandafter\def\csname LT7\endcsname{\color[rgb]{1,0.3,0}}%
      \expandafter\def\csname LT8\endcsname{\color[rgb]{0.5,0.5,0.5}}%
    \else
      % gray
      \def\colorrgb#1{\color{black}}%
      \def\colorgray#1{\color[gray]{#1}}%
      \expandafter\def\csname LTw\endcsname{\color{white}}%
      \expandafter\def\csname LTb\endcsname{\color{black}}%
      \expandafter\def\csname LTa\endcsname{\color{black}}%
      \expandafter\def\csname LT0\endcsname{\color{black}}%
      \expandafter\def\csname LT1\endcsname{\color{black}}%
      \expandafter\def\csname LT2\endcsname{\color{black}}%
      \expandafter\def\csname LT3\endcsname{\color{black}}%
      \expandafter\def\csname LT4\endcsname{\color{black}}%
      \expandafter\def\csname LT5\endcsname{\color{black}}%
      \expandafter\def\csname LT6\endcsname{\color{black}}%
      \expandafter\def\csname LT7\endcsname{\color{black}}%
      \expandafter\def\csname LT8\endcsname{\color{black}}%
    \fi
  \fi
    \setlength{\unitlength}{0.0500bp}%
    \ifx\gptboxheight\undefined%
      \newlength{\gptboxheight}%
      \newlength{\gptboxwidth}%
      \newsavebox{\gptboxtext}%
    \fi%
    \setlength{\fboxrule}{0.5pt}%
    \setlength{\fboxsep}{1pt}%
\begin{picture}(3456.00,2880.00)%
\definecolor{gpBackground}{rgb}{1.000, 1.000, 1.000}%
\put(0,0){\colorbox{gpBackground}{\makebox(3456.00,2880.00)[]{}}}%
    \gplgaddtomacro\gplbacktext{%
      \csname LTb\endcsname%
      \put(490,432){\makebox(0,0)[r]{\strut{}$10^{-3}$}}%
      \put(490,1641){\makebox(0,0)[r]{\strut{}$10^{-2}$}}%
      \put(490,2850){\makebox(0,0)[r]{\strut{}$10^{-1}$}}%
      \put(1169,212){\makebox(0,0){\strut{}$10^{-1}$}}%
      \put(1716,212){\makebox(0,0){\strut{}$10^{0}$}}%
      \put(2263,212){\makebox(0,0){\strut{}$10^{1}$}}%
      \put(2811,212){\makebox(0,0){\strut{}$10^{2}$}}%
      \put(3358,212){\makebox(0,0){\strut{}$10^{3}$}}%
    }%
    \gplgaddtomacro\gplfronttext{%
      \csname LTb\endcsname%
      \put(-82,2850){\rotatebox{0}{\makebox(0,0){\strut{}(a)}}}%
      \put(-82,1641){\rotatebox{-270}{\makebox(0,0){\strut{}$f$}}}%
      \put(2021,-118){\makebox(0,0){\strut{}$Bi$}}%
      \put(2021,2740){\makebox(0,0){\strut{}}}%
    }%
    \gplbacktext
    \put(0,0){\includegraphics{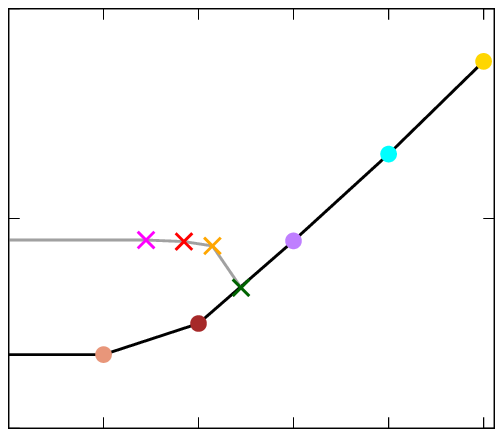}}%
    \gplfronttext
  \end{picture}%
\endgroup

%% file: reynolds.tex
% GNUPLOT: LaTeX picture with Postscript
\begingroup
  \makeatletter
  \providecommand\color[2][]{%
    \GenericError{(gnuplot) \space\space\space\@spaces}{%
      Package color not loaded in conjunction with
      terminal option `colourtext'%
    }{See the gnuplot documentation for explanation.%
    }{Either use 'blacktext' in gnuplot or load the package
      color.sty in LaTeX.}%
    \renewcommand\color[2][]{}%
  }%
  \providecommand\includegraphics[2][]{%
    \GenericError{(gnuplot) \space\space\space\@spaces}{%
      Package graphicx or graphics not loaded%
    }{See the gnuplot documentation for explanation.%
    }{The gnuplot epslatex terminal needs graphicx.sty or graphics.sty.}%
    \renewcommand\includegraphics[2][]{}%
  }%
  \providecommand\rotatebox[2]{#2}%
  \@ifundefined{ifGPcolor}{%
    \newif\ifGPcolor
    \GPcolortrue
  }{}%
  \@ifundefined{ifGPblacktext}{%
    \newif\ifGPblacktext
    \GPblacktexttrue
  }{}%
  % define a \g@addto@macro without @ in the name:
  \let\gplgaddtomacro\g@addto@macro
  % define empty templates for all commands taking text:
  \gdef\gplbacktext{}%
  \gdef\gplfronttext{}%
  \makeatother
  \ifGPblacktext
    % no textcolor at all
    \def\colorrgb#1{}%
    \def\colorgray#1{}%
  \else
    % gray or color?
    \ifGPcolor
      \def\colorrgb#1{\color[rgb]{#1}}%
      \def\colorgray#1{\color[gray]{#1}}%
      \expandafter\def\csname LTw\endcsname{\color{white}}%
      \expandafter\def\csname LTb\endcsname{\color{black}}%
      \expandafter\def\csname LTa\endcsname{\color{black}}%
      \expandafter\def\csname LT0\endcsname{\color[rgb]{1,0,0}}%
      \expandafter\def\csname LT1\endcsname{\color[rgb]{0,1,0}}%
      \expandafter\def\csname LT2\endcsname{\color[rgb]{0,0,1}}%
      \expandafter\def\csname LT3\endcsname{\color[rgb]{1,0,1}}%
      \expandafter\def\csname LT4\endcsname{\color[rgb]{0,1,1}}%
      \expandafter\def\csname LT5\endcsname{\color[rgb]{1,1,0}}%
      \expandafter\def\csname LT6\endcsname{\color[rgb]{0,0,0}}%
      \expandafter\def\csname LT7\endcsname{\color[rgb]{1,0.3,0}}%
      \expandafter\def\csname LT8\endcsname{\color[rgb]{0.5,0.5,0.5}}%
    \else
      % gray
      \def\colorrgb#1{\color{black}}%
      \def\colorgray#1{\color[gray]{#1}}%
      \expandafter\def\csname LTw\endcsname{\color{white}}%
      \expandafter\def\csname LTb\endcsname{\color{black}}%
      \expandafter\def\csname LTa\endcsname{\color{black}}%
      \expandafter\def\csname LT0\endcsname{\color{black}}%
      \expandafter\def\csname LT1\endcsname{\color{black}}%
      \expandafter\def\csname LT2\endcsname{\color{black}}%
      \expandafter\def\csname LT3\endcsname{\color{black}}%
      \expandafter\def\csname LT4\endcsname{\color{black}}%
      \expandafter\def\csname LT5\endcsname{\color{black}}%
      \expandafter\def\csname LT6\endcsname{\color{black}}%
      \expandafter\def\csname LT7\endcsname{\color{black}}%
      \expandafter\def\csname LT8\endcsname{\color{black}}%
    \fi
  \fi
    \setlength{\unitlength}{0.0500bp}%
    \ifx\gptboxheight\undefined%
      \newlength{\gptboxheight}%
      \newlength{\gptboxwidth}%
      \newsavebox{\gptboxtext}%
    \fi%
    \setlength{\fboxrule}{0.5pt}%
    \setlength{\fboxsep}{1pt}%
\begin{picture}(3456.00,2880.00)%
\definecolor{gpBackground}{rgb}{1.000, 1.000, 1.000}%
\put(0,0){\colorbox{gpBackground}{\makebox(3456.00,2880.00)[]{}}}%
    \gplgaddtomacro\gplbacktext{%
      \csname LTb\endcsname%
      \put(490,835){\makebox(0,0)[r]{\strut{}$140$}}%
      \put(490,1641){\makebox(0,0)[r]{\strut{}$160$}}%
      \put(490,2447){\makebox(0,0)[r]{\strut{}$180$}}%
      \put(712,212){\makebox(0,0){\strut{}$0$}}%
      \put(1615,212){\makebox(0,0){\strut{}$1$}}%
      \put(2517,212){\makebox(0,0){\strut{}$2$}}%
      \put(3420,212){\makebox(0,0){\strut{}$3$}}%
    }%
    \gplgaddtomacro\gplfronttext{%
      \csname LTb\endcsname%
      \put(50,2850){\rotatebox{0}{\makebox(0,0){\strut{}(b)}}}%
      \put(50,1641){\rotatebox{-270}{\makebox(0,0){\strut{}$Re_\tau$}}}%
      \put(2021,-118){\makebox(0,0){\strut{}$Bi$}}%
      \put(2021,2740){\makebox(0,0){\strut{}}}%
    }%
    \gplbacktext
    \put(0,0){\includegraphics{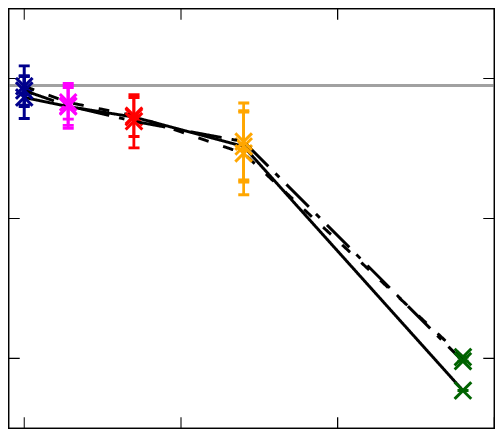}}%
    \gplfronttext
  \end{picture}%
\endgroup

%% file: dpdx.tex
% GNUPLOT: LaTeX picture with Postscript
\begingroup
  \makeatletter
  \providecommand\color[2][]{%
    \GenericError{(gnuplot) \space\space\space\@spaces}{%
      Package color not loaded in conjunction with
      terminal option `colourtext'%
    }{See the gnuplot documentation for explanation.%
    }{Either use 'blacktext' in gnuplot or load the package
      color.sty in LaTeX.}%
    \renewcommand\color[2][]{}%
  }%
  \providecommand\includegraphics[2][]{%
    \GenericError{(gnuplot) \space\space\space\@spaces}{%
      Package graphicx or graphics not loaded%
    }{See the gnuplot documentation for explanation.%
    }{The gnuplot epslatex terminal needs graphicx.sty or graphics.sty.}%
    \renewcommand\includegraphics[2][]{}%
  }%
  \providecommand\rotatebox[2]{#2}%
  \@ifundefined{ifGPcolor}{%
    \newif\ifGPcolor
    \GPcolortrue
  }{}%
  \@ifundefined{ifGPblacktext}{%
    \newif\ifGPblacktext
    \GPblacktexttrue
  }{}%
  % define a \g@addto@macro without @ in the name:
  \let\gplgaddtomacro\g@addto@macro
  % define empty templates for all commands taking text:
  \gdef\gplbacktext{}%
  \gdef\gplfronttext{}%
  \makeatother
  \ifGPblacktext
    % no textcolor at all
    \def\colorrgb#1{}%
    \def\colorgray#1{}%
  \else
    % gray or color?
    \ifGPcolor
      \def\colorrgb#1{\color[rgb]{#1}}%
      \def\colorgray#1{\color[gray]{#1}}%
      \expandafter\def\csname LTw\endcsname{\color{white}}%
      \expandafter\def\csname LTb\endcsname{\color{black}}%
      \expandafter\def\csname LTa\endcsname{\color{black}}%
      \expandafter\def\csname LT0\endcsname{\color[rgb]{1,0,0}}%
      \expandafter\def\csname LT1\endcsname{\color[rgb]{0,1,0}}%
      \expandafter\def\csname LT2\endcsname{\color[rgb]{0,0,1}}%
      \expandafter\def\csname LT3\endcsname{\color[rgb]{1,0,1}}%
      \expandafter\def\csname LT4\endcsname{\color[rgb]{0,1,1}}%
      \expandafter\def\csname LT5\endcsname{\color[rgb]{1,1,0}}%
      \expandafter\def\csname LT6\endcsname{\color[rgb]{0,0,0}}%
      \expandafter\def\csname LT7\endcsname{\color[rgb]{1,0.3,0}}%
      \expandafter\def\csname LT8\endcsname{\color[rgb]{0.5,0.5,0.5}}%
    \else
      % gray
      \def\colorrgb#1{\color{black}}%
      \def\colorgray#1{\color[gray]{#1}}%
      \expandafter\def\csname LTw\endcsname{\color{white}}%
      \expandafter\def\csname LTb\endcsname{\color{black}}%
      \expandafter\def\csname LTa\endcsname{\color{black}}%
      \expandafter\def\csname LT0\endcsname{\color{black}}%
      \expandafter\def\csname LT1\endcsname{\color{black}}%
      \expandafter\def\csname LT2\endcsname{\color{black}}%
      \expandafter\def\csname LT3\endcsname{\color{black}}%
      \expandafter\def\csname LT4\endcsname{\color{black}}%
      \expandafter\def\csname LT5\endcsname{\color{black}}%
      \expandafter\def\csname LT6\endcsname{\color{black}}%
      \expandafter\def\csname LT7\endcsname{\color{black}}%
      \expandafter\def\csname LT8\endcsname{\color{black}}%
    \fi
  \fi
    \setlength{\unitlength}{0.0500bp}%
    \ifx\gptboxheight\undefined%
      \newlength{\gptboxheight}%
      \newlength{\gptboxwidth}%
      \newsavebox{\gptboxtext}%
    \fi%
    \setlength{\fboxrule}{0.5pt}%
    \setlength{\fboxsep}{1pt}%
\begin{picture}(3456.00,2880.00)%
\definecolor{gpBackground}{rgb}{1.000, 1.000, 1.000}%
\put(0,0){\colorbox{gpBackground}{\makebox(3456.00,2880.00)[]{}}}%
    \gplgaddtomacro\gplbacktext{%
      \csname LTb\endcsname%
      \put(490,432){\makebox(0,0)[r]{\strut{}$0.002$}}%
      \put(490,2044){\makebox(0,0)[r]{\strut{}$0.004$}}%
      \put(622,212){\makebox(0,0){\strut{}$0$}}%
      \put(1741,212){\makebox(0,0){\strut{}$200$}}%
      \put(2860,212){\makebox(0,0){\strut{}$400$}}%
    }%
    \gplgaddtomacro\gplfronttext{%
      \csname LTb\endcsname%
      \put(-214,2850){\rotatebox{0}{\makebox(0,0){\strut{}(a)}}}%
      \put(-214,1641){\rotatebox{-270}{\makebox(0,0){\strut{}$dp/dx$}}}%
      \put(2021,-118){\makebox(0,0){\strut{}$t U_b/h$}}%
      \put(2021,2740){\makebox(0,0){\strut{}}}%
    }%
    \gplbacktext
    \put(0,0){\includegraphics{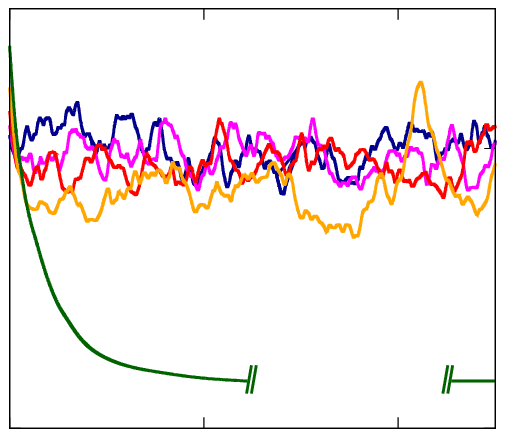}}%
    \gplfronttext
  \end{picture}%
\endgroup

%% file: volPDF.tex
% GNUPLOT: LaTeX picture with Postscript
\begingroup
  \makeatletter
  \providecommand\color[2][]{%
    \GenericError{(gnuplot) \space\space\space\@spaces}{%
      Package color not loaded in conjunction with
      terminal option `colourtext'%
    }{See the gnuplot documentation for explanation.%
    }{Either use 'blacktext' in gnuplot or load the package
      color.sty in LaTeX.}%
    \renewcommand\color[2][]{}%
  }%
  \providecommand\includegraphics[2][]{%
    \GenericError{(gnuplot) \space\space\space\@spaces}{%
      Package graphicx or graphics not loaded%
    }{See the gnuplot documentation for explanation.%
    }{The gnuplot epslatex terminal needs graphicx.sty or graphics.sty.}%
    \renewcommand\includegraphics[2][]{}%
  }%
  \providecommand\rotatebox[2]{#2}%
  \@ifundefined{ifGPcolor}{%
    \newif\ifGPcolor
    \GPcolortrue
  }{}%
  \@ifundefined{ifGPblacktext}{%
    \newif\ifGPblacktext
    \GPblacktexttrue
  }{}%
  % define a \g@addto@macro without @ in the name:
  \let\gplgaddtomacro\g@addto@macro
  % define empty templates for all commands taking text:
  \gdef\gplbacktext{}%
  \gdef\gplfronttext{}%
  \makeatother
  \ifGPblacktext
    % no textcolor at all
    \def\colorrgb#1{}%
    \def\colorgray#1{}%
  \else
    % gray or color?
    \ifGPcolor
      \def\colorrgb#1{\color[rgb]{#1}}%
      \def\colorgray#1{\color[gray]{#1}}%
      \expandafter\def\csname LTw\endcsname{\color{white}}%
      \expandafter\def\csname LTb\endcsname{\color{black}}%
      \expandafter\def\csname LTa\endcsname{\color{black}}%
      \expandafter\def\csname LT0\endcsname{\color[rgb]{1,0,0}}%
      \expandafter\def\csname LT1\endcsname{\color[rgb]{0,1,0}}%
      \expandafter\def\csname LT2\endcsname{\color[rgb]{0,0,1}}%
      \expandafter\def\csname LT3\endcsname{\color[rgb]{1,0,1}}%
      \expandafter\def\csname LT4\endcsname{\color[rgb]{0,1,1}}%
      \expandafter\def\csname LT5\endcsname{\color[rgb]{1,1,0}}%
      \expandafter\def\csname LT6\endcsname{\color[rgb]{0,0,0}}%
      \expandafter\def\csname LT7\endcsname{\color[rgb]{1,0.3,0}}%
      \expandafter\def\csname LT8\endcsname{\color[rgb]{0.5,0.5,0.5}}%
    \else
      % gray
      \def\colorrgb#1{\color{black}}%
      \def\colorgray#1{\color[gray]{#1}}%
      \expandafter\def\csname LTw\endcsname{\color{white}}%
      \expandafter\def\csname LTb\endcsname{\color{black}}%
      \expandafter\def\csname LTa\endcsname{\color{black}}%
      \expandafter\def\csname LT0\endcsname{\color{black}}%
      \expandafter\def\csname LT1\endcsname{\color{black}}%
      \expandafter\def\csname LT2\endcsname{\color{black}}%
      \expandafter\def\csname LT3\endcsname{\color{black}}%
      \expandafter\def\csname LT4\endcsname{\color{black}}%
      \expandafter\def\csname LT5\endcsname{\color{black}}%
      \expandafter\def\csname LT6\endcsname{\color{black}}%
      \expandafter\def\csname LT7\endcsname{\color{black}}%
      \expandafter\def\csname LT8\endcsname{\color{black}}%
    \fi
  \fi
    \setlength{\unitlength}{0.0500bp}%
    \ifx\gptboxheight\undefined%
      \newlength{\gptboxheight}%
      \newlength{\gptboxwidth}%
      \newsavebox{\gptboxtext}%
    \fi%
    \setlength{\fboxrule}{0.5pt}%
    \setlength{\fboxsep}{1pt}%
\begin{picture}(3456.00,2880.00)%
\definecolor{gpBackground}{rgb}{1.000, 1.000, 1.000}%
\put(0,0){\colorbox{gpBackground}{\makebox(3456.00,2880.00)[]{}}}%
    \gplgaddtomacro\gplbacktext{%
      \csname LTb\endcsname%
      \put(490,522){\makebox(0,0)[r]{\strut{}$0$}}%
      \put(490,1641){\makebox(0,0)[r]{\strut{}$0.5$}}%
      \put(490,2760){\makebox(0,0)[r]{\strut{}$1$}}%
      \put(622,212){\makebox(0,0){\strut{}$0$}}%
      \put(2021,212){\makebox(0,0){\strut{}$0.5$}}%
      \put(3420,212){\makebox(0,0){\strut{}$1$}}%
    }%
    \gplgaddtomacro\gplfronttext{%
      \csname LTb\endcsname%
      \put(50,2850){\rotatebox{0}{\makebox(0,0){\strut{}(b)}}}%
      \put(50,1641){\rotatebox{-270}{\makebox(0,0){\strut{}$P_s$}}}%
      \put(2021,-118){\makebox(0,0){\strut{}$y/h$}}%
      \put(2021,2740){\makebox(0,0){\strut{}}}%
      \csname LTb\endcsname%
      \put(1282,2677){\makebox(0,0)[r]{\strut{}$0\%$}}%
      \csname LTb\endcsname%
      \put(1282,2457){\makebox(0,0)[r]{\strut{}$5\%$}}%
      \csname LTb\endcsname%
      \put(1282,2237){\makebox(0,0)[r]{\strut{}$26\%$}}%
      \csname LTb\endcsname%
      \put(1282,2017){\makebox(0,0)[r]{\strut{}$43\%$}}%
      \csname LTb\endcsname%
      \put(1282,1797){\makebox(0,0)[r]{\strut{}$54\%$}}%
    }%
    \gplbacktext
    \put(0,0){\includegraphics{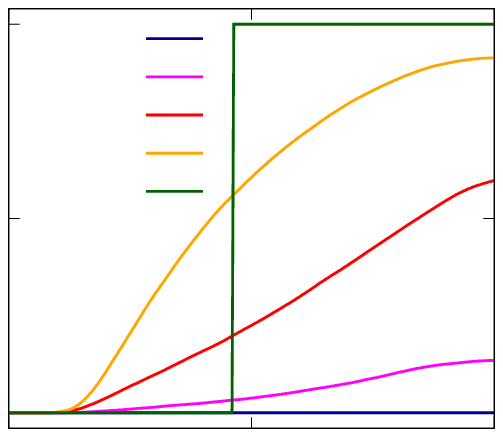}}%
    \gplfronttext
  \end{picture}%
\endgroup

%% file: inter071.tex
% GNUPLOT: LaTeX picture with Postscript
\begingroup
  \makeatletter
  \providecommand\color[2][]{%
    \GenericError{(gnuplot) \space\space\space\@spaces}{%
      Package color not loaded in conjunction with
      terminal option `colourtext'%
    }{See the gnuplot documentation for explanation.%
    }{Either use 'blacktext' in gnuplot or load the package
      color.sty in LaTeX.}%
    \renewcommand\color[2][]{}%
  }%
  \providecommand\includegraphics[2][]{%
    \GenericError{(gnuplot) \space\space\space\@spaces}{%
      Package graphicx or graphics not loaded%
    }{See the gnuplot documentation for explanation.%
    }{The gnuplot epslatex terminal needs graphicx.sty or graphics.sty.}%
    \renewcommand\includegraphics[2][]{}%
  }%
  \providecommand\rotatebox[2]{#2}%
  \@ifundefined{ifGPcolor}{%
    \newif\ifGPcolor
    \GPcolortrue
  }{}%
  \@ifundefined{ifGPblacktext}{%
    \newif\ifGPblacktext
    \GPblacktexttrue
  }{}%
  % define a \g@addto@macro without @ in the name:
  \let\gplgaddtomacro\g@addto@macro
  % define empty templates for all commands taking text:
  \gdef\gplbacktext{}%
  \gdef\gplfronttext{}%
  \makeatother
  \ifGPblacktext
    % no textcolor at all
    \def\colorrgb#1{}%
    \def\colorgray#1{}%
  \else
    % gray or color?
    \ifGPcolor
      \def\colorrgb#1{\color[rgb]{#1}}%
      \def\colorgray#1{\color[gray]{#1}}%
      \expandafter\def\csname LTw\endcsname{\color{white}}%
      \expandafter\def\csname LTb\endcsname{\color{black}}%
      \expandafter\def\csname LTa\endcsname{\color{black}}%
      \expandafter\def\csname LT0\endcsname{\color[rgb]{1,0,0}}%
      \expandafter\def\csname LT1\endcsname{\color[rgb]{0,1,0}}%
      \expandafter\def\csname LT2\endcsname{\color[rgb]{0,0,1}}%
      \expandafter\def\csname LT3\endcsname{\color[rgb]{1,0,1}}%
      \expandafter\def\csname LT4\endcsname{\color[rgb]{0,1,1}}%
      \expandafter\def\csname LT5\endcsname{\color[rgb]{1,1,0}}%
      \expandafter\def\csname LT6\endcsname{\color[rgb]{0,0,0}}%
      \expandafter\def\csname LT7\endcsname{\color[rgb]{1,0.3,0}}%
      \expandafter\def\csname LT8\endcsname{\color[rgb]{0.5,0.5,0.5}}%
    \else
      % gray
      \def\colorrgb#1{\color{black}}%
      \def\colorgray#1{\color[gray]{#1}}%
      \expandafter\def\csname LTw\endcsname{\color{white}}%
      \expandafter\def\csname LTb\endcsname{\color{black}}%
      \expandafter\def\csname LTa\endcsname{\color{black}}%
      \expandafter\def\csname LT0\endcsname{\color{black}}%
      \expandafter\def\csname LT1\endcsname{\color{black}}%
      \expandafter\def\csname LT2\endcsname{\color{black}}%
      \expandafter\def\csname LT3\endcsname{\color{black}}%
      \expandafter\def\csname LT4\endcsname{\color{black}}%
      \expandafter\def\csname LT5\endcsname{\color{black}}%
      \expandafter\def\csname LT6\endcsname{\color{black}}%
      \expandafter\def\csname LT7\endcsname{\color{black}}%
      \expandafter\def\csname LT8\endcsname{\color{black}}%
    \fi
  \fi
    \setlength{\unitlength}{0.0500bp}%
    \ifx\gptboxheight\undefined%
      \newlength{\gptboxheight}%
      \newlength{\gptboxwidth}%
      \newsavebox{\gptboxtext}%
    \fi%
    \setlength{\fboxrule}{0.5pt}%
    \setlength{\fboxsep}{1pt}%
\begin{picture}(1728.00,2880.00)%
\definecolor{gpBackground}{rgb}{1.000, 1.000, 1.000}%
\put(0,0){\colorbox{gpBackground}{\makebox(1728.00,2880.00)[]{}}}%
    \gplgaddtomacro\gplbacktext{%
      \csname LTb\endcsname%
      \put(10,2850){\rotatebox{0}{\makebox(0,0){\strut{}(a)}}}%
      \put(311,212){\makebox(0,0){\strut{}$0$}}%
      \put(1010,212){\makebox(0,0){\strut{}$10$}}%
      \put(1709,212){\makebox(0,0){\strut{}$20$}}%
      \put(171,458){\makebox(0,0)[l]{\strut{}S}}%
      \put(171,721){\makebox(0,0)[l]{\strut{}F}}%
      \put(171,984){\makebox(0,0)[l]{\strut{}S}}%
      \put(171,1247){\makebox(0,0)[l]{\strut{}F}}%
      \put(171,1510){\makebox(0,0)[l]{\strut{}S}}%
      \put(171,1772){\makebox(0,0)[l]{\strut{}F}}%
      \put(171,2035){\makebox(0,0)[l]{\strut{}S}}%
      \put(171,2298){\makebox(0,0)[l]{\strut{}F}}%
      \put(171,2561){\makebox(0,0)[l]{\strut{}S}}%
      \put(171,2824){\makebox(0,0)[l]{\strut{}F}}%
    }%
    \gplgaddtomacro\gplfronttext{%
      \csname LTb\endcsname%
      \put(1010,-118){\makebox(0,0){\strut{}$t U_b/h$}}%
      \put(1010,2740){\makebox(0,0){\strut{}}}%
    }%
    \gplbacktext
    \put(0,0){\includegraphics{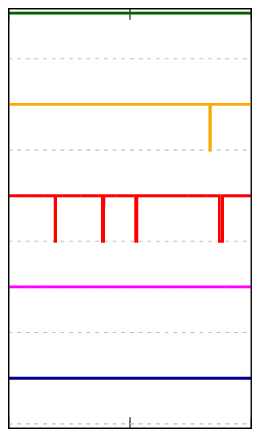}}%
    \gplfronttext
  \end{picture}%
\endgroup

%% file: inter141.tex
% GNUPLOT: LaTeX picture with Postscript
\begingroup
  \makeatletter
  \providecommand\color[2][]{%
    \GenericError{(gnuplot) \space\space\space\@spaces}{%
      Package color not loaded in conjunction with
      terminal option `colourtext'%
    }{See the gnuplot documentation for explanation.%
    }{Either use 'blacktext' in gnuplot or load the package
      color.sty in LaTeX.}%
    \renewcommand\color[2][]{}%
  }%
  \providecommand\includegraphics[2][]{%
    \GenericError{(gnuplot) \space\space\space\@spaces}{%
      Package graphicx or graphics not loaded%
    }{See the gnuplot documentation for explanation.%
    }{The gnuplot epslatex terminal needs graphicx.sty or graphics.sty.}%
    \renewcommand\includegraphics[2][]{}%
  }%
  \providecommand\rotatebox[2]{#2}%
  \@ifundefined{ifGPcolor}{%
    \newif\ifGPcolor
    \GPcolortrue
  }{}%
  \@ifundefined{ifGPblacktext}{%
    \newif\ifGPblacktext
    \GPblacktexttrue
  }{}%
  % define a \g@addto@macro without @ in the name:
  \let\gplgaddtomacro\g@addto@macro
  % define empty templates for all commands taking text:
  \gdef\gplbacktext{}%
  \gdef\gplfronttext{}%
  \makeatother
  \ifGPblacktext
    % no textcolor at all
    \def\colorrgb#1{}%
    \def\colorgray#1{}%
  \else
    % gray or color?
    \ifGPcolor
      \def\colorrgb#1{\color[rgb]{#1}}%
      \def\colorgray#1{\color[gray]{#1}}%
      \expandafter\def\csname LTw\endcsname{\color{white}}%
      \expandafter\def\csname LTb\endcsname{\color{black}}%
      \expandafter\def\csname LTa\endcsname{\color{black}}%
      \expandafter\def\csname LT0\endcsname{\color[rgb]{1,0,0}}%
      \expandafter\def\csname LT1\endcsname{\color[rgb]{0,1,0}}%
      \expandafter\def\csname LT2\endcsname{\color[rgb]{0,0,1}}%
      \expandafter\def\csname LT3\endcsname{\color[rgb]{1,0,1}}%
      \expandafter\def\csname LT4\endcsname{\color[rgb]{0,1,1}}%
      \expandafter\def\csname LT5\endcsname{\color[rgb]{1,1,0}}%
      \expandafter\def\csname LT6\endcsname{\color[rgb]{0,0,0}}%
      \expandafter\def\csname LT7\endcsname{\color[rgb]{1,0.3,0}}%
      \expandafter\def\csname LT8\endcsname{\color[rgb]{0.5,0.5,0.5}}%
    \else
      % gray
      \def\colorrgb#1{\color{black}}%
      \def\colorgray#1{\color[gray]{#1}}%
      \expandafter\def\csname LTw\endcsname{\color{white}}%
      \expandafter\def\csname LTb\endcsname{\color{black}}%
      \expandafter\def\csname LTa\endcsname{\color{black}}%
      \expandafter\def\csname LT0\endcsname{\color{black}}%
      \expandafter\def\csname LT1\endcsname{\color{black}}%
      \expandafter\def\csname LT2\endcsname{\color{black}}%
      \expandafter\def\csname LT3\endcsname{\color{black}}%
      \expandafter\def\csname LT4\endcsname{\color{black}}%
      \expandafter\def\csname LT5\endcsname{\color{black}}%
      \expandafter\def\csname LT6\endcsname{\color{black}}%
      \expandafter\def\csname LT7\endcsname{\color{black}}%
      \expandafter\def\csname LT8\endcsname{\color{black}}%
    \fi
  \fi
    \setlength{\unitlength}{0.0500bp}%
    \ifx\gptboxheight\undefined%
      \newlength{\gptboxheight}%
      \newlength{\gptboxwidth}%
      \newsavebox{\gptboxtext}%
    \fi%
    \setlength{\fboxrule}{0.5pt}%
    \setlength{\fboxsep}{1pt}%
\begin{picture}(1728.00,2880.00)%
\definecolor{gpBackground}{rgb}{1.000, 1.000, 1.000}%
\put(0,0){\colorbox{gpBackground}{\makebox(1728.00,2880.00)[]{}}}%
    \gplgaddtomacro\gplbacktext{%
      \csname LTb\endcsname%
      \put(80,2850){\rotatebox{0}{\makebox(0,0){\strut{}(b)}}}%
      \put(311,212){\makebox(0,0){\strut{}$0$}}%
      \put(1010,212){\makebox(0,0){\strut{}$10$}}%
      \put(1709,212){\makebox(0,0){\strut{}$20$}}%
    }%
    \gplgaddtomacro\gplfronttext{%
      \csname LTb\endcsname%
      \put(1010,-118){\makebox(0,0){\strut{}$t U_b/h$}}%
      \put(1010,2740){\makebox(0,0){\strut{}}}%
    }%
    \gplbacktext
    \put(0,0){\includegraphics{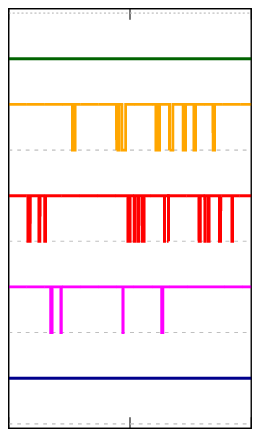}}%
    \gplfronttext
  \end{picture}%
\endgroup

%% file: inter211.tex
% GNUPLOT: LaTeX picture with Postscript
\begingroup
  \makeatletter
  \providecommand\color[2][]{%
    \GenericError{(gnuplot) \space\space\space\@spaces}{%
      Package color not loaded in conjunction with
      terminal option `colourtext'%
    }{See the gnuplot documentation for explanation.%
    }{Either use 'blacktext' in gnuplot or load the package
      color.sty in LaTeX.}%
    \renewcommand\color[2][]{}%
  }%
  \providecommand\includegraphics[2][]{%
    \GenericError{(gnuplot) \space\space\space\@spaces}{%
      Package graphicx or graphics not loaded%
    }{See the gnuplot documentation for explanation.%
    }{The gnuplot epslatex terminal needs graphicx.sty or graphics.sty.}%
    \renewcommand\includegraphics[2][]{}%
  }%
  \providecommand\rotatebox[2]{#2}%
  \@ifundefined{ifGPcolor}{%
    \newif\ifGPcolor
    \GPcolortrue
  }{}%
  \@ifundefined{ifGPblacktext}{%
    \newif\ifGPblacktext
    \GPblacktexttrue
  }{}%
  % define a \g@addto@macro without @ in the name:
  \let\gplgaddtomacro\g@addto@macro
  % define empty templates for all commands taking text:
  \gdef\gplbacktext{}%
  \gdef\gplfronttext{}%
  \makeatother
  \ifGPblacktext
    % no textcolor at all
    \def\colorrgb#1{}%
    \def\colorgray#1{}%
  \else
    % gray or color?
    \ifGPcolor
      \def\colorrgb#1{\color[rgb]{#1}}%
      \def\colorgray#1{\color[gray]{#1}}%
      \expandafter\def\csname LTw\endcsname{\color{white}}%
      \expandafter\def\csname LTb\endcsname{\color{black}}%
      \expandafter\def\csname LTa\endcsname{\color{black}}%
      \expandafter\def\csname LT0\endcsname{\color[rgb]{1,0,0}}%
      \expandafter\def\csname LT1\endcsname{\color[rgb]{0,1,0}}%
      \expandafter\def\csname LT2\endcsname{\color[rgb]{0,0,1}}%
      \expandafter\def\csname LT3\endcsname{\color[rgb]{1,0,1}}%
      \expandafter\def\csname LT4\endcsname{\color[rgb]{0,1,1}}%
      \expandafter\def\csname LT5\endcsname{\color[rgb]{1,1,0}}%
      \expandafter\def\csname LT6\endcsname{\color[rgb]{0,0,0}}%
      \expandafter\def\csname LT7\endcsname{\color[rgb]{1,0.3,0}}%
      \expandafter\def\csname LT8\endcsname{\color[rgb]{0.5,0.5,0.5}}%
    \else
      % gray
      \def\colorrgb#1{\color{black}}%
      \def\colorgray#1{\color[gray]{#1}}%
      \expandafter\def\csname LTw\endcsname{\color{white}}%
      \expandafter\def\csname LTb\endcsname{\color{black}}%
      \expandafter\def\csname LTa\endcsname{\color{black}}%
      \expandafter\def\csname LT0\endcsname{\color{black}}%
      \expandafter\def\csname LT1\endcsname{\color{black}}%
      \expandafter\def\csname LT2\endcsname{\color{black}}%
      \expandafter\def\csname LT3\endcsname{\color{black}}%
      \expandafter\def\csname LT4\endcsname{\color{black}}%
      \expandafter\def\csname LT5\endcsname{\color{black}}%
      \expandafter\def\csname LT6\endcsname{\color{black}}%
      \expandafter\def\csname LT7\endcsname{\color{black}}%
      \expandafter\def\csname LT8\endcsname{\color{black}}%
    \fi
  \fi
    \setlength{\unitlength}{0.0500bp}%
    \ifx\gptboxheight\undefined%
      \newlength{\gptboxheight}%
      \newlength{\gptboxwidth}%
      \newsavebox{\gptboxtext}%
    \fi%
    \setlength{\fboxrule}{0.5pt}%
    \setlength{\fboxsep}{1pt}%
\begin{picture}(1728.00,2880.00)%
\definecolor{gpBackground}{rgb}{1.000, 1.000, 1.000}%
\put(0,0){\colorbox{gpBackground}{\makebox(1728.00,2880.00)[]{}}}%
    \gplgaddtomacro\gplbacktext{%
      \csname LTb\endcsname%
      \put(80,2850){\rotatebox{0}{\makebox(0,0){\strut{}(c)}}}%
      \put(311,212){\makebox(0,0){\strut{}$0$}}%
      \put(1010,212){\makebox(0,0){\strut{}$10$}}%
      \put(1709,212){\makebox(0,0){\strut{}$20$}}%
    }%
    \gplgaddtomacro\gplfronttext{%
      \csname LTb\endcsname%
      \put(1010,-118){\makebox(0,0){\strut{}$t U_b/h$}}%
      \put(1010,2740){\makebox(0,0){\strut{}}}%
    }%
    \gplbacktext
    \put(0,0){\includegraphics{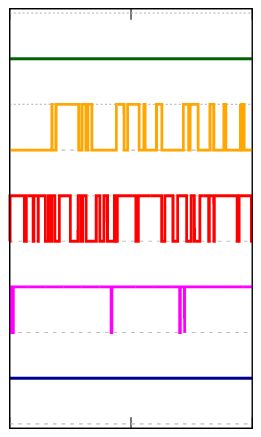}}%
    \gplfronttext
  \end{picture}%
\endgroup

%% file: inter281.tex
% GNUPLOT: LaTeX picture with Postscript
\begingroup
  \makeatletter
  \providecommand\color[2][]{%
    \GenericError{(gnuplot) \space\space\space\@spaces}{%
      Package color not loaded in conjunction with
      terminal option `colourtext'%
    }{See the gnuplot documentation for explanation.%
    }{Either use 'blacktext' in gnuplot or load the package
      color.sty in LaTeX.}%
    \renewcommand\color[2][]{}%
  }%
  \providecommand\includegraphics[2][]{%
    \GenericError{(gnuplot) \space\space\space\@spaces}{%
      Package graphicx or graphics not loaded%
    }{See the gnuplot documentation for explanation.%
    }{The gnuplot epslatex terminal needs graphicx.sty or graphics.sty.}%
    \renewcommand\includegraphics[2][]{}%
  }%
  \providecommand\rotatebox[2]{#2}%
  \@ifundefined{ifGPcolor}{%
    \newif\ifGPcolor
    \GPcolortrue
  }{}%
  \@ifundefined{ifGPblacktext}{%
    \newif\ifGPblacktext
    \GPblacktexttrue
  }{}%
  % define a \g@addto@macro without @ in the name:
  \let\gplgaddtomacro\g@addto@macro
  % define empty templates for all commands taking text:
  \gdef\gplbacktext{}%
  \gdef\gplfronttext{}%
  \makeatother
  \ifGPblacktext
    % no textcolor at all
    \def\colorrgb#1{}%
    \def\colorgray#1{}%
  \else
    % gray or color?
    \ifGPcolor
      \def\colorrgb#1{\color[rgb]{#1}}%
      \def\colorgray#1{\color[gray]{#1}}%
      \expandafter\def\csname LTw\endcsname{\color{white}}%
      \expandafter\def\csname LTb\endcsname{\color{black}}%
      \expandafter\def\csname LTa\endcsname{\color{black}}%
      \expandafter\def\csname LT0\endcsname{\color[rgb]{1,0,0}}%
      \expandafter\def\csname LT1\endcsname{\color[rgb]{0,1,0}}%
      \expandafter\def\csname LT2\endcsname{\color[rgb]{0,0,1}}%
      \expandafter\def\csname LT3\endcsname{\color[rgb]{1,0,1}}%
      \expandafter\def\csname LT4\endcsname{\color[rgb]{0,1,1}}%
      \expandafter\def\csname LT5\endcsname{\color[rgb]{1,1,0}}%
      \expandafter\def\csname LT6\endcsname{\color[rgb]{0,0,0}}%
      \expandafter\def\csname LT7\endcsname{\color[rgb]{1,0.3,0}}%
      \expandafter\def\csname LT8\endcsname{\color[rgb]{0.5,0.5,0.5}}%
    \else
      % gray
      \def\colorrgb#1{\color{black}}%
      \def\colorgray#1{\color[gray]{#1}}%
      \expandafter\def\csname LTw\endcsname{\color{white}}%
      \expandafter\def\csname LTb\endcsname{\color{black}}%
      \expandafter\def\csname LTa\endcsname{\color{black}}%
      \expandafter\def\csname LT0\endcsname{\color{black}}%
      \expandafter\def\csname LT1\endcsname{\color{black}}%
      \expandafter\def\csname LT2\endcsname{\color{black}}%
      \expandafter\def\csname LT3\endcsname{\color{black}}%
      \expandafter\def\csname LT4\endcsname{\color{black}}%
      \expandafter\def\csname LT5\endcsname{\color{black}}%
      \expandafter\def\csname LT6\endcsname{\color{black}}%
      \expandafter\def\csname LT7\endcsname{\color{black}}%
      \expandafter\def\csname LT8\endcsname{\color{black}}%
    \fi
  \fi
    \setlength{\unitlength}{0.0500bp}%
    \ifx\gptboxheight\undefined%
      \newlength{\gptboxheight}%
      \newlength{\gptboxwidth}%
      \newsavebox{\gptboxtext}%
    \fi%
    \setlength{\fboxrule}{0.5pt}%
    \setlength{\fboxsep}{1pt}%
\begin{picture}(1728.00,2880.00)%
\definecolor{gpBackground}{rgb}{1.000, 1.000, 1.000}%
\put(0,0){\colorbox{gpBackground}{\makebox(1728.00,2880.00)[]{}}}%
    \gplgaddtomacro\gplbacktext{%
      \csname LTb\endcsname%
      \put(311,212){\makebox(0,0){\strut{}$0$}}%
      \put(1010,212){\makebox(0,0){\strut{}$10$}}%
      \put(1709,212){\makebox(0,0){\strut{}$20$}}%
    }%
    \gplgaddtomacro\gplfronttext{%
      \csname LTb\endcsname%
      \put(80,2850){\rotatebox{0}{\makebox(0,0){\strut{}(d)}}}%
      \put(1010,-118){\makebox(0,0){\strut{}$t U_b/h$}}%
      \put(1010,2740){\makebox(0,0){\strut{}}}%
    }%
    \gplbacktext
    \put(0,0){\includegraphics{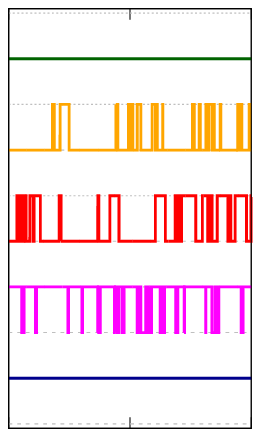}}%
    \gplfronttext
  \end{picture}%
\endgroup

%% file: mean.tex
% GNUPLOT: LaTeX picture with Postscript
\begingroup
  \makeatletter
  \providecommand\color[2][]{%
    \GenericError{(gnuplot) \space\space\space\@spaces}{%
      Package color not loaded in conjunction with
      terminal option `colourtext'%
    }{See the gnuplot documentation for explanation.%
    }{Either use 'blacktext' in gnuplot or load the package
      color.sty in LaTeX.}%
    \renewcommand\color[2][]{}%
  }%
  \providecommand\includegraphics[2][]{%
    \GenericError{(gnuplot) \space\space\space\@spaces}{%
      Package graphicx or graphics not loaded%
    }{See the gnuplot documentation for explanation.%
    }{The gnuplot epslatex terminal needs graphicx.sty or graphics.sty.}%
    \renewcommand\includegraphics[2][]{}%
  }%
  \providecommand\rotatebox[2]{#2}%
  \@ifundefined{ifGPcolor}{%
    \newif\ifGPcolor
    \GPcolortrue
  }{}%
  \@ifundefined{ifGPblacktext}{%
    \newif\ifGPblacktext
    \GPblacktexttrue
  }{}%
  % define a \g@addto@macro without @ in the name:
  \let\gplgaddtomacro\g@addto@macro
  % define empty templates for all commands taking text:
  \gdef\gplbacktext{}%
  \gdef\gplfronttext{}%
  \makeatother
  \ifGPblacktext
    % no textcolor at all
    \def\colorrgb#1{}%
    \def\colorgray#1{}%
  \else
    % gray or color?
    \ifGPcolor
      \def\colorrgb#1{\color[rgb]{#1}}%
      \def\colorgray#1{\color[gray]{#1}}%
      \expandafter\def\csname LTw\endcsname{\color{white}}%
      \expandafter\def\csname LTb\endcsname{\color{black}}%
      \expandafter\def\csname LTa\endcsname{\color{black}}%
      \expandafter\def\csname LT0\endcsname{\color[rgb]{1,0,0}}%
      \expandafter\def\csname LT1\endcsname{\color[rgb]{0,1,0}}%
      \expandafter\def\csname LT2\endcsname{\color[rgb]{0,0,1}}%
      \expandafter\def\csname LT3\endcsname{\color[rgb]{1,0,1}}%
      \expandafter\def\csname LT4\endcsname{\color[rgb]{0,1,1}}%
      \expandafter\def\csname LT5\endcsname{\color[rgb]{1,1,0}}%
      \expandafter\def\csname LT6\endcsname{\color[rgb]{0,0,0}}%
      \expandafter\def\csname LT7\endcsname{\color[rgb]{1,0.3,0}}%
      \expandafter\def\csname LT8\endcsname{\color[rgb]{0.5,0.5,0.5}}%
    \else
      % gray
      \def\colorrgb#1{\color{black}}%
      \def\colorgray#1{\color[gray]{#1}}%
      \expandafter\def\csname LTw\endcsname{\color{white}}%
      \expandafter\def\csname LTb\endcsname{\color{black}}%
      \expandafter\def\csname LTa\endcsname{\color{black}}%
      \expandafter\def\csname LT0\endcsname{\color{black}}%
      \expandafter\def\csname LT1\endcsname{\color{black}}%
      \expandafter\def\csname LT2\endcsname{\color{black}}%
      \expandafter\def\csname LT3\endcsname{\color{black}}%
      \expandafter\def\csname LT4\endcsname{\color{black}}%
      \expandafter\def\csname LT5\endcsname{\color{black}}%
      \expandafter\def\csname LT6\endcsname{\color{black}}%
      \expandafter\def\csname LT7\endcsname{\color{black}}%
      \expandafter\def\csname LT8\endcsname{\color{black}}%
    \fi
  \fi
    \setlength{\unitlength}{0.0500bp}%
    \ifx\gptboxheight\undefined%
      \newlength{\gptboxheight}%
      \newlength{\gptboxwidth}%
      \newsavebox{\gptboxtext}%
    \fi%
    \setlength{\fboxrule}{0.5pt}%
    \setlength{\fboxsep}{1pt}%
\begin{picture}(3456.00,2880.00)%
\definecolor{gpBackground}{rgb}{1.000, 1.000, 1.000}%
\put(0,0){\colorbox{gpBackground}{\makebox(3456.00,2880.00)[]{}}}%
    \gplgaddtomacro\gplbacktext{%
      \csname LTb\endcsname%
      \put(490,432){\makebox(0,0)[r]{\strut{}$0$}}%
      \put(490,1296){\makebox(0,0)[r]{\strut{}$0.5$}}%
      \put(490,2159){\makebox(0,0)[r]{\strut{}$1$}}%
      \put(622,212){\makebox(0,0){\strut{}$0$}}%
      \put(2021,212){\makebox(0,0){\strut{}$0.5$}}%
      \put(3420,212){\makebox(0,0){\strut{}$1$}}%
    }%
    \gplgaddtomacro\gplfronttext{%
      \csname LTb\endcsname%
      \put(50,2850){\rotatebox{0}{\makebox(0,0){\strut{}(a)}}}%
      \put(50,1641){\rotatebox{-270}{\makebox(0,0){\strut{}$\overline{u}/U_b$}}}%
      \put(2021,-118){\makebox(0,0){\strut{}$y/h$}}%
      \put(2021,2740){\makebox(0,0){\strut{}}}%
    }%
    \gplbacktext
    \put(0,0){\includegraphics{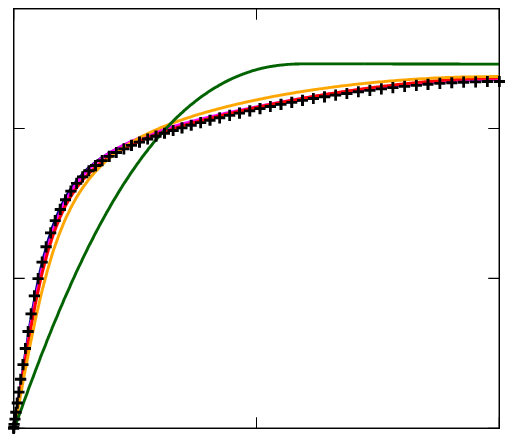}}%
    \gplfronttext
  \end{picture}%
\endgroup

%% file: meanLOG.tex
% GNUPLOT: LaTeX picture with Postscript
\begingroup
  \makeatletter
  \providecommand\color[2][]{%
    \GenericError{(gnuplot) \space\space\space\@spaces}{%
      Package color not loaded in conjunction with
      terminal option `colourtext'%
    }{See the gnuplot documentation for explanation.%
    }{Either use 'blacktext' in gnuplot or load the package
      color.sty in LaTeX.}%
    \renewcommand\color[2][]{}%
  }%
  \providecommand\includegraphics[2][]{%
    \GenericError{(gnuplot) \space\space\space\@spaces}{%
      Package graphicx or graphics not loaded%
    }{See the gnuplot documentation for explanation.%
    }{The gnuplot epslatex terminal needs graphicx.sty or graphics.sty.}%
    \renewcommand\includegraphics[2][]{}%
  }%
  \providecommand\rotatebox[2]{#2}%
  \@ifundefined{ifGPcolor}{%
    \newif\ifGPcolor
    \GPcolortrue
  }{}%
  \@ifundefined{ifGPblacktext}{%
    \newif\ifGPblacktext
    \GPblacktexttrue
  }{}%
  % define a \g@addto@macro without @ in the name:
  \let\gplgaddtomacro\g@addto@macro
  % define empty templates for all commands taking text:
  \gdef\gplbacktext{}%
  \gdef\gplfronttext{}%
  \makeatother
  \ifGPblacktext
    % no textcolor at all
    \def\colorrgb#1{}%
    \def\colorgray#1{}%
  \else
    % gray or color?
    \ifGPcolor
      \def\colorrgb#1{\color[rgb]{#1}}%
      \def\colorgray#1{\color[gray]{#1}}%
      \expandafter\def\csname LTw\endcsname{\color{white}}%
      \expandafter\def\csname LTb\endcsname{\color{black}}%
      \expandafter\def\csname LTa\endcsname{\color{black}}%
      \expandafter\def\csname LT0\endcsname{\color[rgb]{1,0,0}}%
      \expandafter\def\csname LT1\endcsname{\color[rgb]{0,1,0}}%
      \expandafter\def\csname LT2\endcsname{\color[rgb]{0,0,1}}%
      \expandafter\def\csname LT3\endcsname{\color[rgb]{1,0,1}}%
      \expandafter\def\csname LT4\endcsname{\color[rgb]{0,1,1}}%
      \expandafter\def\csname LT5\endcsname{\color[rgb]{1,1,0}}%
      \expandafter\def\csname LT6\endcsname{\color[rgb]{0,0,0}}%
      \expandafter\def\csname LT7\endcsname{\color[rgb]{1,0.3,0}}%
      \expandafter\def\csname LT8\endcsname{\color[rgb]{0.5,0.5,0.5}}%
    \else
      % gray
      \def\colorrgb#1{\color{black}}%
      \def\colorgray#1{\color[gray]{#1}}%
      \expandafter\def\csname LTw\endcsname{\color{white}}%
      \expandafter\def\csname LTb\endcsname{\color{black}}%
      \expandafter\def\csname LTa\endcsname{\color{black}}%
      \expandafter\def\csname LT0\endcsname{\color{black}}%
      \expandafter\def\csname LT1\endcsname{\color{black}}%
      \expandafter\def\csname LT2\endcsname{\color{black}}%
      \expandafter\def\csname LT3\endcsname{\color{black}}%
      \expandafter\def\csname LT4\endcsname{\color{black}}%
      \expandafter\def\csname LT5\endcsname{\color{black}}%
      \expandafter\def\csname LT6\endcsname{\color{black}}%
      \expandafter\def\csname LT7\endcsname{\color{black}}%
      \expandafter\def\csname LT8\endcsname{\color{black}}%
    \fi
  \fi
    \setlength{\unitlength}{0.0500bp}%
    \ifx\gptboxheight\undefined%
      \newlength{\gptboxheight}%
      \newlength{\gptboxwidth}%
      \newsavebox{\gptboxtext}%
    \fi%
    \setlength{\fboxrule}{0.5pt}%
    \setlength{\fboxsep}{1pt}%
\begin{picture}(3456.00,2880.00)%
\definecolor{gpBackground}{rgb}{1.000, 1.000, 1.000}%
\put(0,0){\colorbox{gpBackground}{\makebox(3456.00,2880.00)[]{}}}%
    \gplgaddtomacro\gplbacktext{%
      \csname LTb\endcsname%
      \put(490,432){\makebox(0,0)[r]{\strut{}$0$}}%
      \put(490,1399){\makebox(0,0)[r]{\strut{}$10$}}%
      \put(490,2366){\makebox(0,0)[r]{\strut{}$20$}}%
      \put(622,212){\makebox(0,0){\strut{}$1$}}%
      \put(1838,212){\makebox(0,0){\strut{}$10$}}%
      \put(3054,212){\makebox(0,0){\strut{}$100$}}%
    }%
    \gplgaddtomacro\gplfronttext{%
      \csname LTb\endcsname%
      \put(182,2850){\rotatebox{0}{\makebox(0,0){\strut{}(b)}}}%
      \put(182,1641){\rotatebox{-270}{\makebox(0,0){\strut{}$\overline{u}/u_\tau$}}}%
      \put(2021,-118){\makebox(0,0){\strut{}$y^+$}}%
      \put(2021,2740){\makebox(0,0){\strut{}}}%
    }%
    \gplbacktext
    \put(0,0){\includegraphics{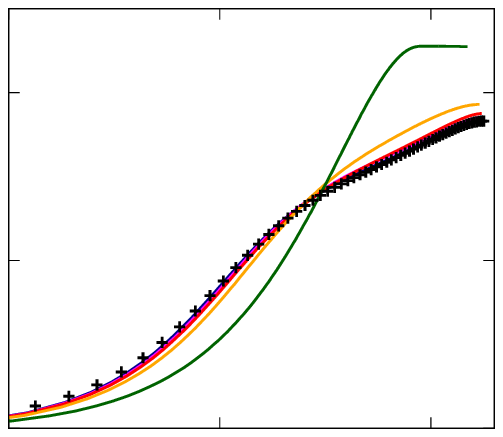}}%
    \gplfronttext
  \end{picture}%
\endgroup

%% file: reyU.tex
% GNUPLOT: LaTeX picture with Postscript
\begingroup
  \makeatletter
  \providecommand\color[2][]{%
    \GenericError{(gnuplot) \space\space\space\@spaces}{%
      Package color not loaded in conjunction with
      terminal option `colourtext'%
    }{See the gnuplot documentation for explanation.%
    }{Either use 'blacktext' in gnuplot or load the package
      color.sty in LaTeX.}%
    \renewcommand\color[2][]{}%
  }%
  \providecommand\includegraphics[2][]{%
    \GenericError{(gnuplot) \space\space\space\@spaces}{%
      Package graphicx or graphics not loaded%
    }{See the gnuplot documentation for explanation.%
    }{The gnuplot epslatex terminal needs graphicx.sty or graphics.sty.}%
    \renewcommand\includegraphics[2][]{}%
  }%
  \providecommand\rotatebox[2]{#2}%
  \@ifundefined{ifGPcolor}{%
    \newif\ifGPcolor
    \GPcolortrue
  }{}%
  \@ifundefined{ifGPblacktext}{%
    \newif\ifGPblacktext
    \GPblacktexttrue
  }{}%
  % define a \g@addto@macro without @ in the name:
  \let\gplgaddtomacro\g@addto@macro
  % define empty templates for all commands taking text:
  \gdef\gplbacktext{}%
  \gdef\gplfronttext{}%
  \makeatother
  \ifGPblacktext
    % no textcolor at all
    \def\colorrgb#1{}%
    \def\colorgray#1{}%
  \else
    % gray or color?
    \ifGPcolor
      \def\colorrgb#1{\color[rgb]{#1}}%
      \def\colorgray#1{\color[gray]{#1}}%
      \expandafter\def\csname LTw\endcsname{\color{white}}%
      \expandafter\def\csname LTb\endcsname{\color{black}}%
      \expandafter\def\csname LTa\endcsname{\color{black}}%
      \expandafter\def\csname LT0\endcsname{\color[rgb]{1,0,0}}%
      \expandafter\def\csname LT1\endcsname{\color[rgb]{0,1,0}}%
      \expandafter\def\csname LT2\endcsname{\color[rgb]{0,0,1}}%
      \expandafter\def\csname LT3\endcsname{\color[rgb]{1,0,1}}%
      \expandafter\def\csname LT4\endcsname{\color[rgb]{0,1,1}}%
      \expandafter\def\csname LT5\endcsname{\color[rgb]{1,1,0}}%
      \expandafter\def\csname LT6\endcsname{\color[rgb]{0,0,0}}%
      \expandafter\def\csname LT7\endcsname{\color[rgb]{1,0.3,0}}%
      \expandafter\def\csname LT8\endcsname{\color[rgb]{0.5,0.5,0.5}}%
    \else
      % gray
      \def\colorrgb#1{\color{black}}%
      \def\colorgray#1{\color[gray]{#1}}%
      \expandafter\def\csname LTw\endcsname{\color{white}}%
      \expandafter\def\csname LTb\endcsname{\color{black}}%
      \expandafter\def\csname LTa\endcsname{\color{black}}%
      \expandafter\def\csname LT0\endcsname{\color{black}}%
      \expandafter\def\csname LT1\endcsname{\color{black}}%
      \expandafter\def\csname LT2\endcsname{\color{black}}%
      \expandafter\def\csname LT3\endcsname{\color{black}}%
      \expandafter\def\csname LT4\endcsname{\color{black}}%
      \expandafter\def\csname LT5\endcsname{\color{black}}%
      \expandafter\def\csname LT6\endcsname{\color{black}}%
      \expandafter\def\csname LT7\endcsname{\color{black}}%
      \expandafter\def\csname LT8\endcsname{\color{black}}%
    \fi
  \fi
    \setlength{\unitlength}{0.0500bp}%
    \ifx\gptboxheight\undefined%
      \newlength{\gptboxheight}%
      \newlength{\gptboxwidth}%
      \newsavebox{\gptboxtext}%
    \fi%
    \setlength{\fboxrule}{0.5pt}%
    \setlength{\fboxsep}{1pt}%
\begin{picture}(3456.00,2880.00)%
\definecolor{gpBackground}{rgb}{1.000, 1.000, 1.000}%
\put(0,0){\colorbox{gpBackground}{\makebox(3456.00,2880.00)[]{}}}%
    \gplgaddtomacro\gplbacktext{%
      \csname LTb\endcsname%
      \put(490,547){\makebox(0,0)[r]{\strut{}$0$}}%
      \put(490,1699){\makebox(0,0)[r]{\strut{}$5$}}%
      \put(490,2850){\makebox(0,0)[r]{\strut{}$10$}}%
      \put(622,212){\makebox(0,0){\strut{}$1$}}%
      \put(1863,212){\makebox(0,0){\strut{}$10$}}%
      \put(3103,212){\makebox(0,0){\strut{}$100$}}%
    }%
    \gplgaddtomacro\gplfronttext{%
      \csname LTb\endcsname%
      \put(182,2850){\rotatebox{0}{\makebox(0,0){\strut{}(a)}}}%
      \put(182,1641){\rotatebox{-270}{\makebox(0,0){\strut{}$\overline{u'u'}/u_\tau^2$}}}%
      \put(2021,-118){\makebox(0,0){\strut{}$y^+$}}%
      \put(2021,2740){\makebox(0,0){\strut{}}}%
    }%
    \gplbacktext
    \put(0,0){\includegraphics{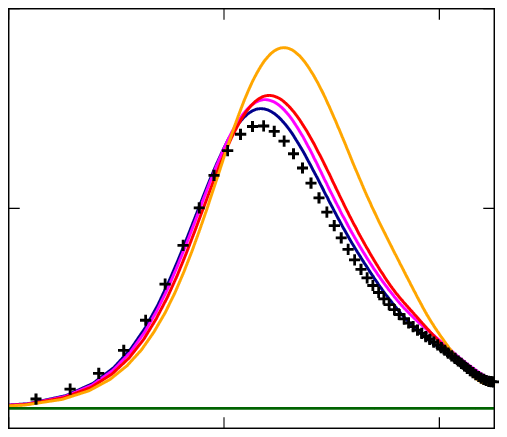}}%
    \gplfronttext
  \end{picture}%
\endgroup

%% file: reyV.tex
% GNUPLOT: LaTeX picture with Postscript
\begingroup
  \makeatletter
  \providecommand\color[2][]{%
    \GenericError{(gnuplot) \space\space\space\@spaces}{%
      Package color not loaded in conjunction with
      terminal option `colourtext'%
    }{See the gnuplot documentation for explanation.%
    }{Either use 'blacktext' in gnuplot or load the package
      color.sty in LaTeX.}%
    \renewcommand\color[2][]{}%
  }%
  \providecommand\includegraphics[2][]{%
    \GenericError{(gnuplot) \space\space\space\@spaces}{%
      Package graphicx or graphics not loaded%
    }{See the gnuplot documentation for explanation.%
    }{The gnuplot epslatex terminal needs graphicx.sty or graphics.sty.}%
    \renewcommand\includegraphics[2][]{}%
  }%
  \providecommand\rotatebox[2]{#2}%
  \@ifundefined{ifGPcolor}{%
    \newif\ifGPcolor
    \GPcolortrue
  }{}%
  \@ifundefined{ifGPblacktext}{%
    \newif\ifGPblacktext
    \GPblacktexttrue
  }{}%
  % define a \g@addto@macro without @ in the name:
  \let\gplgaddtomacro\g@addto@macro
  % define empty templates for all commands taking text:
  \gdef\gplbacktext{}%
  \gdef\gplfronttext{}%
  \makeatother
  \ifGPblacktext
    % no textcolor at all
    \def\colorrgb#1{}%
    \def\colorgray#1{}%
  \else
    % gray or color?
    \ifGPcolor
      \def\colorrgb#1{\color[rgb]{#1}}%
      \def\colorgray#1{\color[gray]{#1}}%
      \expandafter\def\csname LTw\endcsname{\color{white}}%
      \expandafter\def\csname LTb\endcsname{\color{black}}%
      \expandafter\def\csname LTa\endcsname{\color{black}}%
      \expandafter\def\csname LT0\endcsname{\color[rgb]{1,0,0}}%
      \expandafter\def\csname LT1\endcsname{\color[rgb]{0,1,0}}%
      \expandafter\def\csname LT2\endcsname{\color[rgb]{0,0,1}}%
      \expandafter\def\csname LT3\endcsname{\color[rgb]{1,0,1}}%
      \expandafter\def\csname LT4\endcsname{\color[rgb]{0,1,1}}%
      \expandafter\def\csname LT5\endcsname{\color[rgb]{1,1,0}}%
      \expandafter\def\csname LT6\endcsname{\color[rgb]{0,0,0}}%
      \expandafter\def\csname LT7\endcsname{\color[rgb]{1,0.3,0}}%
      \expandafter\def\csname LT8\endcsname{\color[rgb]{0.5,0.5,0.5}}%
    \else
      % gray
      \def\colorrgb#1{\color{black}}%
      \def\colorgray#1{\color[gray]{#1}}%
      \expandafter\def\csname LTw\endcsname{\color{white}}%
      \expandafter\def\csname LTb\endcsname{\color{black}}%
      \expandafter\def\csname LTa\endcsname{\color{black}}%
      \expandafter\def\csname LT0\endcsname{\color{black}}%
      \expandafter\def\csname LT1\endcsname{\color{black}}%
      \expandafter\def\csname LT2\endcsname{\color{black}}%
      \expandafter\def\csname LT3\endcsname{\color{black}}%
      \expandafter\def\csname LT4\endcsname{\color{black}}%
      \expandafter\def\csname LT5\endcsname{\color{black}}%
      \expandafter\def\csname LT6\endcsname{\color{black}}%
      \expandafter\def\csname LT7\endcsname{\color{black}}%
      \expandafter\def\csname LT8\endcsname{\color{black}}%
    \fi
  \fi
    \setlength{\unitlength}{0.0500bp}%
    \ifx\gptboxheight\undefined%
      \newlength{\gptboxheight}%
      \newlength{\gptboxwidth}%
      \newsavebox{\gptboxtext}%
    \fi%
    \setlength{\fboxrule}{0.5pt}%
    \setlength{\fboxsep}{1pt}%
\begin{picture}(3456.00,2880.00)%
\definecolor{gpBackground}{rgb}{1.000, 1.000, 1.000}%
\put(0,0){\colorbox{gpBackground}{\makebox(3456.00,2880.00)[]{}}}%
    \gplgaddtomacro\gplbacktext{%
      \csname LTb\endcsname%
      \put(490,574){\makebox(0,0)[r]{\strut{}$0$}}%
      \put(490,1712){\makebox(0,0)[r]{\strut{}$0.4$}}%
      \put(490,2850){\makebox(0,0)[r]{\strut{}$0.8$}}%
      \put(622,212){\makebox(0,0){\strut{}$1$}}%
      \put(1863,212){\makebox(0,0){\strut{}$10$}}%
      \put(3103,212){\makebox(0,0){\strut{}$100$}}%
    }%
    \gplgaddtomacro\gplfronttext{%
      \csname LTb\endcsname%
      \put(50,2850){\rotatebox{0}{\makebox(0,0){\strut{}(b)}}}%
      \put(50,1641){\rotatebox{-270}{\makebox(0,0){\strut{}$\overline{v'v'}/u_\tau^2$}}}%
      \put(2021,-118){\makebox(0,0){\strut{}$y^+$}}%
      \put(2021,2740){\makebox(0,0){\strut{}}}%
    }%
    \gplbacktext
    \put(0,0){\includegraphics{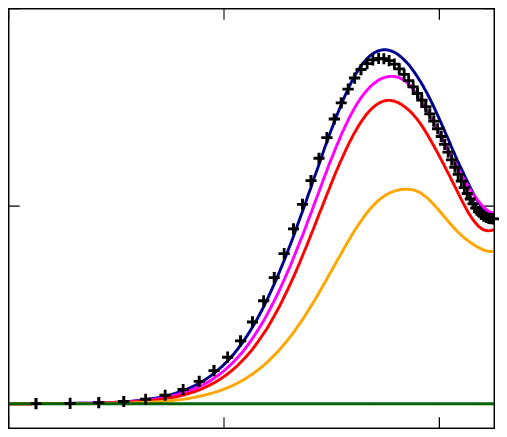}}%
    \gplfronttext
  \end{picture}%
\endgroup

%% file: reyW.tex
% GNUPLOT: LaTeX picture with Postscript
\begingroup
  \makeatletter
  \providecommand\color[2][]{%
    \GenericError{(gnuplot) \space\space\space\@spaces}{%
      Package color not loaded in conjunction with
      terminal option `colourtext'%
    }{See the gnuplot documentation for explanation.%
    }{Either use 'blacktext' in gnuplot or load the package
      color.sty in LaTeX.}%
    \renewcommand\color[2][]{}%
  }%
  \providecommand\includegraphics[2][]{%
    \GenericError{(gnuplot) \space\space\space\@spaces}{%
      Package graphicx or graphics not loaded%
    }{See the gnuplot documentation for explanation.%
    }{The gnuplot epslatex terminal needs graphicx.sty or graphics.sty.}%
    \renewcommand\includegraphics[2][]{}%
  }%
  \providecommand\rotatebox[2]{#2}%
  \@ifundefined{ifGPcolor}{%
    \newif\ifGPcolor
    \GPcolortrue
  }{}%
  \@ifundefined{ifGPblacktext}{%
    \newif\ifGPblacktext
    \GPblacktexttrue
  }{}%
  % define a \g@addto@macro without @ in the name:
  \let\gplgaddtomacro\g@addto@macro
  % define empty templates for all commands taking text:
  \gdef\gplbacktext{}%
  \gdef\gplfronttext{}%
  \makeatother
  \ifGPblacktext
    % no textcolor at all
    \def\colorrgb#1{}%
    \def\colorgray#1{}%
  \else
    % gray or color?
    \ifGPcolor
      \def\colorrgb#1{\color[rgb]{#1}}%
      \def\colorgray#1{\color[gray]{#1}}%
      \expandafter\def\csname LTw\endcsname{\color{white}}%
      \expandafter\def\csname LTb\endcsname{\color{black}}%
      \expandafter\def\csname LTa\endcsname{\color{black}}%
      \expandafter\def\csname LT0\endcsname{\color[rgb]{1,0,0}}%
      \expandafter\def\csname LT1\endcsname{\color[rgb]{0,1,0}}%
      \expandafter\def\csname LT2\endcsname{\color[rgb]{0,0,1}}%
      \expandafter\def\csname LT3\endcsname{\color[rgb]{1,0,1}}%
      \expandafter\def\csname LT4\endcsname{\color[rgb]{0,1,1}}%
      \expandafter\def\csname LT5\endcsname{\color[rgb]{1,1,0}}%
      \expandafter\def\csname LT6\endcsname{\color[rgb]{0,0,0}}%
      \expandafter\def\csname LT7\endcsname{\color[rgb]{1,0.3,0}}%
      \expandafter\def\csname LT8\endcsname{\color[rgb]{0.5,0.5,0.5}}%
    \else
      % gray
      \def\colorrgb#1{\color{black}}%
      \def\colorgray#1{\color[gray]{#1}}%
      \expandafter\def\csname LTw\endcsname{\color{white}}%
      \expandafter\def\csname LTb\endcsname{\color{black}}%
      \expandafter\def\csname LTa\endcsname{\color{black}}%
      \expandafter\def\csname LT0\endcsname{\color{black}}%
      \expandafter\def\csname LT1\endcsname{\color{black}}%
      \expandafter\def\csname LT2\endcsname{\color{black}}%
      \expandafter\def\csname LT3\endcsname{\color{black}}%
      \expandafter\def\csname LT4\endcsname{\color{black}}%
      \expandafter\def\csname LT5\endcsname{\color{black}}%
      \expandafter\def\csname LT6\endcsname{\color{black}}%
      \expandafter\def\csname LT7\endcsname{\color{black}}%
      \expandafter\def\csname LT8\endcsname{\color{black}}%
    \fi
  \fi
    \setlength{\unitlength}{0.0500bp}%
    \ifx\gptboxheight\undefined%
      \newlength{\gptboxheight}%
      \newlength{\gptboxwidth}%
      \newsavebox{\gptboxtext}%
    \fi%
    \setlength{\fboxrule}{0.5pt}%
    \setlength{\fboxsep}{1pt}%
\begin{picture}(3456.00,2880.00)%
\definecolor{gpBackground}{rgb}{1.000, 1.000, 1.000}%
\put(0,0){\colorbox{gpBackground}{\makebox(3456.00,2880.00)[]{}}}%
    \gplgaddtomacro\gplbacktext{%
      \csname LTb\endcsname%
      \put(490,529){\makebox(0,0)[r]{\strut{}$0$}}%
      \put(490,1496){\makebox(0,0)[r]{\strut{}$0.5$}}%
      \put(490,2463){\makebox(0,0)[r]{\strut{}$1$}}%
      \put(622,212){\makebox(0,0){\strut{}$1$}}%
      \put(1863,212){\makebox(0,0){\strut{}$10$}}%
      \put(3103,212){\makebox(0,0){\strut{}$100$}}%
    }%
    \gplgaddtomacro\gplfronttext{%
      \csname LTb\endcsname%
      \put(50,2850){\rotatebox{0}{\makebox(0,0){\strut{}(c)}}}%
      \put(50,1641){\rotatebox{-270}{\makebox(0,0){\strut{}$\overline{w'w'}/u_\tau^2$}}}%
      \put(2021,-118){\makebox(0,0){\strut{}$y^+$}}%
      \put(2021,2740){\makebox(0,0){\strut{}}}%
    }%
    \gplbacktext
    \put(0,0){\includegraphics{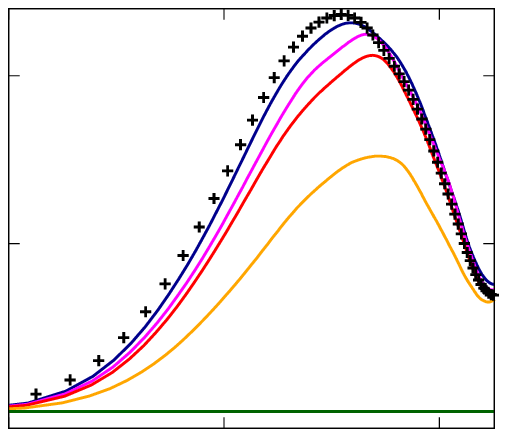}}%
    \gplfronttext
  \end{picture}%
\endgroup

%% file: reyUV.tex
% GNUPLOT: LaTeX picture with Postscript
\begingroup
  \makeatletter
  \providecommand\color[2][]{%
    \GenericError{(gnuplot) \space\space\space\@spaces}{%
      Package color not loaded in conjunction with
      terminal option `colourtext'%
    }{See the gnuplot documentation for explanation.%
    }{Either use 'blacktext' in gnuplot or load the package
      color.sty in LaTeX.}%
    \renewcommand\color[2][]{}%
  }%
  \providecommand\includegraphics[2][]{%
    \GenericError{(gnuplot) \space\space\space\@spaces}{%
      Package graphicx or graphics not loaded%
    }{See the gnuplot documentation for explanation.%
    }{The gnuplot epslatex terminal needs graphicx.sty or graphics.sty.}%
    \renewcommand\includegraphics[2][]{}%
  }%
  \providecommand\rotatebox[2]{#2}%
  \@ifundefined{ifGPcolor}{%
    \newif\ifGPcolor
    \GPcolortrue
  }{}%
  \@ifundefined{ifGPblacktext}{%
    \newif\ifGPblacktext
    \GPblacktexttrue
  }{}%
  % define a \g@addto@macro without @ in the name:
  \let\gplgaddtomacro\g@addto@macro
  % define empty templates for all commands taking text:
  \gdef\gplbacktext{}%
  \gdef\gplfronttext{}%
  \makeatother
  \ifGPblacktext
    % no textcolor at all
    \def\colorrgb#1{}%
    \def\colorgray#1{}%
  \else
    % gray or color?
    \ifGPcolor
      \def\colorrgb#1{\color[rgb]{#1}}%
      \def\colorgray#1{\color[gray]{#1}}%
      \expandafter\def\csname LTw\endcsname{\color{white}}%
      \expandafter\def\csname LTb\endcsname{\color{black}}%
      \expandafter\def\csname LTa\endcsname{\color{black}}%
      \expandafter\def\csname LT0\endcsname{\color[rgb]{1,0,0}}%
      \expandafter\def\csname LT1\endcsname{\color[rgb]{0,1,0}}%
      \expandafter\def\csname LT2\endcsname{\color[rgb]{0,0,1}}%
      \expandafter\def\csname LT3\endcsname{\color[rgb]{1,0,1}}%
      \expandafter\def\csname LT4\endcsname{\color[rgb]{0,1,1}}%
      \expandafter\def\csname LT5\endcsname{\color[rgb]{1,1,0}}%
      \expandafter\def\csname LT6\endcsname{\color[rgb]{0,0,0}}%
      \expandafter\def\csname LT7\endcsname{\color[rgb]{1,0.3,0}}%
      \expandafter\def\csname LT8\endcsname{\color[rgb]{0.5,0.5,0.5}}%
    \else
      % gray
      \def\colorrgb#1{\color{black}}%
      \def\colorgray#1{\color[gray]{#1}}%
      \expandafter\def\csname LTw\endcsname{\color{white}}%
      \expandafter\def\csname LTb\endcsname{\color{black}}%
      \expandafter\def\csname LTa\endcsname{\color{black}}%
      \expandafter\def\csname LT0\endcsname{\color{black}}%
      \expandafter\def\csname LT1\endcsname{\color{black}}%
      \expandafter\def\csname LT2\endcsname{\color{black}}%
      \expandafter\def\csname LT3\endcsname{\color{black}}%
      \expandafter\def\csname LT4\endcsname{\color{black}}%
      \expandafter\def\csname LT5\endcsname{\color{black}}%
      \expandafter\def\csname LT6\endcsname{\color{black}}%
      \expandafter\def\csname LT7\endcsname{\color{black}}%
      \expandafter\def\csname LT8\endcsname{\color{black}}%
    \fi
  \fi
    \setlength{\unitlength}{0.0500bp}%
    \ifx\gptboxheight\undefined%
      \newlength{\gptboxheight}%
      \newlength{\gptboxwidth}%
      \newsavebox{\gptboxtext}%
    \fi%
    \setlength{\fboxrule}{0.5pt}%
    \setlength{\fboxsep}{1pt}%
\begin{picture}(3456.00,2880.00)%
\definecolor{gpBackground}{rgb}{1.000, 1.000, 1.000}%
\put(0,0){\colorbox{gpBackground}{\makebox(3456.00,2880.00)[]{}}}%
    \gplgaddtomacro\gplbacktext{%
      \csname LTb\endcsname%
      \put(490,574){\makebox(0,0)[r]{\strut{}$0$}}%
      \put(490,1712){\makebox(0,0)[r]{\strut{}$0.4$}}%
      \put(490,2850){\makebox(0,0)[r]{\strut{}$0.8$}}%
      \put(622,212){\makebox(0,0){\strut{}$1$}}%
      \put(1863,212){\makebox(0,0){\strut{}$10$}}%
      \put(3103,212){\makebox(0,0){\strut{}$100$}}%
    }%
    \gplgaddtomacro\gplfronttext{%
      \csname LTb\endcsname%
      \put(50,2850){\rotatebox{0}{\makebox(0,0){\strut{}(d)}}}%
      \put(50,1641){\rotatebox{-270}{\makebox(0,0){\strut{}$-\overline{u'v'}/u_\tau^2$}}}%
      \put(2021,-118){\makebox(0,0){\strut{}$y^+$}}%
      \put(2021,2740){\makebox(0,0){\strut{}}}%
    }%
    \gplbacktext
    \put(0,0){\includegraphics{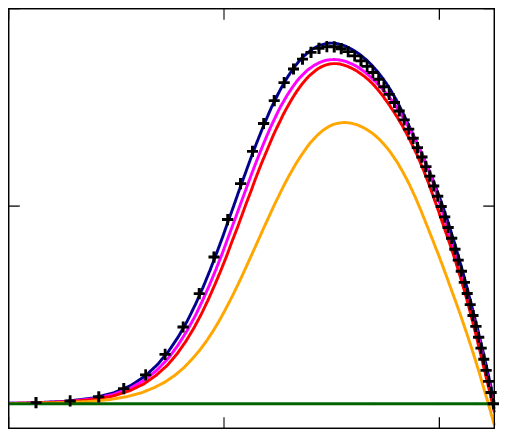}}%
    \gplfronttext
  \end{picture}%
\endgroup

%% file: reyK.tex
% GNUPLOT: LaTeX picture with Postscript
\begingroup
  \makeatletter
  \providecommand\color[2][]{%
    \GenericError{(gnuplot) \space\space\space\@spaces}{%
      Package color not loaded in conjunction with
      terminal option `colourtext'%
    }{See the gnuplot documentation for explanation.%
    }{Either use 'blacktext' in gnuplot or load the package
      color.sty in LaTeX.}%
    \renewcommand\color[2][]{}%
  }%
  \providecommand\includegraphics[2][]{%
    \GenericError{(gnuplot) \space\space\space\@spaces}{%
      Package graphicx or graphics not loaded%
    }{See the gnuplot documentation for explanation.%
    }{The gnuplot epslatex terminal needs graphicx.sty or graphics.sty.}%
    \renewcommand\includegraphics[2][]{}%
  }%
  \providecommand\rotatebox[2]{#2}%
  \@ifundefined{ifGPcolor}{%
    \newif\ifGPcolor
    \GPcolortrue
  }{}%
  \@ifundefined{ifGPblacktext}{%
    \newif\ifGPblacktext
    \GPblacktexttrue
  }{}%
  % define a \g@addto@macro without @ in the name:
  \let\gplgaddtomacro\g@addto@macro
  % define empty templates for all commands taking text:
  \gdef\gplbacktext{}%
  \gdef\gplfronttext{}%
  \makeatother
  \ifGPblacktext
    % no textcolor at all
    \def\colorrgb#1{}%
    \def\colorgray#1{}%
  \else
    % gray or color?
    \ifGPcolor
      \def\colorrgb#1{\color[rgb]{#1}}%
      \def\colorgray#1{\color[gray]{#1}}%
      \expandafter\def\csname LTw\endcsname{\color{white}}%
      \expandafter\def\csname LTb\endcsname{\color{black}}%
      \expandafter\def\csname LTa\endcsname{\color{black}}%
      \expandafter\def\csname LT0\endcsname{\color[rgb]{1,0,0}}%
      \expandafter\def\csname LT1\endcsname{\color[rgb]{0,1,0}}%
      \expandafter\def\csname LT2\endcsname{\color[rgb]{0,0,1}}%
      \expandafter\def\csname LT3\endcsname{\color[rgb]{1,0,1}}%
      \expandafter\def\csname LT4\endcsname{\color[rgb]{0,1,1}}%
      \expandafter\def\csname LT5\endcsname{\color[rgb]{1,1,0}}%
      \expandafter\def\csname LT6\endcsname{\color[rgb]{0,0,0}}%
      \expandafter\def\csname LT7\endcsname{\color[rgb]{1,0.3,0}}%
      \expandafter\def\csname LT8\endcsname{\color[rgb]{0.5,0.5,0.5}}%
    \else
      % gray
      \def\colorrgb#1{\color{black}}%
      \def\colorgray#1{\color[gray]{#1}}%
      \expandafter\def\csname LTw\endcsname{\color{white}}%
      \expandafter\def\csname LTb\endcsname{\color{black}}%
      \expandafter\def\csname LTa\endcsname{\color{black}}%
      \expandafter\def\csname LT0\endcsname{\color{black}}%
      \expandafter\def\csname LT1\endcsname{\color{black}}%
      \expandafter\def\csname LT2\endcsname{\color{black}}%
      \expandafter\def\csname LT3\endcsname{\color{black}}%
      \expandafter\def\csname LT4\endcsname{\color{black}}%
      \expandafter\def\csname LT5\endcsname{\color{black}}%
      \expandafter\def\csname LT6\endcsname{\color{black}}%
      \expandafter\def\csname LT7\endcsname{\color{black}}%
      \expandafter\def\csname LT8\endcsname{\color{black}}%
    \fi
  \fi
    \setlength{\unitlength}{0.0500bp}%
    \ifx\gptboxheight\undefined%
      \newlength{\gptboxheight}%
      \newlength{\gptboxwidth}%
      \newsavebox{\gptboxtext}%
    \fi%
    \setlength{\fboxrule}{0.5pt}%
    \setlength{\fboxsep}{1pt}%
\begin{picture}(3456.00,2880.00)%
\definecolor{gpBackground}{rgb}{1.000, 1.000, 1.000}%
\put(0,0){\colorbox{gpBackground}{\makebox(3456.00,2880.00)[]{}}}%
    \gplgaddtomacro\gplbacktext{%
      \csname LTb\endcsname%
      \put(490,618){\makebox(0,0)[r]{\strut{}$0$}}%
      \put(490,1362){\makebox(0,0)[r]{\strut{}$2$}}%
      \put(490,2106){\makebox(0,0)[r]{\strut{}$4$}}%
      \put(490,2850){\makebox(0,0)[r]{\strut{}$6$}}%
      \put(622,212){\makebox(0,0){\strut{}$1$}}%
      \put(1863,212){\makebox(0,0){\strut{}$10$}}%
      \put(3103,212){\makebox(0,0){\strut{}$100$}}%
    }%
    \gplgaddtomacro\gplfronttext{%
      \csname LTb\endcsname%
      \put(214,2850){\rotatebox{0}{\makebox(0,0){\strut{}(a)}}}%
      \put(214,1641){\rotatebox{-270}{\makebox(0,0){\strut{}$\mathcal{K}/u_\tau^2$}}}%
      \put(2021,-118){\makebox(0,0){\strut{}$y^+$}}%
      \put(2021,2740){\makebox(0,0){\strut{}}}%
    }%
    \gplbacktext
    \put(0,0){\includegraphics{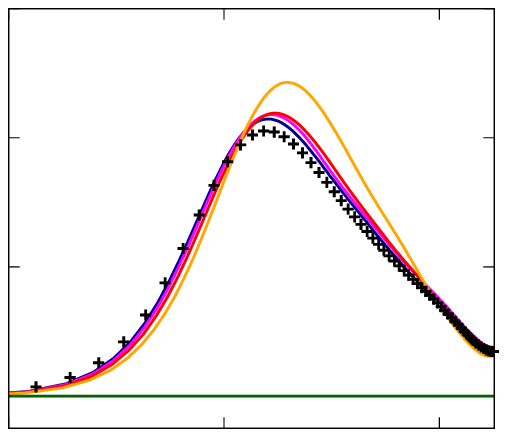}}%
    \gplfronttext
  \end{picture}%
\endgroup

%% file: prod.tex
% GNUPLOT: LaTeX picture with Postscript
\begingroup
  \makeatletter
  \providecommand\color[2][]{%
    \GenericError{(gnuplot) \space\space\space\@spaces}{%
      Package color not loaded in conjunction with
      terminal option `colourtext'%
    }{See the gnuplot documentation for explanation.%
    }{Either use 'blacktext' in gnuplot or load the package
      color.sty in LaTeX.}%
    \renewcommand\color[2][]{}%
  }%
  \providecommand\includegraphics[2][]{%
    \GenericError{(gnuplot) \space\space\space\@spaces}{%
      Package graphicx or graphics not loaded%
    }{See the gnuplot documentation for explanation.%
    }{The gnuplot epslatex terminal needs graphicx.sty or graphics.sty.}%
    \renewcommand\includegraphics[2][]{}%
  }%
  \providecommand\rotatebox[2]{#2}%
  \@ifundefined{ifGPcolor}{%
    \newif\ifGPcolor
    \GPcolortrue
  }{}%
  \@ifundefined{ifGPblacktext}{%
    \newif\ifGPblacktext
    \GPblacktexttrue
  }{}%
  % define a \g@addto@macro without @ in the name:
  \let\gplgaddtomacro\g@addto@macro
  % define empty templates for all commands taking text:
  \gdef\gplbacktext{}%
  \gdef\gplfronttext{}%
  \makeatother
  \ifGPblacktext
    % no textcolor at all
    \def\colorrgb#1{}%
    \def\colorgray#1{}%
  \else
    % gray or color?
    \ifGPcolor
      \def\colorrgb#1{\color[rgb]{#1}}%
      \def\colorgray#1{\color[gray]{#1}}%
      \expandafter\def\csname LTw\endcsname{\color{white}}%
      \expandafter\def\csname LTb\endcsname{\color{black}}%
      \expandafter\def\csname LTa\endcsname{\color{black}}%
      \expandafter\def\csname LT0\endcsname{\color[rgb]{1,0,0}}%
      \expandafter\def\csname LT1\endcsname{\color[rgb]{0,1,0}}%
      \expandafter\def\csname LT2\endcsname{\color[rgb]{0,0,1}}%
      \expandafter\def\csname LT3\endcsname{\color[rgb]{1,0,1}}%
      \expandafter\def\csname LT4\endcsname{\color[rgb]{0,1,1}}%
      \expandafter\def\csname LT5\endcsname{\color[rgb]{1,1,0}}%
      \expandafter\def\csname LT6\endcsname{\color[rgb]{0,0,0}}%
      \expandafter\def\csname LT7\endcsname{\color[rgb]{1,0.3,0}}%
      \expandafter\def\csname LT8\endcsname{\color[rgb]{0.5,0.5,0.5}}%
    \else
      % gray
      \def\colorrgb#1{\color{black}}%
      \def\colorgray#1{\color[gray]{#1}}%
      \expandafter\def\csname LTw\endcsname{\color{white}}%
      \expandafter\def\csname LTb\endcsname{\color{black}}%
      \expandafter\def\csname LTa\endcsname{\color{black}}%
      \expandafter\def\csname LT0\endcsname{\color{black}}%
      \expandafter\def\csname LT1\endcsname{\color{black}}%
      \expandafter\def\csname LT2\endcsname{\color{black}}%
      \expandafter\def\csname LT3\endcsname{\color{black}}%
      \expandafter\def\csname LT4\endcsname{\color{black}}%
      \expandafter\def\csname LT5\endcsname{\color{black}}%
      \expandafter\def\csname LT6\endcsname{\color{black}}%
      \expandafter\def\csname LT7\endcsname{\color{black}}%
      \expandafter\def\csname LT8\endcsname{\color{black}}%
    \fi
  \fi
    \setlength{\unitlength}{0.0500bp}%
    \ifx\gptboxheight\undefined%
      \newlength{\gptboxheight}%
      \newlength{\gptboxwidth}%
      \newsavebox{\gptboxtext}%
    \fi%
    \setlength{\fboxrule}{0.5pt}%
    \setlength{\fboxsep}{1pt}%
\begin{picture}(3456.00,2880.00)%
\definecolor{gpBackground}{rgb}{1.000, 1.000, 1.000}%
\put(0,0){\colorbox{gpBackground}{\makebox(3456.00,2880.00)[]{}}}%
    \gplgaddtomacro\gplbacktext{%
      \csname LTb\endcsname%
      \put(490,485){\makebox(0,0)[r]{\strut{}$0$}}%
      \put(490,1536){\makebox(0,0)[r]{\strut{}$20$}}%
      \put(490,2587){\makebox(0,0)[r]{\strut{}$40$}}%
      \put(622,212){\makebox(0,0){\strut{}$1$}}%
      \put(1863,212){\makebox(0,0){\strut{}$10$}}%
      \put(3103,212){\makebox(0,0){\strut{}$100$}}%
    }%
    \gplgaddtomacro\gplfronttext{%
      \csname LTb\endcsname%
      \put(182,2850){\rotatebox{0}{\makebox(0,0){\strut{}(b)}}}%
      \put(182,1641){\rotatebox{-270}{\makebox(0,0){\strut{}$\mathcal{P}$}}}%
      \put(2021,-118){\makebox(0,0){\strut{}$y^+$}}%
      \put(2021,2740){\makebox(0,0){\strut{}}}%
    }%
    \gplbacktext
    \put(0,0){\includegraphics{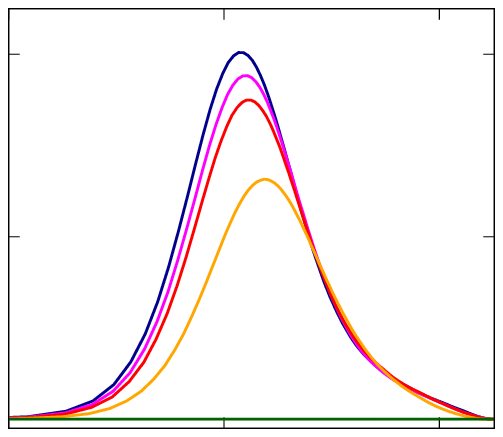}}%
    \gplfronttext
  \end{picture}%
\endgroup

%% file: diss.tex
% GNUPLOT: LaTeX picture with Postscript
\begingroup
  \makeatletter
  \providecommand\color[2][]{%
    \GenericError{(gnuplot) \space\space\space\@spaces}{%
      Package color not loaded in conjunction with
      terminal option `colourtext'%
    }{See the gnuplot documentation for explanation.%
    }{Either use 'blacktext' in gnuplot or load the package
      color.sty in LaTeX.}%
    \renewcommand\color[2][]{}%
  }%
  \providecommand\includegraphics[2][]{%
    \GenericError{(gnuplot) \space\space\space\@spaces}{%
      Package graphicx or graphics not loaded%
    }{See the gnuplot documentation for explanation.%
    }{The gnuplot epslatex terminal needs graphicx.sty or graphics.sty.}%
    \renewcommand\includegraphics[2][]{}%
  }%
  \providecommand\rotatebox[2]{#2}%
  \@ifundefined{ifGPcolor}{%
    \newif\ifGPcolor
    \GPcolortrue
  }{}%
  \@ifundefined{ifGPblacktext}{%
    \newif\ifGPblacktext
    \GPblacktexttrue
  }{}%
  % define a \g@addto@macro without @ in the name:
  \let\gplgaddtomacro\g@addto@macro
  % define empty templates for all commands taking text:
  \gdef\gplbacktext{}%
  \gdef\gplfronttext{}%
  \makeatother
  \ifGPblacktext
    % no textcolor at all
    \def\colorrgb#1{}%
    \def\colorgray#1{}%
  \else
    % gray or color?
    \ifGPcolor
      \def\colorrgb#1{\color[rgb]{#1}}%
      \def\colorgray#1{\color[gray]{#1}}%
      \expandafter\def\csname LTw\endcsname{\color{white}}%
      \expandafter\def\csname LTb\endcsname{\color{black}}%
      \expandafter\def\csname LTa\endcsname{\color{black}}%
      \expandafter\def\csname LT0\endcsname{\color[rgb]{1,0,0}}%
      \expandafter\def\csname LT1\endcsname{\color[rgb]{0,1,0}}%
      \expandafter\def\csname LT2\endcsname{\color[rgb]{0,0,1}}%
      \expandafter\def\csname LT3\endcsname{\color[rgb]{1,0,1}}%
      \expandafter\def\csname LT4\endcsname{\color[rgb]{0,1,1}}%
      \expandafter\def\csname LT5\endcsname{\color[rgb]{1,1,0}}%
      \expandafter\def\csname LT6\endcsname{\color[rgb]{0,0,0}}%
      \expandafter\def\csname LT7\endcsname{\color[rgb]{1,0.3,0}}%
      \expandafter\def\csname LT8\endcsname{\color[rgb]{0.5,0.5,0.5}}%
    \else
      % gray
      \def\colorrgb#1{\color{black}}%
      \def\colorgray#1{\color[gray]{#1}}%
      \expandafter\def\csname LTw\endcsname{\color{white}}%
      \expandafter\def\csname LTb\endcsname{\color{black}}%
      \expandafter\def\csname LTa\endcsname{\color{black}}%
      \expandafter\def\csname LT0\endcsname{\color{black}}%
      \expandafter\def\csname LT1\endcsname{\color{black}}%
      \expandafter\def\csname LT2\endcsname{\color{black}}%
      \expandafter\def\csname LT3\endcsname{\color{black}}%
      \expandafter\def\csname LT4\endcsname{\color{black}}%
      \expandafter\def\csname LT5\endcsname{\color{black}}%
      \expandafter\def\csname LT6\endcsname{\color{black}}%
      \expandafter\def\csname LT7\endcsname{\color{black}}%
      \expandafter\def\csname LT8\endcsname{\color{black}}%
    \fi
  \fi
    \setlength{\unitlength}{0.0500bp}%
    \ifx\gptboxheight\undefined%
      \newlength{\gptboxheight}%
      \newlength{\gptboxwidth}%
      \newsavebox{\gptboxtext}%
    \fi%
    \setlength{\fboxrule}{0.5pt}%
    \setlength{\fboxsep}{1pt}%
\begin{picture}(3456.00,2880.00)%
\definecolor{gpBackground}{rgb}{1.000, 1.000, 1.000}%
\put(0,0){\colorbox{gpBackground}{\makebox(3456.00,2880.00)[]{}}}%
    \gplgaddtomacro\gplbacktext{%
      \csname LTb\endcsname%
      \put(490,485){\makebox(0,0)[r]{\strut{}$0$}}%
      \put(490,1536){\makebox(0,0)[r]{\strut{}$20$}}%
      \put(490,2587){\makebox(0,0)[r]{\strut{}$40$}}%
      \put(622,212){\makebox(0,0){\strut{}$1$}}%
      \put(1863,212){\makebox(0,0){\strut{}$10$}}%
      \put(3103,212){\makebox(0,0){\strut{}$100$}}%
    }%
    \gplgaddtomacro\gplfronttext{%
      \csname LTb\endcsname%
      \put(182,2850){\rotatebox{0}{\makebox(0,0){\strut{}(a)}}}%
      \put(182,1641){\rotatebox{-270}{\makebox(0,0){\strut{}$\varepsilon$}}}%
      \put(2021,-118){\makebox(0,0){\strut{}$y^+$}}%
      \put(2021,2740){\makebox(0,0){\strut{}}}%
    }%
    \gplbacktext
    \put(0,0){\includegraphics{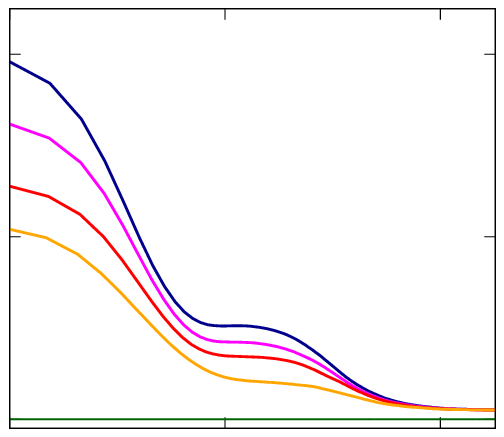}}%
    \gplfronttext
  \end{picture}%
\endgroup

%% file: stressVisc.tex
% GNUPLOT: LaTeX picture with Postscript
\begingroup
  \makeatletter
  \providecommand\color[2][]{%
    \GenericError{(gnuplot) \space\space\space\@spaces}{%
      Package color not loaded in conjunction with
      terminal option `colourtext'%
    }{See the gnuplot documentation for explanation.%
    }{Either use 'blacktext' in gnuplot or load the package
      color.sty in LaTeX.}%
    \renewcommand\color[2][]{}%
  }%
  \providecommand\includegraphics[2][]{%
    \GenericError{(gnuplot) \space\space\space\@spaces}{%
      Package graphicx or graphics not loaded%
    }{See the gnuplot documentation for explanation.%
    }{The gnuplot epslatex terminal needs graphicx.sty or graphics.sty.}%
    \renewcommand\includegraphics[2][]{}%
  }%
  \providecommand\rotatebox[2]{#2}%
  \@ifundefined{ifGPcolor}{%
    \newif\ifGPcolor
    \GPcolortrue
  }{}%
  \@ifundefined{ifGPblacktext}{%
    \newif\ifGPblacktext
    \GPblacktexttrue
  }{}%
  % define a \g@addto@macro without @ in the name:
  \let\gplgaddtomacro\g@addto@macro
  % define empty templates for all commands taking text:
  \gdef\gplbacktext{}%
  \gdef\gplfronttext{}%
  \makeatother
  \ifGPblacktext
    % no textcolor at all
    \def\colorrgb#1{}%
    \def\colorgray#1{}%
  \else
    % gray or color?
    \ifGPcolor
      \def\colorrgb#1{\color[rgb]{#1}}%
      \def\colorgray#1{\color[gray]{#1}}%
      \expandafter\def\csname LTw\endcsname{\color{white}}%
      \expandafter\def\csname LTb\endcsname{\color{black}}%
      \expandafter\def\csname LTa\endcsname{\color{black}}%
      \expandafter\def\csname LT0\endcsname{\color[rgb]{1,0,0}}%
      \expandafter\def\csname LT1\endcsname{\color[rgb]{0,1,0}}%
      \expandafter\def\csname LT2\endcsname{\color[rgb]{0,0,1}}%
      \expandafter\def\csname LT3\endcsname{\color[rgb]{1,0,1}}%
      \expandafter\def\csname LT4\endcsname{\color[rgb]{0,1,1}}%
      \expandafter\def\csname LT5\endcsname{\color[rgb]{1,1,0}}%
      \expandafter\def\csname LT6\endcsname{\color[rgb]{0,0,0}}%
      \expandafter\def\csname LT7\endcsname{\color[rgb]{1,0.3,0}}%
      \expandafter\def\csname LT8\endcsname{\color[rgb]{0.5,0.5,0.5}}%
    \else
      % gray
      \def\colorrgb#1{\color{black}}%
      \def\colorgray#1{\color[gray]{#1}}%
      \expandafter\def\csname LTw\endcsname{\color{white}}%
      \expandafter\def\csname LTb\endcsname{\color{black}}%
      \expandafter\def\csname LTa\endcsname{\color{black}}%
      \expandafter\def\csname LT0\endcsname{\color{black}}%
      \expandafter\def\csname LT1\endcsname{\color{black}}%
      \expandafter\def\csname LT2\endcsname{\color{black}}%
      \expandafter\def\csname LT3\endcsname{\color{black}}%
      \expandafter\def\csname LT4\endcsname{\color{black}}%
      \expandafter\def\csname LT5\endcsname{\color{black}}%
      \expandafter\def\csname LT6\endcsname{\color{black}}%
      \expandafter\def\csname LT7\endcsname{\color{black}}%
      \expandafter\def\csname LT8\endcsname{\color{black}}%
    \fi
  \fi
    \setlength{\unitlength}{0.0500bp}%
    \ifx\gptboxheight\undefined%
      \newlength{\gptboxheight}%
      \newlength{\gptboxwidth}%
      \newsavebox{\gptboxtext}%
    \fi%
    \setlength{\fboxrule}{0.5pt}%
    \setlength{\fboxsep}{1pt}%
\begin{picture}(3456.00,2880.00)%
\definecolor{gpBackground}{rgb}{1.000, 1.000, 1.000}%
\put(0,0){\colorbox{gpBackground}{\makebox(3456.00,2880.00)[]{}}}%
    \gplgaddtomacro\gplbacktext{%
      \csname LTb\endcsname%
      \put(490,432){\makebox(0,0)[r]{\strut{}$0$}}%
      \put(490,1641){\makebox(0,0)[r]{\strut{}$3$}}%
      \put(490,2850){\makebox(0,0)[r]{\strut{}$6$}}%
      \put(622,212){\makebox(0,0){\strut{}$1$}}%
      \put(1838,212){\makebox(0,0){\strut{}$10$}}%
      \put(3054,212){\makebox(0,0){\strut{}$100$}}%
    }%
    \gplgaddtomacro\gplfronttext{%
      \csname LTb\endcsname%
      \put(214,2850){\rotatebox{0}{\makebox(0,0){\strut{}(b)}}}%
      \put(214,1641){\rotatebox{-270}{\makebox(0,0){\strut{}$\mu_e/\mu_0$}}}%
      \put(2021,-118){\makebox(0,0){\strut{}$y^+$}}%
      \put(2021,2740){\makebox(0,0){\strut{}}}%
    }%
    \gplgaddtomacro\gplbacktext{%
      \csname LTb\endcsname%
      \put(1077,1872){\makebox(0,0)[r]{\strut{}$0$}}%
      \put(1077,2304){\makebox(0,0)[r]{\strut{}$3$}}%
      \put(1077,2735){\makebox(0,0)[r]{\strut{}$6$}}%
      \put(1727,1652){\makebox(0,0){\strut{}$0.1$}}%
      \put(2246,1652){\makebox(0,0){\strut{}$1$}}%
      \put(2764,1652){\makebox(0,0){\strut{}$10$}}%
    }%
    \gplgaddtomacro\gplfronttext{%
      \csname LTb\endcsname%
      \put(801,2303){\rotatebox{-270}{\makebox(0,0){\strut{}$\mu_e/\mu_0$}}}%
      \put(1986,1322){\makebox(0,0){\strut{}$\dot{\gamma}$}}%
      \put(1986,2625){\makebox(0,0){\strut{}}}%
    }%
    \gplbacktext
    \put(0,0){\includegraphics{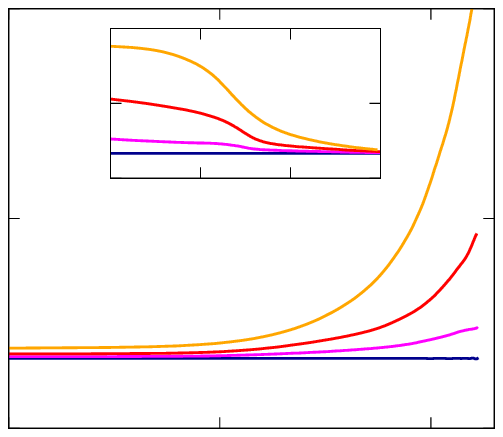}}%
    \gplfronttext
  \end{picture}%
\endgroup

%% file: meanSii.tex
% GNUPLOT: LaTeX picture with Postscript
\begingroup
  \makeatletter
  \providecommand\color[2][]{%
    \GenericError{(gnuplot) \space\space\space\@spaces}{%
      Package color not loaded in conjunction with
      terminal option `colourtext'%
    }{See the gnuplot documentation for explanation.%
    }{Either use 'blacktext' in gnuplot or load the package
      color.sty in LaTeX.}%
    \renewcommand\color[2][]{}%
  }%
  \providecommand\includegraphics[2][]{%
    \GenericError{(gnuplot) \space\space\space\@spaces}{%
      Package graphicx or graphics not loaded%
    }{See the gnuplot documentation for explanation.%
    }{The gnuplot epslatex terminal needs graphicx.sty or graphics.sty.}%
    \renewcommand\includegraphics[2][]{}%
  }%
  \providecommand\rotatebox[2]{#2}%
  \@ifundefined{ifGPcolor}{%
    \newif\ifGPcolor
    \GPcolortrue
  }{}%
  \@ifundefined{ifGPblacktext}{%
    \newif\ifGPblacktext
    \GPblacktexttrue
  }{}%
  % define a \g@addto@macro without @ in the name:
  \let\gplgaddtomacro\g@addto@macro
  % define empty templates for all commands taking text:
  \gdef\gplbacktext{}%
  \gdef\gplfronttext{}%
  \makeatother
  \ifGPblacktext
    % no textcolor at all
    \def\colorrgb#1{}%
    \def\colorgray#1{}%
  \else
    % gray or color?
    \ifGPcolor
      \def\colorrgb#1{\color[rgb]{#1}}%
      \def\colorgray#1{\color[gray]{#1}}%
      \expandafter\def\csname LTw\endcsname{\color{white}}%
      \expandafter\def\csname LTb\endcsname{\color{black}}%
      \expandafter\def\csname LTa\endcsname{\color{black}}%
      \expandafter\def\csname LT0\endcsname{\color[rgb]{1,0,0}}%
      \expandafter\def\csname LT1\endcsname{\color[rgb]{0,1,0}}%
      \expandafter\def\csname LT2\endcsname{\color[rgb]{0,0,1}}%
      \expandafter\def\csname LT3\endcsname{\color[rgb]{1,0,1}}%
      \expandafter\def\csname LT4\endcsname{\color[rgb]{0,1,1}}%
      \expandafter\def\csname LT5\endcsname{\color[rgb]{1,1,0}}%
      \expandafter\def\csname LT6\endcsname{\color[rgb]{0,0,0}}%
      \expandafter\def\csname LT7\endcsname{\color[rgb]{1,0.3,0}}%
      \expandafter\def\csname LT8\endcsname{\color[rgb]{0.5,0.5,0.5}}%
    \else
      % gray
      \def\colorrgb#1{\color{black}}%
      \def\colorgray#1{\color[gray]{#1}}%
      \expandafter\def\csname LTw\endcsname{\color{white}}%
      \expandafter\def\csname LTb\endcsname{\color{black}}%
      \expandafter\def\csname LTa\endcsname{\color{black}}%
      \expandafter\def\csname LT0\endcsname{\color{black}}%
      \expandafter\def\csname LT1\endcsname{\color{black}}%
      \expandafter\def\csname LT2\endcsname{\color{black}}%
      \expandafter\def\csname LT3\endcsname{\color{black}}%
      \expandafter\def\csname LT4\endcsname{\color{black}}%
      \expandafter\def\csname LT5\endcsname{\color{black}}%
      \expandafter\def\csname LT6\endcsname{\color{black}}%
      \expandafter\def\csname LT7\endcsname{\color{black}}%
      \expandafter\def\csname LT8\endcsname{\color{black}}%
    \fi
  \fi
    \setlength{\unitlength}{0.0500bp}%
    \ifx\gptboxheight\undefined%
      \newlength{\gptboxheight}%
      \newlength{\gptboxwidth}%
      \newsavebox{\gptboxtext}%
    \fi%
    \setlength{\fboxrule}{0.5pt}%
    \setlength{\fboxsep}{1pt}%
\begin{picture}(3456.00,2880.00)%
\definecolor{gpBackground}{rgb}{1.000, 1.000, 1.000}%
\put(0,0){\colorbox{gpBackground}{\makebox(3456.00,2880.00)[]{}}}%
    \gplgaddtomacro\gplbacktext{%
      \csname LTb\endcsname%
      \put(490,456){\makebox(0,0)[r]{\strut{}$0$}}%
      \put(490,1653){\makebox(0,0)[r]{\strut{}$5$}}%
      \put(490,2850){\makebox(0,0)[r]{\strut{}$10$}}%
      \put(622,212){\makebox(0,0){\strut{}$0$}}%
      \put(2021,212){\makebox(0,0){\strut{}$0.5$}}%
      \put(3420,212){\makebox(0,0){\strut{}$1$}}%
    }%
    \gplgaddtomacro\gplfronttext{%
      \csname LTb\endcsname%
      \put(-49,2850){\rotatebox{0}{\makebox(0,0){\strut{}(a)}}}%
      \put(-49,1641){\rotatebox{-270}{\makebox(0,0){\strut{}$10^4 \times \overline{\tau}_{ii}$}}}%
      \put(2021,-118){\makebox(0,0){\strut{}$y/h$}}%
      \put(2021,2740){\makebox(0,0){\strut{}}}%
    }%
    \gplbacktext
    \put(0,0){\includegraphics{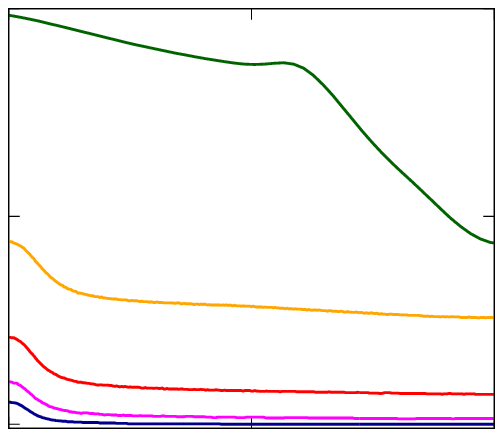}}%
    \gplfronttext
  \end{picture}%
\endgroup

%% file: meanS12.tex
% GNUPLOT: LaTeX picture with Postscript
\begingroup
  \makeatletter
  \providecommand\color[2][]{%
    \GenericError{(gnuplot) \space\space\space\@spaces}{%
      Package color not loaded in conjunction with
      terminal option `colourtext'%
    }{See the gnuplot documentation for explanation.%
    }{Either use 'blacktext' in gnuplot or load the package
      color.sty in LaTeX.}%
    \renewcommand\color[2][]{}%
  }%
  \providecommand\includegraphics[2][]{%
    \GenericError{(gnuplot) \space\space\space\@spaces}{%
      Package graphicx or graphics not loaded%
    }{See the gnuplot documentation for explanation.%
    }{The gnuplot epslatex terminal needs graphicx.sty or graphics.sty.}%
    \renewcommand\includegraphics[2][]{}%
  }%
  \providecommand\rotatebox[2]{#2}%
  \@ifundefined{ifGPcolor}{%
    \newif\ifGPcolor
    \GPcolortrue
  }{}%
  \@ifundefined{ifGPblacktext}{%
    \newif\ifGPblacktext
    \GPblacktexttrue
  }{}%
  % define a \g@addto@macro without @ in the name:
  \let\gplgaddtomacro\g@addto@macro
  % define empty templates for all commands taking text:
  \gdef\gplbacktext{}%
  \gdef\gplfronttext{}%
  \makeatother
  \ifGPblacktext
    % no textcolor at all
    \def\colorrgb#1{}%
    \def\colorgray#1{}%
  \else
    % gray or color?
    \ifGPcolor
      \def\colorrgb#1{\color[rgb]{#1}}%
      \def\colorgray#1{\color[gray]{#1}}%
      \expandafter\def\csname LTw\endcsname{\color{white}}%
      \expandafter\def\csname LTb\endcsname{\color{black}}%
      \expandafter\def\csname LTa\endcsname{\color{black}}%
      \expandafter\def\csname LT0\endcsname{\color[rgb]{1,0,0}}%
      \expandafter\def\csname LT1\endcsname{\color[rgb]{0,1,0}}%
      \expandafter\def\csname LT2\endcsname{\color[rgb]{0,0,1}}%
      \expandafter\def\csname LT3\endcsname{\color[rgb]{1,0,1}}%
      \expandafter\def\csname LT4\endcsname{\color[rgb]{0,1,1}}%
      \expandafter\def\csname LT5\endcsname{\color[rgb]{1,1,0}}%
      \expandafter\def\csname LT6\endcsname{\color[rgb]{0,0,0}}%
      \expandafter\def\csname LT7\endcsname{\color[rgb]{1,0.3,0}}%
      \expandafter\def\csname LT8\endcsname{\color[rgb]{0.5,0.5,0.5}}%
    \else
      % gray
      \def\colorrgb#1{\color{black}}%
      \def\colorgray#1{\color[gray]{#1}}%
      \expandafter\def\csname LTw\endcsname{\color{white}}%
      \expandafter\def\csname LTb\endcsname{\color{black}}%
      \expandafter\def\csname LTa\endcsname{\color{black}}%
      \expandafter\def\csname LT0\endcsname{\color{black}}%
      \expandafter\def\csname LT1\endcsname{\color{black}}%
      \expandafter\def\csname LT2\endcsname{\color{black}}%
      \expandafter\def\csname LT3\endcsname{\color{black}}%
      \expandafter\def\csname LT4\endcsname{\color{black}}%
      \expandafter\def\csname LT5\endcsname{\color{black}}%
      \expandafter\def\csname LT6\endcsname{\color{black}}%
      \expandafter\def\csname LT7\endcsname{\color{black}}%
      \expandafter\def\csname LT8\endcsname{\color{black}}%
    \fi
  \fi
    \setlength{\unitlength}{0.0500bp}%
    \ifx\gptboxheight\undefined%
      \newlength{\gptboxheight}%
      \newlength{\gptboxwidth}%
      \newsavebox{\gptboxtext}%
    \fi%
    \setlength{\fboxrule}{0.5pt}%
    \setlength{\fboxsep}{1pt}%
\begin{picture}(3456.00,2880.00)%
\definecolor{gpBackground}{rgb}{1.000, 1.000, 1.000}%
\put(0,0){\colorbox{gpBackground}{\makebox(3456.00,2880.00)[]{}}}%
    \gplgaddtomacro\gplbacktext{%
      \csname LTb\endcsname%
      \put(490,456){\makebox(0,0)[r]{\strut{}$0$}}%
      \put(490,1653){\makebox(0,0)[r]{\strut{}$5$}}%
      \put(490,2850){\makebox(0,0)[r]{\strut{}$10$}}%
      \put(622,212){\makebox(0,0){\strut{}$0$}}%
      \put(2021,212){\makebox(0,0){\strut{}$0.5$}}%
      \put(3420,212){\makebox(0,0){\strut{}$1$}}%
    }%
    \gplgaddtomacro\gplfronttext{%
      \csname LTb\endcsname%
      \put(182,2850){\rotatebox{0}{\makebox(0,0){\strut{}(b)}}}%
      \put(182,1641){\rotatebox{-270}{\makebox(0,0){\strut{}$10^4 \times \overline{\tau}_{12}$}}}%
      \put(2021,-118){\makebox(0,0){\strut{}$y/h$}}%
      \put(2021,2740){\makebox(0,0){\strut{}}}%
    }%
    \gplbacktext
    \put(0,0){\includegraphics{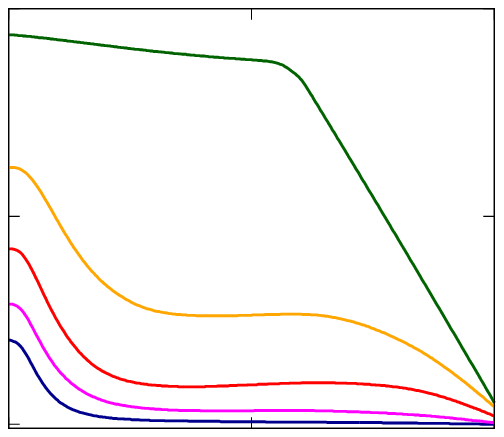}}%
    \gplfronttext
  \end{picture}%
\endgroup

%% file: stress01.tex
% GNUPLOT: LaTeX picture with Postscript
\begingroup
  \makeatletter
  \providecommand\color[2][]{%
    \GenericError{(gnuplot) \space\space\space\@spaces}{%
      Package color not loaded in conjunction with
      terminal option `colourtext'%
    }{See the gnuplot documentation for explanation.%
    }{Either use 'blacktext' in gnuplot or load the package
      color.sty in LaTeX.}%
    \renewcommand\color[2][]{}%
  }%
  \providecommand\includegraphics[2][]{%
    \GenericError{(gnuplot) \space\space\space\@spaces}{%
      Package graphicx or graphics not loaded%
    }{See the gnuplot documentation for explanation.%
    }{The gnuplot epslatex terminal needs graphicx.sty or graphics.sty.}%
    \renewcommand\includegraphics[2][]{}%
  }%
  \providecommand\rotatebox[2]{#2}%
  \@ifundefined{ifGPcolor}{%
    \newif\ifGPcolor
    \GPcolortrue
  }{}%
  \@ifundefined{ifGPblacktext}{%
    \newif\ifGPblacktext
    \GPblacktexttrue
  }{}%
  % define a \g@addto@macro without @ in the name:
  \let\gplgaddtomacro\g@addto@macro
  % define empty templates for all commands taking text:
  \gdef\gplbacktext{}%
  \gdef\gplfronttext{}%
  \makeatother
  \ifGPblacktext
    % no textcolor at all
    \def\colorrgb#1{}%
    \def\colorgray#1{}%
  \else
    % gray or color?
    \ifGPcolor
      \def\colorrgb#1{\color[rgb]{#1}}%
      \def\colorgray#1{\color[gray]{#1}}%
      \expandafter\def\csname LTw\endcsname{\color{white}}%
      \expandafter\def\csname LTb\endcsname{\color{black}}%
      \expandafter\def\csname LTa\endcsname{\color{black}}%
      \expandafter\def\csname LT0\endcsname{\color[rgb]{1,0,0}}%
      \expandafter\def\csname LT1\endcsname{\color[rgb]{0,1,0}}%
      \expandafter\def\csname LT2\endcsname{\color[rgb]{0,0,1}}%
      \expandafter\def\csname LT3\endcsname{\color[rgb]{1,0,1}}%
      \expandafter\def\csname LT4\endcsname{\color[rgb]{0,1,1}}%
      \expandafter\def\csname LT5\endcsname{\color[rgb]{1,1,0}}%
      \expandafter\def\csname LT6\endcsname{\color[rgb]{0,0,0}}%
      \expandafter\def\csname LT7\endcsname{\color[rgb]{1,0.3,0}}%
      \expandafter\def\csname LT8\endcsname{\color[rgb]{0.5,0.5,0.5}}%
    \else
      % gray
      \def\colorrgb#1{\color{black}}%
      \def\colorgray#1{\color[gray]{#1}}%
      \expandafter\def\csname LTw\endcsname{\color{white}}%
      \expandafter\def\csname LTb\endcsname{\color{black}}%
      \expandafter\def\csname LTa\endcsname{\color{black}}%
      \expandafter\def\csname LT0\endcsname{\color{black}}%
      \expandafter\def\csname LT1\endcsname{\color{black}}%
      \expandafter\def\csname LT2\endcsname{\color{black}}%
      \expandafter\def\csname LT3\endcsname{\color{black}}%
      \expandafter\def\csname LT4\endcsname{\color{black}}%
      \expandafter\def\csname LT5\endcsname{\color{black}}%
      \expandafter\def\csname LT6\endcsname{\color{black}}%
      \expandafter\def\csname LT7\endcsname{\color{black}}%
      \expandafter\def\csname LT8\endcsname{\color{black}}%
    \fi
  \fi
    \setlength{\unitlength}{0.0500bp}%
    \ifx\gptboxheight\undefined%
      \newlength{\gptboxheight}%
      \newlength{\gptboxwidth}%
      \newsavebox{\gptboxtext}%
    \fi%
    \setlength{\fboxrule}{0.5pt}%
    \setlength{\fboxsep}{1pt}%
\begin{picture}(3456.00,2880.00)%
\definecolor{gpBackground}{rgb}{1.000, 1.000, 1.000}%
\put(0,0){\colorbox{gpBackground}{\makebox(3456.00,2880.00)[]{}}}%
    \gplgaddtomacro\gplbacktext{%
      \csname LTb\endcsname%
      \put(490,479){\makebox(0,0)[r]{\strut{}$0$}}%
      \put(490,1641){\makebox(0,0)[r]{\strut{}$0.5$}}%
      \put(490,2804){\makebox(0,0)[r]{\strut{}$1$}}%
      \put(622,212){\makebox(0,0){\strut{}$0$}}%
      \put(2021,212){\makebox(0,0){\strut{}$0.5$}}%
      \put(3420,212){\makebox(0,0){\strut{}$1$}}%
    }%
    \gplgaddtomacro\gplfronttext{%
      \csname LTb\endcsname%
      \put(50,2850){\rotatebox{0}{\makebox(0,0){\strut{}(a)}}}%
      \put(50,1641){\rotatebox{-270}{\makebox(0,0){\strut{}$\tau/\tau_w$}}}%
      \put(2021,-118){\makebox(0,0){\strut{}$y/h$}}%
      \put(2021,2740){\makebox(0,0){\strut{}}}%
    }%
    \gplbacktext
    \put(0,0){\includegraphics{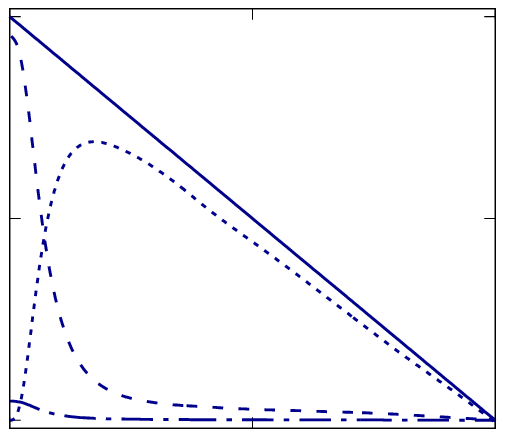}}%
    \gplfronttext
  \end{picture}%
\endgroup

%% file: stress04.tex
% GNUPLOT: LaTeX picture with Postscript
\begingroup
  \makeatletter
  \providecommand\color[2][]{%
    \GenericError{(gnuplot) \space\space\space\@spaces}{%
      Package color not loaded in conjunction with
      terminal option `colourtext'%
    }{See the gnuplot documentation for explanation.%
    }{Either use 'blacktext' in gnuplot or load the package
      color.sty in LaTeX.}%
    \renewcommand\color[2][]{}%
  }%
  \providecommand\includegraphics[2][]{%
    \GenericError{(gnuplot) \space\space\space\@spaces}{%
      Package graphicx or graphics not loaded%
    }{See the gnuplot documentation for explanation.%
    }{The gnuplot epslatex terminal needs graphicx.sty or graphics.sty.}%
    \renewcommand\includegraphics[2][]{}%
  }%
  \providecommand\rotatebox[2]{#2}%
  \@ifundefined{ifGPcolor}{%
    \newif\ifGPcolor
    \GPcolortrue
  }{}%
  \@ifundefined{ifGPblacktext}{%
    \newif\ifGPblacktext
    \GPblacktexttrue
  }{}%
  % define a \g@addto@macro without @ in the name:
  \let\gplgaddtomacro\g@addto@macro
  % define empty templates for all commands taking text:
  \gdef\gplbacktext{}%
  \gdef\gplfronttext{}%
  \makeatother
  \ifGPblacktext
    % no textcolor at all
    \def\colorrgb#1{}%
    \def\colorgray#1{}%
  \else
    % gray or color?
    \ifGPcolor
      \def\colorrgb#1{\color[rgb]{#1}}%
      \def\colorgray#1{\color[gray]{#1}}%
      \expandafter\def\csname LTw\endcsname{\color{white}}%
      \expandafter\def\csname LTb\endcsname{\color{black}}%
      \expandafter\def\csname LTa\endcsname{\color{black}}%
      \expandafter\def\csname LT0\endcsname{\color[rgb]{1,0,0}}%
      \expandafter\def\csname LT1\endcsname{\color[rgb]{0,1,0}}%
      \expandafter\def\csname LT2\endcsname{\color[rgb]{0,0,1}}%
      \expandafter\def\csname LT3\endcsname{\color[rgb]{1,0,1}}%
      \expandafter\def\csname LT4\endcsname{\color[rgb]{0,1,1}}%
      \expandafter\def\csname LT5\endcsname{\color[rgb]{1,1,0}}%
      \expandafter\def\csname LT6\endcsname{\color[rgb]{0,0,0}}%
      \expandafter\def\csname LT7\endcsname{\color[rgb]{1,0.3,0}}%
      \expandafter\def\csname LT8\endcsname{\color[rgb]{0.5,0.5,0.5}}%
    \else
      % gray
      \def\colorrgb#1{\color{black}}%
      \def\colorgray#1{\color[gray]{#1}}%
      \expandafter\def\csname LTw\endcsname{\color{white}}%
      \expandafter\def\csname LTb\endcsname{\color{black}}%
      \expandafter\def\csname LTa\endcsname{\color{black}}%
      \expandafter\def\csname LT0\endcsname{\color{black}}%
      \expandafter\def\csname LT1\endcsname{\color{black}}%
      \expandafter\def\csname LT2\endcsname{\color{black}}%
      \expandafter\def\csname LT3\endcsname{\color{black}}%
      \expandafter\def\csname LT4\endcsname{\color{black}}%
      \expandafter\def\csname LT5\endcsname{\color{black}}%
      \expandafter\def\csname LT6\endcsname{\color{black}}%
      \expandafter\def\csname LT7\endcsname{\color{black}}%
      \expandafter\def\csname LT8\endcsname{\color{black}}%
    \fi
  \fi
    \setlength{\unitlength}{0.0500bp}%
    \ifx\gptboxheight\undefined%
      \newlength{\gptboxheight}%
      \newlength{\gptboxwidth}%
      \newsavebox{\gptboxtext}%
    \fi%
    \setlength{\fboxrule}{0.5pt}%
    \setlength{\fboxsep}{1pt}%
\begin{picture}(3456.00,2880.00)%
\definecolor{gpBackground}{rgb}{1.000, 1.000, 1.000}%
\put(0,0){\colorbox{gpBackground}{\makebox(3456.00,2880.00)[]{}}}%
    \gplgaddtomacro\gplbacktext{%
      \csname LTb\endcsname%
      \put(490,479){\makebox(0,0)[r]{\strut{}$0$}}%
      \put(490,1641){\makebox(0,0)[r]{\strut{}$0.5$}}%
      \put(490,2804){\makebox(0,0)[r]{\strut{}$1$}}%
      \put(622,212){\makebox(0,0){\strut{}$0$}}%
      \put(2021,212){\makebox(0,0){\strut{}$0.5$}}%
      \put(3420,212){\makebox(0,0){\strut{}$1$}}%
    }%
    \gplgaddtomacro\gplfronttext{%
      \csname LTb\endcsname%
      \put(75,2850){\rotatebox{0}{\makebox(0,0){\strut{}(b)}}}%
      \put(75,1641){\rotatebox{-270}{\makebox(0,0){\strut{}$\tau/\tau_w$}}}%
      \put(2021,-118){\makebox(0,0){\strut{}$y/h$}}%
      \put(2021,2740){\makebox(0,0){\strut{}}}%
    }%
    \gplbacktext
    \put(0,0){\includegraphics{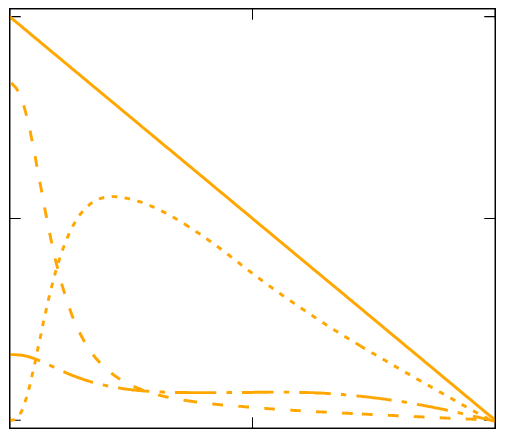}}%
    \gplfronttext
  \end{picture}%
\endgroup

%% file: dubi01.tex
% GNUPLOT: LaTeX picture with Postscript
\begingroup
  \makeatletter
  \providecommand\color[2][]{%
    \GenericError{(gnuplot) \space\space\space\@spaces}{%
      Package color not loaded in conjunction with
      terminal option `colourtext'%
    }{See the gnuplot documentation for explanation.%
    }{Either use 'blacktext' in gnuplot or load the package
      color.sty in LaTeX.}%
    \renewcommand\color[2][]{}%
  }%
  \providecommand\includegraphics[2][]{%
    \GenericError{(gnuplot) \space\space\space\@spaces}{%
      Package graphicx or graphics not loaded%
    }{See the gnuplot documentation for explanation.%
    }{The gnuplot epslatex terminal needs graphicx.sty or graphics.sty.}%
    \renewcommand\includegraphics[2][]{}%
  }%
  \providecommand\rotatebox[2]{#2}%
  \@ifundefined{ifGPcolor}{%
    \newif\ifGPcolor
    \GPcolortrue
  }{}%
  \@ifundefined{ifGPblacktext}{%
    \newif\ifGPblacktext
    \GPblacktexttrue
  }{}%
  % define a \g@addto@macro without @ in the name:
  \let\gplgaddtomacro\g@addto@macro
  % define empty templates for all commands taking text:
  \gdef\gplbacktext{}%
  \gdef\gplfronttext{}%
  \makeatother
  \ifGPblacktext
    % no textcolor at all
    \def\colorrgb#1{}%
    \def\colorgray#1{}%
  \else
    % gray or color?
    \ifGPcolor
      \def\colorrgb#1{\color[rgb]{#1}}%
      \def\colorgray#1{\color[gray]{#1}}%
      \expandafter\def\csname LTw\endcsname{\color{white}}%
      \expandafter\def\csname LTb\endcsname{\color{black}}%
      \expandafter\def\csname LTa\endcsname{\color{black}}%
      \expandafter\def\csname LT0\endcsname{\color[rgb]{1,0,0}}%
      \expandafter\def\csname LT1\endcsname{\color[rgb]{0,1,0}}%
      \expandafter\def\csname LT2\endcsname{\color[rgb]{0,0,1}}%
      \expandafter\def\csname LT3\endcsname{\color[rgb]{1,0,1}}%
      \expandafter\def\csname LT4\endcsname{\color[rgb]{0,1,1}}%
      \expandafter\def\csname LT5\endcsname{\color[rgb]{1,1,0}}%
      \expandafter\def\csname LT6\endcsname{\color[rgb]{0,0,0}}%
      \expandafter\def\csname LT7\endcsname{\color[rgb]{1,0.3,0}}%
      \expandafter\def\csname LT8\endcsname{\color[rgb]{0.5,0.5,0.5}}%
    \else
      % gray
      \def\colorrgb#1{\color{black}}%
      \def\colorgray#1{\color[gray]{#1}}%
      \expandafter\def\csname LTw\endcsname{\color{white}}%
      \expandafter\def\csname LTb\endcsname{\color{black}}%
      \expandafter\def\csname LTa\endcsname{\color{black}}%
      \expandafter\def\csname LT0\endcsname{\color{black}}%
      \expandafter\def\csname LT1\endcsname{\color{black}}%
      \expandafter\def\csname LT2\endcsname{\color{black}}%
      \expandafter\def\csname LT3\endcsname{\color{black}}%
      \expandafter\def\csname LT4\endcsname{\color{black}}%
      \expandafter\def\csname LT5\endcsname{\color{black}}%
      \expandafter\def\csname LT6\endcsname{\color{black}}%
      \expandafter\def\csname LT7\endcsname{\color{black}}%
      \expandafter\def\csname LT8\endcsname{\color{black}}%
    \fi
  \fi
    \setlength{\unitlength}{0.0500bp}%
    \ifx\gptboxheight\undefined%
      \newlength{\gptboxheight}%
      \newlength{\gptboxwidth}%
      \newsavebox{\gptboxtext}%
    \fi%
    \setlength{\fboxrule}{0.5pt}%
    \setlength{\fboxsep}{1pt}%
\begin{picture}(3456.00,2880.00)%
\definecolor{gpBackground}{rgb}{1.000, 1.000, 1.000}%
\put(0,0){\colorbox{gpBackground}{\makebox(3456.00,2880.00)[]{}}}%
    \gplgaddtomacro\gplbacktext{%
      \csname LTb\endcsname%
      \put(490,432){\makebox(0,0)[r]{\strut{}$-1$}}%
      \put(490,1641){\makebox(0,0)[r]{\strut{}$0$}}%
      \put(490,2850){\makebox(0,0)[r]{\strut{}$1$}}%
      \put(622,212){\makebox(0,0){\strut{}$0$}}%
      \put(2021,212){\makebox(0,0){\strut{}$0.5$}}%
      \put(3420,212){\makebox(0,0){\strut{}$1$}}%
    }%
    \gplgaddtomacro\gplfronttext{%
      \csname LTb\endcsname%
      \put(182,2850){\rotatebox{0}{\makebox(0,0){\strut{}(a)}}}%
      \put(182,1641){\rotatebox{-270}{\makebox(0,0){\strut{}$\rho_{i}$}}}%
      \put(2021,-118){\makebox(0,0){\strut{}$y/h$}}%
      \put(2021,2740){\makebox(0,0){\strut{}}}%
    }%
    \gplbacktext
    \put(0,0){\includegraphics{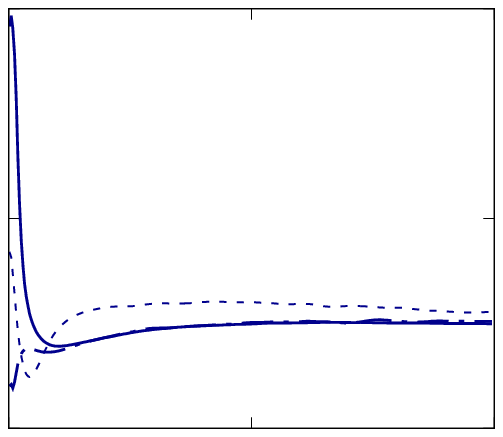}}%
    \gplfronttext
  \end{picture}%
\endgroup

%% file: dubi04.tex
% GNUPLOT: LaTeX picture with Postscript
\begingroup
  \makeatletter
  \providecommand\color[2][]{%
    \GenericError{(gnuplot) \space\space\space\@spaces}{%
      Package color not loaded in conjunction with
      terminal option `colourtext'%
    }{See the gnuplot documentation for explanation.%
    }{Either use 'blacktext' in gnuplot or load the package
      color.sty in LaTeX.}%
    \renewcommand\color[2][]{}%
  }%
  \providecommand\includegraphics[2][]{%
    \GenericError{(gnuplot) \space\space\space\@spaces}{%
      Package graphicx or graphics not loaded%
    }{See the gnuplot documentation for explanation.%
    }{The gnuplot epslatex terminal needs graphicx.sty or graphics.sty.}%
    \renewcommand\includegraphics[2][]{}%
  }%
  \providecommand\rotatebox[2]{#2}%
  \@ifundefined{ifGPcolor}{%
    \newif\ifGPcolor
    \GPcolortrue
  }{}%
  \@ifundefined{ifGPblacktext}{%
    \newif\ifGPblacktext
    \GPblacktexttrue
  }{}%
  % define a \g@addto@macro without @ in the name:
  \let\gplgaddtomacro\g@addto@macro
  % define empty templates for all commands taking text:
  \gdef\gplbacktext{}%
  \gdef\gplfronttext{}%
  \makeatother
  \ifGPblacktext
    % no textcolor at all
    \def\colorrgb#1{}%
    \def\colorgray#1{}%
  \else
    % gray or color?
    \ifGPcolor
      \def\colorrgb#1{\color[rgb]{#1}}%
      \def\colorgray#1{\color[gray]{#1}}%
      \expandafter\def\csname LTw\endcsname{\color{white}}%
      \expandafter\def\csname LTb\endcsname{\color{black}}%
      \expandafter\def\csname LTa\endcsname{\color{black}}%
      \expandafter\def\csname LT0\endcsname{\color[rgb]{1,0,0}}%
      \expandafter\def\csname LT1\endcsname{\color[rgb]{0,1,0}}%
      \expandafter\def\csname LT2\endcsname{\color[rgb]{0,0,1}}%
      \expandafter\def\csname LT3\endcsname{\color[rgb]{1,0,1}}%
      \expandafter\def\csname LT4\endcsname{\color[rgb]{0,1,1}}%
      \expandafter\def\csname LT5\endcsname{\color[rgb]{1,1,0}}%
      \expandafter\def\csname LT6\endcsname{\color[rgb]{0,0,0}}%
      \expandafter\def\csname LT7\endcsname{\color[rgb]{1,0.3,0}}%
      \expandafter\def\csname LT8\endcsname{\color[rgb]{0.5,0.5,0.5}}%
    \else
      % gray
      \def\colorrgb#1{\color{black}}%
      \def\colorgray#1{\color[gray]{#1}}%
      \expandafter\def\csname LTw\endcsname{\color{white}}%
      \expandafter\def\csname LTb\endcsname{\color{black}}%
      \expandafter\def\csname LTa\endcsname{\color{black}}%
      \expandafter\def\csname LT0\endcsname{\color{black}}%
      \expandafter\def\csname LT1\endcsname{\color{black}}%
      \expandafter\def\csname LT2\endcsname{\color{black}}%
      \expandafter\def\csname LT3\endcsname{\color{black}}%
      \expandafter\def\csname LT4\endcsname{\color{black}}%
      \expandafter\def\csname LT5\endcsname{\color{black}}%
      \expandafter\def\csname LT6\endcsname{\color{black}}%
      \expandafter\def\csname LT7\endcsname{\color{black}}%
      \expandafter\def\csname LT8\endcsname{\color{black}}%
    \fi
  \fi
    \setlength{\unitlength}{0.0500bp}%
    \ifx\gptboxheight\undefined%
      \newlength{\gptboxheight}%
      \newlength{\gptboxwidth}%
      \newsavebox{\gptboxtext}%
    \fi%
    \setlength{\fboxrule}{0.5pt}%
    \setlength{\fboxsep}{1pt}%
\begin{picture}(3456.00,2880.00)%
\definecolor{gpBackground}{rgb}{1.000, 1.000, 1.000}%
\put(0,0){\colorbox{gpBackground}{\makebox(3456.00,2880.00)[]{}}}%
    \gplgaddtomacro\gplbacktext{%
      \csname LTb\endcsname%
      \put(490,432){\makebox(0,0)[r]{\strut{}$-1$}}%
      \put(490,1641){\makebox(0,0)[r]{\strut{}$0$}}%
      \put(490,2850){\makebox(0,0)[r]{\strut{}$1$}}%
      \put(622,212){\makebox(0,0){\strut{}$0$}}%
      \put(2021,212){\makebox(0,0){\strut{}$0.5$}}%
      \put(3420,212){\makebox(0,0){\strut{}$1$}}%
    }%
    \gplgaddtomacro\gplfronttext{%
      \csname LTb\endcsname%
      \put(182,2850){\rotatebox{0}{\makebox(0,0){\strut{}(b)}}}%
      \put(182,1641){\rotatebox{-270}{\makebox(0,0){\strut{}$\rho_{i}$}}}%
      \put(2021,-118){\makebox(0,0){\strut{}$y/h$}}%
      \put(2021,2740){\makebox(0,0){\strut{}}}%
    }%
    \gplbacktext
    \put(0,0){\includegraphics{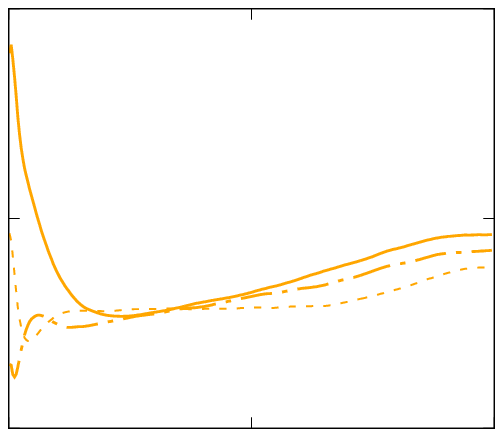}}%
    \gplfronttext
  \end{picture}%
\endgroup